\documentclass{iopjournal}

\usepackage{graphicx}
\usepackage{dcolumn}
\usepackage{float}
\usepackage{subcaption}

\usepackage{bm}
\usepackage{multirow}
\usepackage{hyperref}
\usepackage{booktabs}
\usepackage{siunitx}
\usepackage{tabularx}
\usepackage{caption} 

\begin{document}

\articletype{Paper} 

\title{Poissonian Analysis of Glitches Observed in the LIGO Gravitational Wave Interferometers }

\author{Giovanna Souza Rodrigues Costa$^1$\orcid{0000-0000-0000-0000}, Julio César Martins$^2$\orcid{0000-0000-0000-0000} and Odylio Denys Aguiar$^{2,*}$\orcid{0000-0000-0000-0000}}

\affil{$^1$Department of Astronomy, Institute of Astronomy, Geophysics and Atmospheric Sciences of University of São Paulo, São Paulo, Brazil}

\affil{$^2$Department of Astrophysics, Instituto Nacional de Pesquisas Espaciais (INPE), São José dos Campos, SP, Brazil}

\affil{$^*$Author to whom any correspondence should be addressed.}

\email{rodriguesgiovanna902@gmail.com}

\keywords{Gravitational Waves, Glitches, Poissonian Statistics, LIGO, LLO, LHO}

\begin{abstract}
This work investigates the temporal distribution of glitches detected by LIGO, focusing on the morphological classification provided by the Gravity Spy project. Starting from the hypothesis that these events follow a Poisson process, we developed a statistical methodology to evaluate the agreement between the empirical distribution of glitches and an ideal Poisson model, using the coefficient of determination (R²) as the main metric. The analysis was applied to real data from the LIGO detectors in Livingston and Hanford throughout the O3 run, as well as to synthetic datasets generated from pure Poisson distributions. The results show that while several morphologies exhibit good agreement with the proposed model, classes such as 1400Ripples, Fast Scattering, and Power Line display significant deviations (R² < 0.6), suggesting that their origins do not strictly follow Poissonian statistics. In some cases, a dependence on the detector or the observing run was also observed. This analysis provides a quantitative basis for distinguishing glitch classes based on their degree of "Poissonness", potentially supporting the development of more effective glitch mitigation strategies in gravitational wave detector data.
\end{abstract}

\section{\label{sec:level1} Introduction}

Gravitational waves arise as a direct consequence of the General Theory of Relativity \cite{Einstein1916,Einstein1918}, which describes gravity as a deformation of space-time caused by concentrations of energy and matter. When a mass distribution is accelerated, perturbations are triggered in the fabric of space-time that propagate throughout the Universe at the speed of light — being more intense in the case of compact objects, such as neutron stars or black holes, due to their concentration of large amounts of mass in a small volume. These perturbations allow us to study the most energetic and extreme phenomena in the cosmos. The detection of gravitational waves represents one of the most significant advances in modern astrophysics, ushering in a new era of observing the universe.

Since the first detection by the Laser Interferometer Gravitational-Wave Observatory (LIGO) in September 2015 \cite{PhysRevLett.116.061102}, multiple extreme cosmic events, such as mergers of black holes and/or neutron stars, have been observed, with the end of the third observational run (O3) bringing the number of gravitational wave detections to 90 detections \cite{ligo2024} \cite{Abbott:2024} \cite{Abbott:2023}. This offers unique opportunities to explore high-energy phenomena and to test the limits of the General Theory of Relativity \cite{generalrelativity} under extreme conditions (strong-field and relativistic velocities).

However, the statistical significance of these signals is often challenged by non-Gaussian transient noise events known as glitches \cite{glanzer2023data} \cite{astroglitchstad341}. These aperiodic events, captured by the interferometers, can mask or even mimic real gravitational wave signals, reducing the statistical confidence of detections and increasing the false alarm rate \cite{Buikema_2020}. As a result, it becomes essential to investigate the origin, nature, and distribution of these glitches in order to minimize their impact on observational data. This phenomenon has already been widely investigated and characterized in several studies in the literature \cite{arnaud2022ligovirgodetectorcharacterizationdata}, aiming to understand their mechanisms and develop strategies for their mitigation.

The Poisson distribution models the probability of occurrence of discrete and statistically independent events within a fixed interval, under the assumption of a constant average rate. This distribution is widely used to describe processes such as photon detection, radioactive decay, and shot noise, the latter being particularly relevant in optical and electronic systems such as the gravitational-wave interferometers used by LIGO \cite{shotnoise}.

In contrast, many environmental or instrumental noise phenomena deviate from the Poissonian model. Thermal noise, for instance, arises from thermodynamic fluctuations within materials and is typically modeled as Gaussian white noise. Its continuous nature is justified by the Central Limit Theorem, which states that the sum of many independent random contributions tends to converge toward a normal distribution. For this reason, in radio-frequency (RF) communication systems, background noise is predominantly treated as a continuous Gaussian process \cite{PerezCruz2008}.

However, the distinction between discrete and continuous domains becomes subtler in the context of data sampling. When an intrinsically continuous noise process, such as thermal noise, is acquired by a data acquisition system, it is converted into a sequence of discrete samples. Under these conditions, the process can be interpreted as the counting of energy fluctuations within each sampling interval. Consequently, thermal-origin glitches in interferometers may be associated with significant energetic variations of the noise within the acquisition window. Thus, a continuous physical process, when observed through discrete sampling, may exhibit Poisson-like statistical features, potentially altering data analysis and interpretation.

This behavior was observed in the case of resonant-bar gravitational-wave detectors \cite{Hamilton:1988mh}. Under ideal conditions where thermal noise was the dominant source, the event listings approximated a Poissonian statistical pattern. The experience gained from the analysis of resonant-bar detector data is now being applied to the study of glitches in the LIGO interferometers.

This study addresses the Poissonian analysis of glitches recorded during the third observational run of the LIGO interferometers \cite{Davis_2021}, with the goal of characterizing their temporal distribution and identifying patterns that may provide clues about their physical causes. In this context, the use of computational methods and detailed statistical analyses enables a quantitative approach to assess whether certain types of glitches follow a Poisson distribution, offering insights into their origin and potentially assisting in the mitigation of such noise.

\section{Methodology}
It is essential to characterize the LIGO interferometers and monitor their behavior regarding the occurrence of glitches. The classification of these glitches can provide important clues about their potential sources, especially when morphological similarities between events are observed. One such classification methodology was developed by Gravity Spy \cite{glanzer2023data} \cite{Zevin2024}, which employs a machine learning model using time–frequency spectrograms to categorize transient noise events into 23 predefined morphological classes. The model relies on inputs generated by the Omicron pipeline \cite{omicron}, which detects glitches by identifying excess power relative to the background noise in a multi-resolution spectrogram. While Omicron itself can cover a broader frequency band and characterizes events through Q-transform spectrograms, Gravity Spy restricts its analysis to Omicron triggers within the 10–2048 Hz range. For this study, only glitches classified by Gravity Spy with at least 90\% confidence were selected.

In order to study the most recent versions of the publicly available data \cite{opendata}, glitches from the O3a and O3b observational runs - the first and second halves of the LIGO interferometers third observing run - were analyzed. The runs were treated separately to avoid mixing distinct experimental conditions, which could compromise the consistency of the results.

\subsection{Statistical Model} 

To characterize the temporal behavior of glitches, we used the Poisson distribution \cite{Durran1970}, as described in Equation 1, an approach commonly employed to model events that occur independently over time. The choice of this distribution is based on the hypothesis that, if the probability of glitch occurrence is constant within any observation interval, their occurrence rate will naturally follow a Poisson distribution. If a given glitch morphology shows significant deviations from this expected behavior, it suggests that its underlying source likely does not follow a Poissonian process \cite{Hamilton1987Gravitational}.

\begin{equation}
N = e^{-\mu} \cdot \frac{\mu^r}{r!} \cdot G
\end{equation}

Where in the equation above:

\begin{itemize}
    \item \(N\) is the expected number of intervals (or time boxes) containing exactly \(r\) glitches;
    \item \(\mu\) represents the expected number of glitches per box, calculated as \(\mu = \lambda \cdot t_0\);
    \item \(\lambda\) is the average glitch rate per unit time, defined as
    \[
    \lambda = \frac{\text{total number of glitches}}{\text{total sample time}};
    \]
    \item \(t_0\) is the size of the time interval (box) used to segment the data;
    \item \(r\) is the number of glitches contained in a specific box;
    \item \(G\) is the total number of glitches observed in the analyzed data.
\end{itemize}

Initially, the data were divided into time boxes of size \(t_0\). For each box, the number of observed glitches (\(r\)) was counted and their occurrence frequency was recorded \cite{Evans1955}.

One of the conditions to be addressed in the analysis concerns the consideration of the interferometer’s duty cycle. That is, to ensure the robustness of the analysis, only observation intervals during which the interferometer was effectively collecting data were considered. Thus, only continuous data segments that could be divided into complete boxes of \(t_0\) seconds were selected. To ensure statistical consistency, any residual intervals shorter than \(t_0\) at the end of a segment were discarded, preserving only integer multiples of \(t_0\) for analysis. Periods when the interferometer was not collecting valid data were also excluded. This filtering consequently implies that some glitch data present in the discarded time segments were rendered unusable; however, this approach is necessary to ensure that data segmentation respects statistical integrity.

To ensure that the distribution correctly described the probability of glitch occurrence in each interval \(r\), the Poisson distribution was defined such that the sum of \(P(r)\) over all boxes equaled 1. This resulted in Equation 2 below.

\begin{equation}
P(r) = e^{-\mu} \cdot \frac{\mu^r}{r!} \cdot 
\end{equation}

Thus, a Python script was developed to perform this analysis. The script implements the segmentation of the data into time boxes of size \(t_0\), calculates the average glitch rate \(\lambda\), and applies the Poisson distribution to model the expected behavior of the events, allowing comparison between the observed data and the fitted theoretical values. This enables the identification of potential statistical deviations from the model.

\subsection{Coefficient of Determination}

To evaluate the goodness-of-fit of the Poisson distribution to the observed data, we used the coefficient of determination \(R^2\), as shown in Equation 3. The model was tested using both real and synthetic data, the latter generated using a Poisson function from the NumPy library \cite{nelli2023numpy}. These synthetic data served as a control, ensuring model validation in an ideal scenario where the Poisson distribution was guaranteed under the same conditions—total observation time, glitch occurrence rate, and time box size \(t_0\).

\begin{equation}
R^2 = 1 - \frac{\sum \left( N_{\text{obs}} - N_{\text{exp}} \right)^2}{\sum \left( N_{\text{obs}} - \bar{N}_{\text{obs}} \right)^2}
\end{equation}

where:

\begin{itemize}
    \item \( N_{\text{obs}} \) represents the number of glitches observed in each time box;
    \item \( N_{\text{exp}} \) is the expected number of glitches per box, according to the Poisson model;
    \item \( \bar{N}_{\text{obs}} \) corresponds to the average number of glitches observed across all time boxes;
    \item The numerator \( \sum (N_{\text{obs}} - N_{\text{exp}})^2 \) measures the sum of squared errors between the observed values and those predicted by the model, indicating the fitting error;
    \item The denominator \( \sum (N_{\text{obs}} - \bar{N}_{\text{obs}})^2 \) represents the total variability of the data relative to the observed mean;
\end{itemize}

In this case, values of the coefficient of determination \(R^2\) close to 1 indicate that the fitted model explains nearly all the variation observed in the data. This suggests a strong agreement between the observed values \(N_{\text{obs}}\) and the expected values \(N_{\text{exp}}\) from the Poisson distribution, implying that the behavior of the glitches is well represented by the proposed model. On the other hand, significantly lower \(R^2\) values indicate discrepancies, suggesting that the Poisson distribution may not be suitable for describing certain data sets or specific glitch morphologies.

\subsection{Determination of Box Size}

\begin{figure}[H]
    \centering
    \begin{subfigure}{0.48\linewidth}
        \centering
        \includegraphics[width=\linewidth]{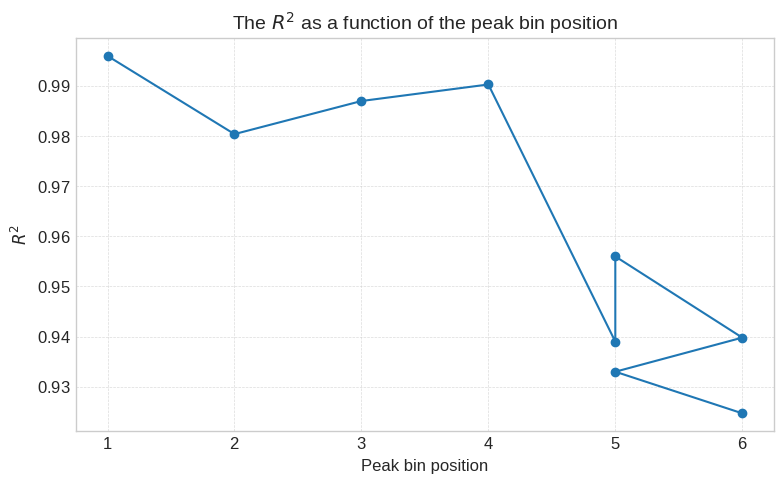}
        \caption{Variation of \(R^2\) with peak bin position in the Poisson distribution.}
        \label{analisecaixas}
    \end{subfigure}
    \hfill
    \begin{subfigure}{0.48\linewidth}
        \centering
        \includegraphics[width=\linewidth]{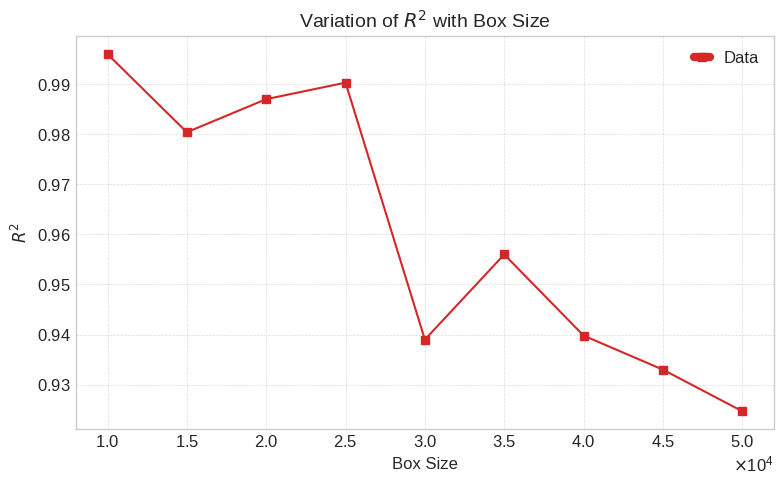}
        \caption{Variation of \(R^2\) in terms of the box length \(t_0\) in seconds.}
        \label{analise2}
    \end{subfigure}
    \caption{Behavior of \(R^2\) as a function of the Poisson peak position and box length \(t_0\).}
    \label{fig:r2analysis}
\end{figure}

The appropriate size to be used for the boxes should not be arbitrary; it must be large enough to adequately capture the counting statistics, but not so large as to mask relevant fluctuations. Therefore, an analysis of the behavior and variation of \(R^2\) was carried out as a function of the position of the peak bin of the Poisson curve, based on synthetic data for each glitch morphology for different values to \(t_0\). With this, it was possible to develop Figures \ref{analisecaixas} and \ref{analise2}, showing that the coefficient of determination becomes more sensitive as the peak of the Poisson distribution approaches \(r=1\).

Thus, according to the figures above, in order to maximize the representativeness of the comparison, we seek the regime in which the probability of observing a single event, that is, the bin with (\(r=1\)), is maximal, such that only the bin with (\(r=0\)) lies to the left of the peak. This is achieved when the derivative of the Poisson function is zero at \(r=1\).

By calculating the derivative of \( P(r; \mu) \), treating \( r \) as a continuous variable, we obtain:

\[
\frac{dP}{dr} = P(r; \mu) \left( \ln(\mu) - \psi(r+1) \right)
\]

where \( \psi \) is the digamma function, defined as the derivative of the logarithm of the gamma function.

Setting the derivative to zero at \( r = 1 \) yields:

\[
\ln(\mu) - \psi(2) = 0 \quad \Rightarrow \quad \mu = e^{\psi(2)}
\]

Using the known value \( \psi(2) \approx 0.42278 \), we obtain:

\[
\mu \approx 1.526
\]

Thus, the expected number of events per box should be approximately \( \mu \approx 1.526 \) to optimize the fit to a Poisson distribution centered around one event per interval.

Finally, since \( \mu = \lambda \times t_0 \), the ideal box size \( t_0 \) is calculated as shown in Equation \ref{eq:derivada_poisson}.

\begin{equation}
    t_0 = \frac{1.526}{\lambda}
    \label{eq:derivada_poisson}
\end{equation}

This process therefore indicates a choice of interval that tends to maximize the quality of the fit to the Poisson model, which is reflected in a high value of the coefficient of determination \( R^2 \), thus providing a more accurate assessment of the Poisson nature of the glitches.

\section{Glitch Morphologies}

The dataset analyzed contains all classifications performed by the Gravity Spy machine learning model for glitches in the LIGO interferometers during the O3a and O3b observation runs \cite{sec3}. Gravity Spy classified all noise events identified by the Omicron trigger pipeline, considering events with a signal-to-noise ratio greater than 7.5 and a peak frequency between 10 Hz and 2048 Hz.

The analyzed glitch classes are the 23 classes considered by Gravity Spy, namely: 1080Lines, 1400Ripples, Air Compressor, Blip, Blip Low Frequency, Chirp, Extremely Loud, Helix, Fast Scattering, Koi Fish, Light Modulation, Low Frequency Burst, Low Frequency Lines, No Glitch, Paired Doves, Power Line, Repeating Blips, Scattered Light, Scratchy, Tomte, Violin Mode, Wandering Line, and Whistle.

The data used and the pipeline, as well as the \textit{Omega scans} of each glitch, can be obtained through the GWpy\cite{MACLEOD2021100657} tool for gravitational-wave data analysis. The available CSV files contain detailed metadata of the events, including time of occurrence, corresponding interferometer, peak frequency, duration, amplitude, and signal-to-noise ratio, as well as the classification probabilities assigned by the Gravity Spy machine learning model.

It is worth noting that some glitch morphologies do not have sufficient data for the statistical distribution analyzed to exhibit a well-defined behavior. In such cases, the estimation of the coefficient of determination \( R^2 \) may be compromised due to the small sample size, impacting the validity of the modeling through Poisson fitting. Therefore, to ensure that the relative uncertainty is less than or equal to 5\%, we use the formula for the relative uncertainty associated with event counting:

\[
\frac{\sigma}{\mu} = \frac{1}{\sqrt{N}} \leq 0.05
\]

Squaring both sides:

\[
\frac{1}{N} \leq 0.0025 \quad \Rightarrow \quad N \geq \frac{1}{0.0025} = 400
\]

Therefore, a minimum of \( N \geq 400 \) events is required for the statistical uncertainty associated with the count to be less than or equal to 5\%.

Among the morphologies categorized by Gravity Spy, the 1400Ripples are observed mainly in Livingston, with durations of about 0.05 seconds or longer, occurring around 1400 Hz and appearing either isolated or multiple within the same time image.

Glitches at 50 Hz are also common, such as the Air Compressor glitches, which were related to air compressor motors at LIGO Hanford. These events were almost completely resolved on September 29, 2015, when the compressors' vibration isolators were replaced \cite{berger2018identification}.

Blip glitches are short and wide pulses in time, with frequencies between 30 Hz and 300 Hz. In spectrograms, they appear as isolated, concentrated drops, similar to Blip Low Frequency glitches. The main difference lies in the frequency range: as the name suggests, Blip Low Frequency glitches occur at lower frequencies. \cite{cabero2019blip}.

Extremely Loud glitches are high-intensity events that span wide frequency bands. They occur due to disturbances in the detector, such as an actuator reaching its range limit or a photo-diode saturating \cite{gravityloud}.

Fast Scattering glitches occur when a small fraction of stray light strikes a moving surface, gets reflected back towards the point of scattering, and rejoins the main laser beam \cite{soni2024modeling}.

Koi Fish glitches, in turn, resemble a fish with a head at the low-frequency end and a thin tail that can reach frequencies above 256 Hz. They are believed to be a subclass of Blip glitches, but their origin is still not understood.

Low Frequency Bursts are bursts of noise below 30 Hz, generally associated with seismic movements and atmospheric fluctuations that affect the interferometer’s stability.

Low Frequency Lines appear as horizontal lines at low frequencies and can be confused with Scattered Light glitches, but scattered light usually shows some curvature on the timescale.

When a spectrogram shows no visible anomaly, it is classified as No Glitch.

Another type of glitch is the Power Line glitch, caused by the 60 Hz alternating current from the U.S. power grid. These events occur when large equipment, such as heaters or compressors, turns on or off, generating oscillations detectable in the spectrogram.

Repeating Blips are Blips that repeat at regular intervals of 0.25 or 0.5 seconds. Scattered Light glitches appear as arches in the time-frequency spectrograms \cite{soni2020reducing}. This type of interference can generate signals that resemble real gravitational waves.

Tomte glitches resemble Blip glitches but tend to occur at slightly higher frequencies and are more common at LIGO Hanford.

Finally, Whistle glitches appear as curved traces with increasing frequency over time. These glitches typically last a few seconds and appear in higher frequency bands \cite{glanzer2023data}.

Additionally, the remaining types of glitches were not analyzed in this work due to the scarcity of data in the O3a and O3b runs, both at LIGO Livingston and LIGO Hanford, making it unfeasible to obtain statistically reliable models for them.

\begin{itemize}
    \item \textbf{1080Lines}: observed at LIGO Hanford during the O2 run, they appeared as points near 1080 Hz. After January 2017, these events became rare due to adjustments in the control system.

    \item \textbf{Chirp Glitches}: gravitational waves from the coalescence of binary compact objects, or injected simulations of these. \cite{gravitychirp}

    \item \textbf{Paired Doves}: low-frequency glitches with oscillatory variation, possibly linked to the movement of the \textit{beamsplitter} at Hanford.

    \item \textbf{Helix Glitches}: have a helical pattern and may be related to failures in auxiliary calibration lasers.

    \item \textbf{Light Modulation Glitches}: exhibit periodic oscillations in signal intensity.

    \item \textbf{Scratchy Glitches}: noise with a wavy pattern, frequent at LIGO Hanford and known to have limited the detection of mergers during the early observing runs.

    \item \textbf{Violin Mode Glitches}: appear as signals centered around 500 Hz and its harmonics.

    \item \textbf{Wandering Line Glitches}: sinuous lines in spectrograms, associated with mechanical vibrations such as motors affecting the optical system.
\end{itemize}

\begin{figure*}[htbp]
    \centering
    \includegraphics[width=0.99\linewidth]{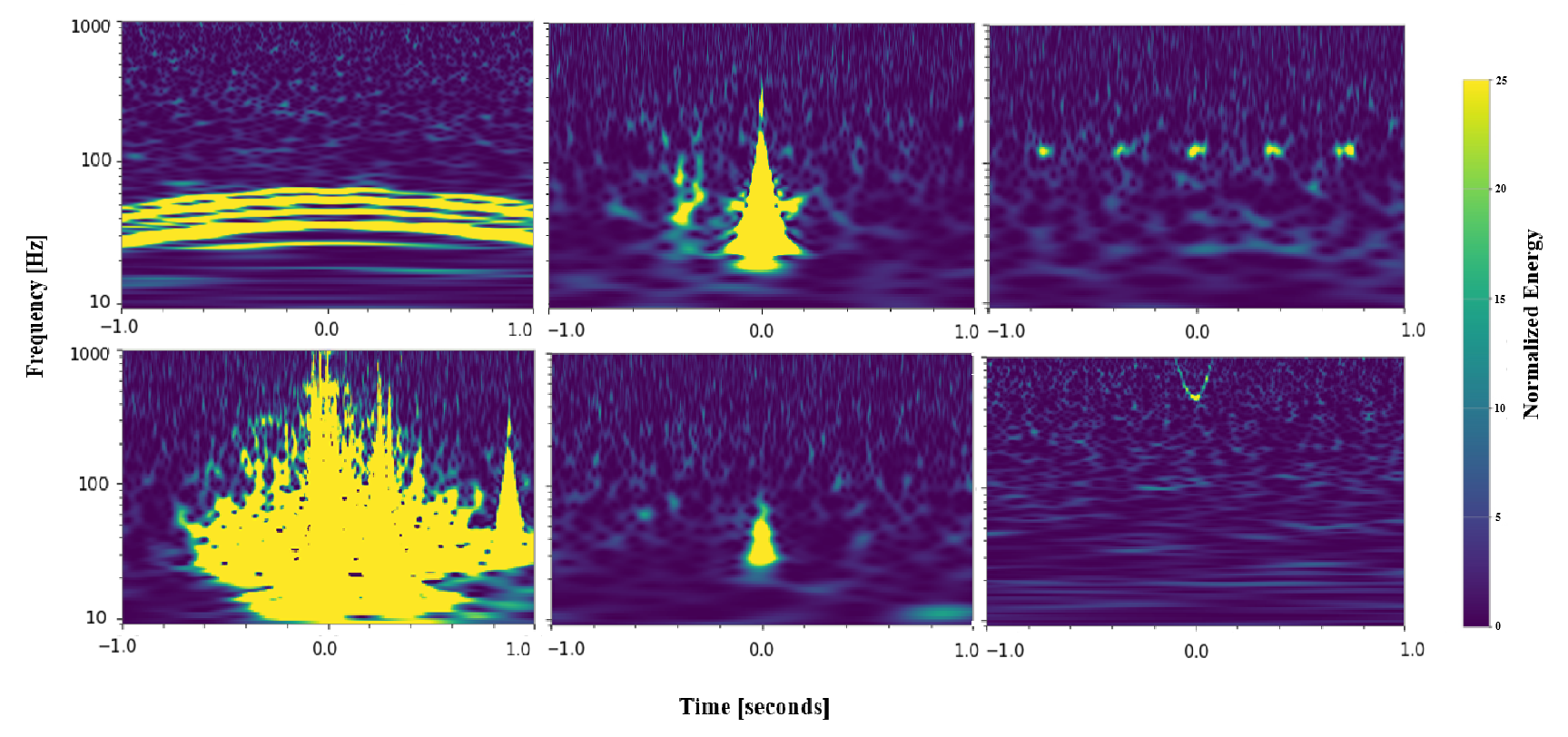}
    \caption{Spectrograms of some of the analyzed glitch morphologies, corresponding respectively to the classes Scattered Light, Koi Fish, Scratchy, Extremely Loud, Tomte and Whistle.}
    \label{compilado}
\end{figure*}

Figure \ref{compilado} shows examples of spectrograms of some of the morphologies mentioned above.

\section{Input Data}
The data used exhibit a distinct frequency structure for each glitch morphology. Below are plots for each of them, where on the left side, there is a scatter plot showing the temporal distribution of glitches with confidence greater than 90\%. The horizontal axis represents GPS time, while the vertical axis indicates the central frequency of the glitches. On the right side, a histogram displays the count of glitches in different central frequency ranges for the O3a run of the Livingston interferometer, allowing visualization of which frequency intervals concentrate the majority of events.

\begin{figure}[H]
    \centering

    \begin{subfigure}{0.48\linewidth}
        \centering
        \includegraphics[width=\linewidth]{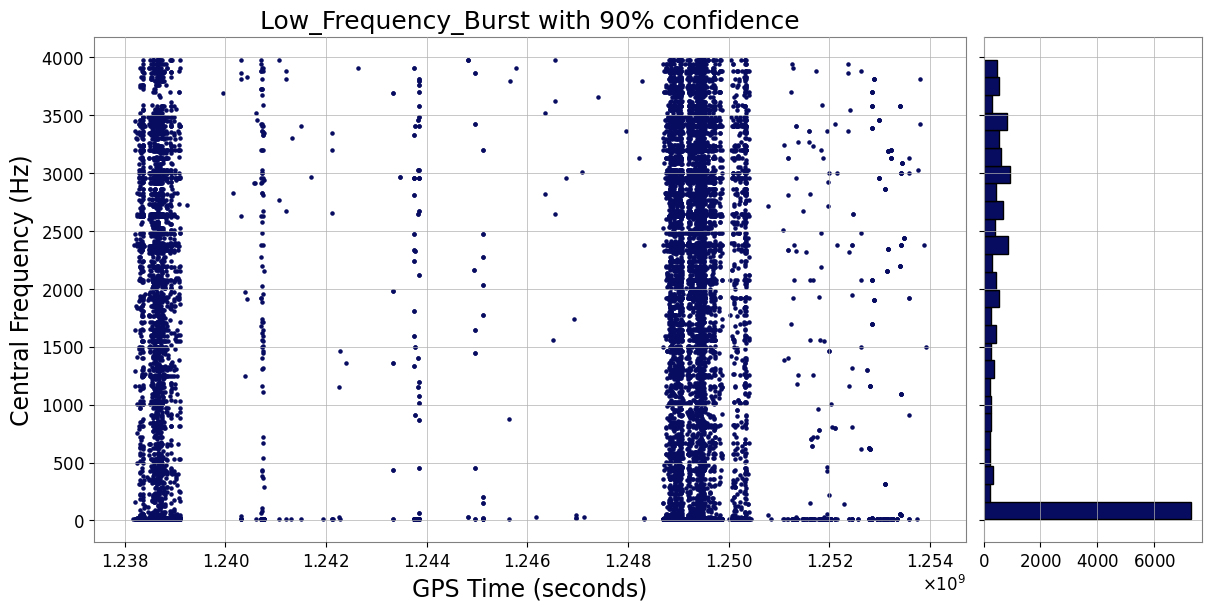}
        \caption{Low Frequency Burst glitch data from the Hanford interferometer (O3a).}
        \label{burst}
    \end{subfigure}
    \hfill
    \begin{subfigure}{0.48\linewidth}
        \centering
        \includegraphics[width=\linewidth]{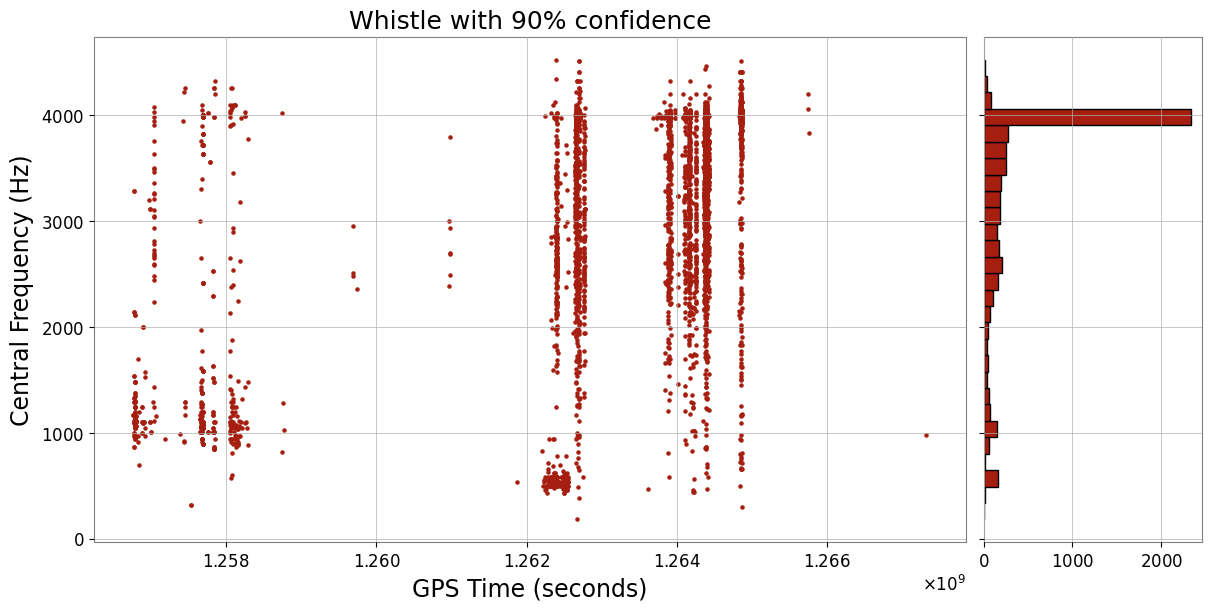}
        \caption{Whistle glitch data from the Livingston interferometer (O3b).}
        \label{whistle}
    \end{subfigure}

    \vspace{0.5cm} 

    \begin{subfigure}{0.48\linewidth}
        \centering
        \includegraphics[width=\linewidth]{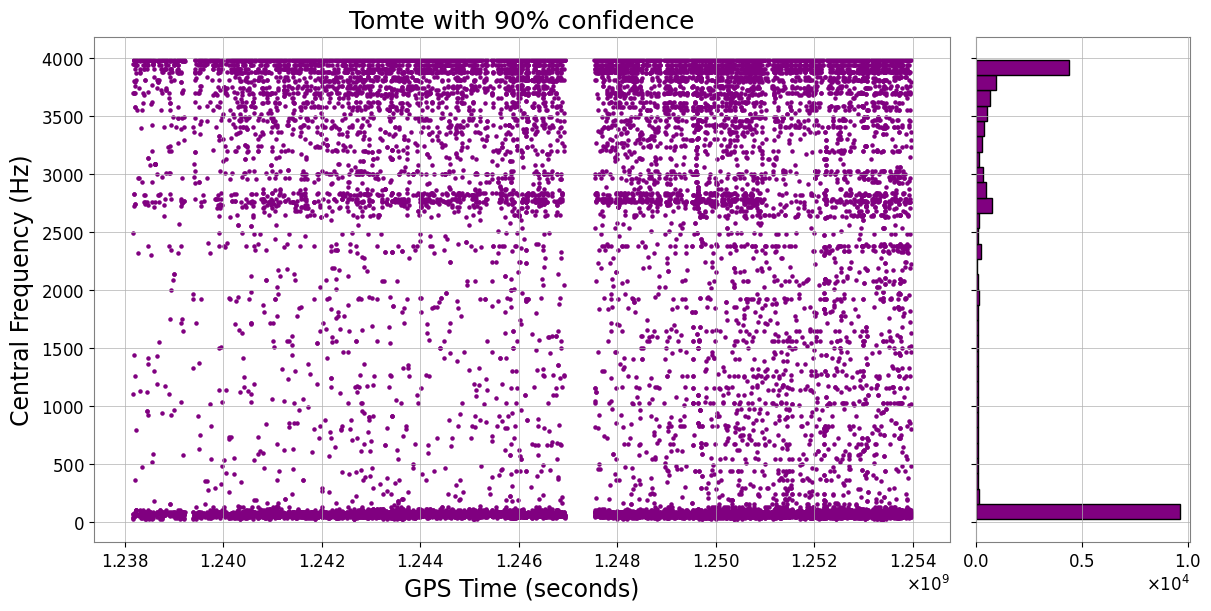}
        \caption{Tomte glitch data from the Livingston interferometer (O3a).}
        \label{tomte}
    \end{subfigure}
    \hfill
    \begin{subfigure}{0.48\linewidth}
        \centering
        \includegraphics[width=\linewidth]{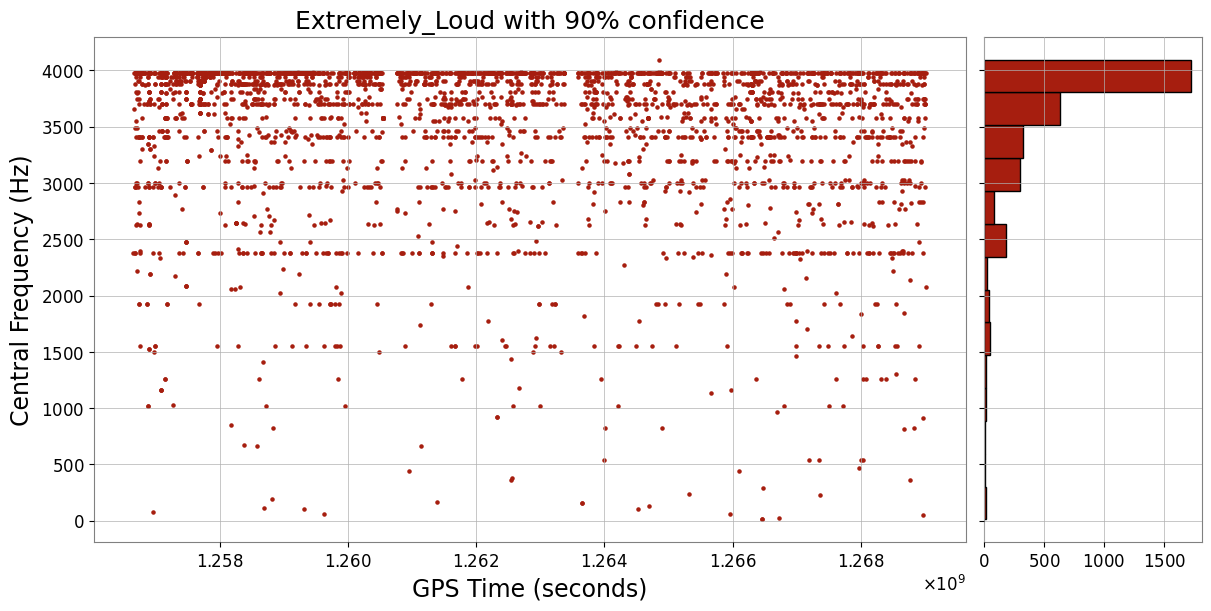}
        \caption{Extremely Loud glitch data from the Livingston interferometer (O3b).}
        \label{loud}
    \end{subfigure}

    \caption{Frequency distributions of selected glitch morphologies observed in different interferometers and observing runs.}
    \label{fig:freq_glitches}
\end{figure}

The Figures \ref{burst} e \ref{whistle} from Low Frequency Burst and Whistle categories show well-defined intervals with a noticeably higher occurrence rate, indicating the presence of non-uniform temporal clustering in these glitches.

In contrast, the Extremely Loud and Tomte categories display an approximately uniform occurrence rate throughout the entire observation period, without clear trends or concentrations in specific intervals, as shown in Figures \ref{loud} and \ref{tomte}.

These lines may provide useful information about environmental or instrumental conditions around the interferometer, helping to understand what caused the number of glitches observed during this time interval.

The other graphs developed for O3a and O3b of the Livingston and Handford interferometers for these and other glitches morphologies can be found in the Appendix \ref{dataapendice}.

\section{Analysis and Results}

In this section, we present the analysis of glitches detected during the O3a and O3b observation runs of the LIGO interferometers. The analysis was carried out considering data from both detectors, LIGO Hanford (H1) and LIGO Livingston (L1).

The analysis yields histograms of the temporal distribution of glitches compared to the frequency of observed events, with the expectation based on the Poisson statistical model. The observed curves are shown in blue, and the expected Poisson distributions for each glitch are indicated in red. The agreements or discrepancies between the observed data and the proposed statistical model are quantitatively assessed using the correlation coefficient, \(R^2\).

\begin{figure}[H]
    \centering
    \begin{subfigure}[b]{0.48\linewidth}
        \centering
        \includegraphics[width=\linewidth]{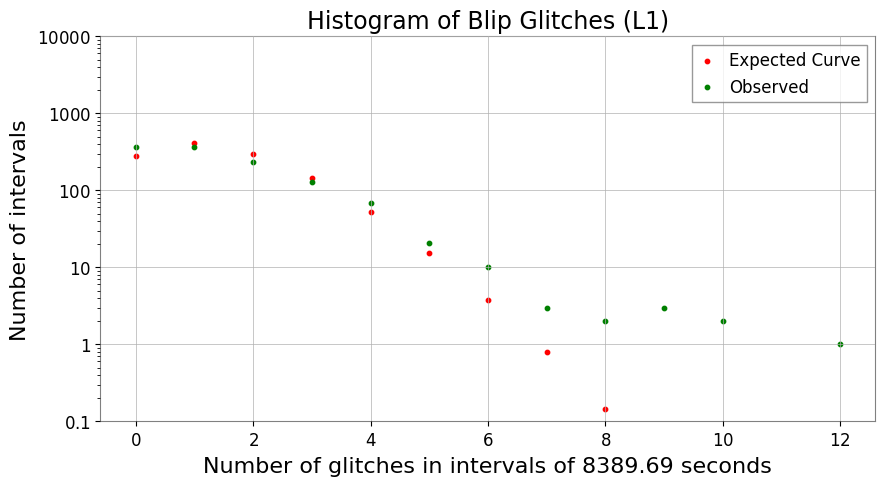}
        \caption{Blip glitches in intervals of 8,389.69 seconds during O3a in the L1 detector.}
        \label{blipo3aliv}
    \end{subfigure}
    \hfill
    \begin{subfigure}[b]{0.48\linewidth}
        \centering
        \includegraphics[width=\linewidth]{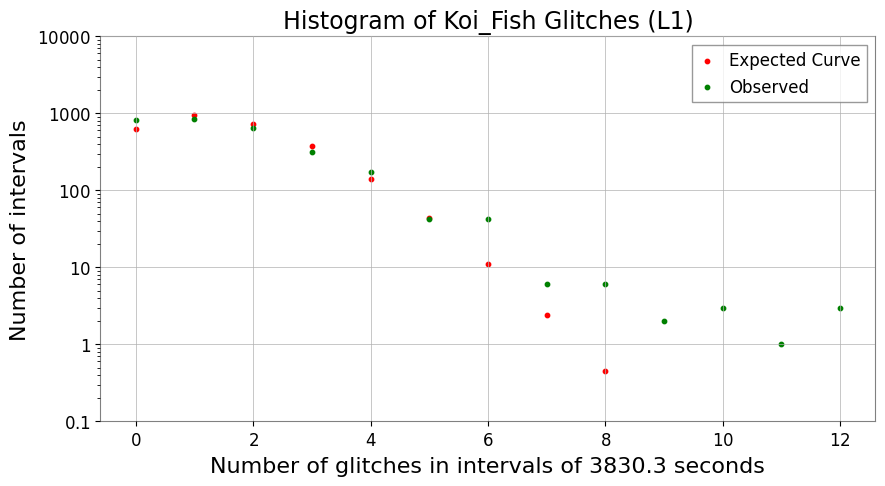}
        \caption{Koi Fish glitches in intervals of 3,830.30 seconds during O3a in the L1 detector.}
        \label{koio3aliv}
    \end{subfigure}
    \caption{Histograms of Blip and Koi Fish glitches during O3a in the L1 detector.}
    \label{allhistograms}
\end{figure}

To illustrate the results of the statistical classification, we present representative cases of both Poisson-like and non-Poisson-like behaviors.
Among the morphologies studied, some classes follow the expected Poissonian distribution. The Blip glitches (Figure \ref{blipo3aliv}) yield coefficients of determination above 0.9 across multiple detectors and observing runs, values that are consistent with the synthetic simulations. This same behavior is observed for other glitch classes, such as Koi Fish (Figure \ref{koio3aliv}), which also exhibit high \(R²\) values.

\begin{figure}[H]
    \centering
    \begin{subfigure}[b]{0.48\linewidth}
        \centering
        \includegraphics[width=\linewidth]{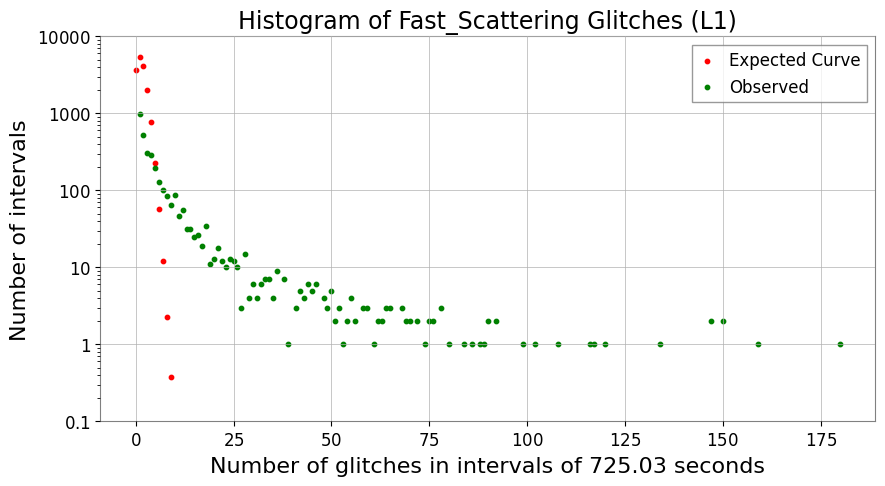}
        \caption{Fast Scattering glitches in intervals of 725.03 seconds during O3a (L1).}
        \label{fasto3aliv}
    \end{subfigure}
    \hfill
    \begin{subfigure}[b]{0.48\linewidth}
        \centering
        \includegraphics[width=\linewidth]{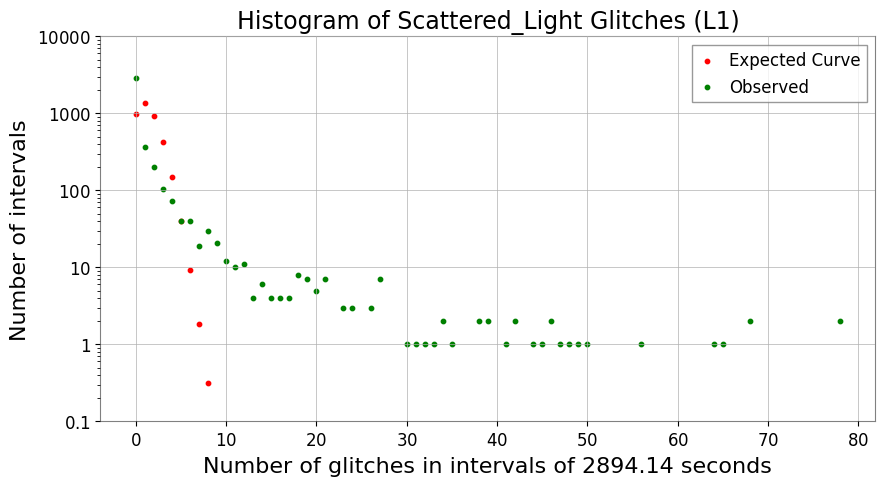}
        \caption{Scattered Light glitches in intervals of 2,894.14 seconds during O3a (L1).}
        \label{scatteredo3aliv}
    \end{subfigure}
    \caption{Histograms of Fast Scattering and Scattered Light glitches during O3a in the L1 detector.}
    \label{fast_scattered_histograms}
\end{figure}

In contrast, Fast Scattering (Figure \ref{fasto3aliv}) and Scattered Light glitches (\ref{scatteredo3aliv}) provide a clear example of a class with strong deviations from Poisson statistics, with \(R^2\) values well around 0.3 despite simulations predicting near-ideal Poissonian behavior.

A summary of the results obtained was compiled in the Tables \ref{tab:resumo_glitches_L1} and \ref{tab:resumo_glitches_H1} below, describing, for each run, interferometer and glitch morphology, the number of events considered, size of the temporal box used \(t_0\) and the observed and synthetically generated \(R^2\) values.

\begin{table*}[htbp]
\centering
\caption {Summary of glitch morphologies for the Livingston interferometer during the O3a and O3b runs.}
\label{tab:resumo_glitches_L1}
\setlength\tabcolsep{4pt}
\renewcommand{\arraystretch}{1.1}
\scriptsize
\resizebox{0.9\textwidth}{!}{%
\begin{tabularx}{\textwidth}{>{\raggedright\arraybackslash}X
 S[table-format=5.0] S[table-format=5.2] S[table-format=1.3] S[table-format=1.3]
 S[table-format=5.0] S[table-format=5.2] S[table-format=1.3] S[table-format=1.3]}
\toprule
 & \multicolumn{4}{c}{\textbf{O3a}} & \multicolumn{4}{c}{\textbf{O3b}}\\
\cmidrule(lr){2-5}\cmidrule(lr){6-9}
\textbf{Glitch Type} & {\(N\)} & {\textbf{\(t_0\)}} & {\(\mathbf{R^2}\)} & {\(\mathbf{R^2_{\text{sint}}}\)} 
                & {\(N\)} & {\textbf{\(t_0\)}} & {\(\mathbf{R^2}\)} & {\(\mathbf{R^2_{\text{sint}}}\)}\\
\midrule
1400Ripples           & 2248  & 6935.60 & 0.508 & 0.995 & \multicolumn{1}{c}{--} & \multicolumn{1}{c}{--} & \multicolumn{1}{c}{--} & \multicolumn{1}{c}{--}\\
Air Compressor        & \multicolumn{1}{c}{--} & \multicolumn{1}{c}{--} & \multicolumn{1}{c}{--} & \multicolumn{1}{c}{--} & 785   & 15310.14 & 0.200 & 0.985\\
Blip                  & 1763  & 8389.69 & 0.940 & 0.998 & 2240  & 5839.82  & 0.932 & 0.996\\
Blip Low Frequency    & 7820  & 2194.27 & 0.957 & 0.999 & 8286  & 1715.54  & 0.903 & 0.999\\
Extremely Loud        & 4362  & 3785.80 & 0.970 & 0.994 & 2796  & 4326.29  & 0.977 & 0.995\\
Fast Scattering       & 24325 & 725.03  & 0.269 & 0.999 & 35816 & 405.58   & 0.277 & 0.999\\
Koi Fish              & 4414  & 3830.30 & 0.953 & 0.998 & 2667  & 4974.52  & 0.952 & 0.995\\
Low Frequency Burst   & 841   & 11949.95& 0.562 & 0.974 & 2444  & 5131.36  & 0.602 & 0.996\\
Low Frequency Lines   & 488   & 20665.43& 0.552 & 0.989 & 2749  & 4693.82  & 0.605 & 0.998\\
No Glitch             & 5657  & 3010.65 & 0.490 & 0.999 & 956   & 7181.44  & 0.675 & 0.999\\
Power Line            & 1014  & 14069.06& 0.204 & 0.991 & \multicolumn{1}{c}{--} & \multicolumn{1}{c}{--} & \multicolumn{1}{c}{--} & \multicolumn{1}{c}{--}\\
Scattered Light       & 5355  & 2894.14 & 0.370 & 0.999 & 33471 & 335.10   & 0.481 & 0.999\\
Tomte                 & 20411 & 875.43  & 0.929 & 0.999 & 21771 & 669.61   & 0.901 & 0.999\\
Whistle               & 457   & 18265.84& 0.623 & 0.999 & 3853  & 2782.62  & 0.379 & 0.999\\
\bottomrule
\end{tabularx}}
\end{table*}

\begin{table*}[htbp]
\centering
\caption{Summary of glitch morphologies for the Hanford interferometer during the O3a and O3b runs.}
\label{tab:resumo_glitches_H1}
\setlength\tabcolsep{4pt}
\renewcommand{\arraystretch}{1.1}
\scriptsize
\resizebox{0.9\textwidth}{!}{%
\begin{tabularx}{\textwidth}{>{\raggedright\arraybackslash}X
 S[table-format=5.0] S[table-format=5.2] S[table-format=1.3] S[table-format=1.3]
 S[table-format=5.0] S[table-format=5.2] S[table-format=1.3] S[table-format=1.3]}
\toprule
 & \multicolumn{4}{c}{\textbf{O3a}} & \multicolumn{4}{c}{\textbf{O3b}}\\
\cmidrule(lr){2-5}\cmidrule(lr){6-9}
\textbf{Glitch Type} & {\(N\)} & {\textbf{\(t_0\)}} & {\(\mathbf{R^2}\)} & {\(\mathbf{R^2_{\text{sint}}}\)} 
                & {\(N\)} & {\textbf{\(t_0\)}} & {\(\mathbf{R^2}\)} & {\(\mathbf{R^2_{\text{sint}}}\)}\\
\midrule
Blip                 & 3778  & 4230.78  & 0.909 & 0.999 & 3055  & 4460.99  & 0.887 & 0.998\\
Blip Low Frequency   & 1102  & 12186.47 & 0.920 & 0.971 & 1572  & 7890.03  & 0.698 & 0.997\\
Extremely Loud       & 7866  & 1711.68  & 0.916 & 0.999 & 5046  & 2714.48  & 0.962 & 0.998\\
Fast Scattering      & \multicolumn{1}{c}{--} & \multicolumn{1}{c}{--} & \multicolumn{1}{c}{--} & \multicolumn{1}{c}{--} & 934   & 11665.63 & 0.322 & 0.998\\
Koi Fish             & 6675  & 2408.49  & 0.922 & 0.994 & 3420  & 4100.26  & 0.963 & 0.995\\
Low Frequency Burst  & 17656 & 943.20   & 0.209 & 0.999 & 1502  & 8070.01  & 0.290 & 0.997\\
Low Frequency Lines  & 673   & 18099.36 & 0.364 & 0.999 & 695   & 14943.00 & 0.504 & 0.991\\
No Glitch            & 5048  & 3063.51  & 0.315 & 0.999 & \multicolumn{1}{c}{--} & \multicolumn{1}{c}{--} & \multicolumn{1}{c}{--} & \multicolumn{1}{c}{--}\\
Repeating Blips      & 508   & 20779.12 & 0.850 & 0.991 & \multicolumn{1}{c}{--} & \multicolumn{1}{c}{--} & \multicolumn{1}{c}{--} & \multicolumn{1}{c}{--}\\
Scattered Light      & 8468  & 1814.53  & 0.279 & 0.999 & 695   & 258.97   & 0.236 & 0.999\\
Tomte                & 406   & 24600.57 & 0.851 & 0.943 & 1094  & 11210.01 & 0.503 & 0.994\\
Whistle              & 5325  & 2737.76  & 0.294 & 0.999 & 734   & 12092.19 & 0.415 & 0.998\\
\bottomrule
\end{tabularx}
}
\end{table*}

In addition, the other histograms developed for O3a and O3b of the Livingston and Handford interferometers for these and other glitches morphologies can be found in the Appendix \ref{histo}.

\section{Conclusion}

The statistical analysis of different types of glitches indicates that only a fraction of them are compatible with a Poisson distribution, suggesting stochastic behavior consistent with random physical sources such as thermal noise, quantum background fluctuations and shot noise. Glitches of the types Blip, Koi Fish, and Extremely Loud exhibited temporal patterns consistent with a constant occurrence rate, thus agreeing with the Poisson distribution. This statistical adherence suggests that their causes are directly related to processes that are also Poissonian in nature.

On the other hand, glitches with non-Poissonian signatures, such as Scattered Light and Whistle, exhibit inconsistent temporal patterns, indicating the influence of external sources as their origin. In particular, scattered light arises from small surface imperfections on the test mass mirrors that cause a fraction of the laser beam to scatter, reflect off moving surfaces such as chamber walls, and recombine with the main beam, thereby introducing phase noise in the interferometer signal \cite{scattered}. While Whistle glitches, in turn, are associated with radio-frequency signals that couple to the detector electronics through beating with the Voltage Controlled Oscillators of LIGO \cite{GravitySpyWikiWhistle}.

These results, summarized in Table \ref{tab:glitch_poisson_all}, suggest that the statistical classification of glitches may serve as an auxiliary tool for identifying their physical causes and, consequently, for their mitigation.

\begin{table}[H]
  \centering
  \caption{Classification of glitch types according to their adherence to the Poisson distribution for different interferometers and observing runs}
  \tiny
  \begin{tabular}{lccccccccc}
    \hline\hline
    \multirow{2}{*}{\textbf{Glitch Type}} & \multicolumn{1}{c}{\textbf{L1 O3a}} & \multicolumn{1}{c}{\textbf{L1 O3b}} & \multicolumn{1}{c}{\textbf{H1 O3a}} & \multicolumn{1}{c}{\textbf{H1 O3b}} & \multirow{2}{*}{\textbf{Likely}} \\
    & \(R^2\) & \(R^2\) & \(R^2\) & \(R^2\) & \\
    \hline
    1400Ripples         & 0.508 & —      & —      & —      & Non-Poissonian \\
    Air Compressor      & —      & 0.200 & —      & —      & Non-Poissonian \\
    Blip                & 0.940 & 0.932 & 0.909 & 0.887 & Poissonian \\
    Blip Low Frequency  & 0.957 & 0.903 & 0.920 & 0.698 & obs (1) \\
    Extremely Loud      & 0.970 & 0.977 & 0.916 & 0.962 & Poissonian \\
    Fast Scattering     & 0.269 & 0.277 & —      & 0.322 & Non-Poissonian \\
    Koi Fish            & 0.953 & 0.952 & 0.922 & 0.963 & Poissonian \\
    Low Frequency Burst & 0.562 & 0.602 & 0.209 & 0.290 & obs (2) \\
    Low Frequency Lines & 0.552 & 0.605 & 0.364 & 0.504 & obs (3) \\
    No Glitch           & 0.490 & 0.675 & 0.315 & —      & Non-Poissonian\\
    Power Line          & 0.204 & —      & —      & —      & Non-Poissonian \\
    Repeating Blips     & —      & —      & 0.850 & —      & Poissonian \\
    Scattered Light     & 0.370 & 0.481 & 0.279 & 0.236 & Non-Poissonian \\
    Tomte               & 0.929 & 0.901 & 0.851 & 0.503 & obs (4) \\
    Whistle             & 0.623 & 0.379 & 0.294 & 0.415 & Non-Poissonian \\
    \hline\hline
  \end{tabular}
  \label{tab:glitch_poisson_all}
\end{table}

\begin{itemize}
    \item\textbf{Obs 1:} The \textit{Blip Low Frequency} category exhibits \(R^2\) coefficients consistent with Poisson-like behavior in the Livingston and Hanford interferometers during O3a; however, a significant deviation from this behavior is observed during O3b in Handford.
    
    \item\textbf{Obs 2:} The \textit{Low Frequency Burst} category presents \(R^2\) coefficients around 0.5 and 0.6 for the Livingston interferometer. Although relatively low, these values suggest that the class likely does not follow a purely Poissonian statistic. In the Hanford interferometer, the \(R^2\) values drop considerably, reaching values around 0.2 to 0.3.
    
    \item\textbf{Obs 3:} In the case of the \textit{Low Frequency Lines} category, the situation is similar to that of the \textit{Low Frequency Burst}. The \(R^2\) coefficients are approximately 0.5 to 0.6 for the Livingston interferometer and for O3b at Hanford. However, the \(R^2\) drops significantly during O3a at Hanford, reaching values on the order of 0.3.
    
    \item\textbf{Obs 4:} The \textit{Tomte} category shows \(R^2\) coefficients consistent with Poissonian behavior in the Livingston interferometer and only slight deviation in Hanford during O3a. However, a significant deviation is observed during O3b, where \(R^2 = 0.503\). It is noteworthy that the number of Tomte glitches identified at Hanford is approximately 20 times lower than in Livingston, which may help explain this statistical variation.

    In future work, we aim to use this analysis to investigate the physical origin of the glitches.
    
\end{itemize}

\thispagestyle{empty}
\mbox{} 

\section*{Acknowledgments}
This research made use of data or software obtained from the Gravitational Wave Open Science Center (gwosc.org), a service of the LIGO Scientific Collaboration, the Virgo Collaboration, and the KAGRA Collaboration. This material is based upon work supported by the LIGO Laboratory, a major facility fully funded by the National Science Foundation (NSF). Additional support for the construction of Advanced LIGO and for the construction and operation of the GEO600 detector has been provided by the Science and Technology Facilities Council (STFC) of the United Kingdom, the Max Planck Society (MPS), and the State of Lower Saxony/Germany. Further support for Advanced LIGO was provided by the Australian Research Council.

Virgo is funded through the European Gravitational Observatory (EGO) by the French Centre National de la Recherche Scientifique (CNRS), the Italian Istituto Nazionale di Fisica Nucleare (INFN), and the Dutch Nikhef, with contributions from institutions in Belgium, Germany, Greece, Hungary, Ireland, Japan, Monaco, Poland, Portugal, and Spain.

KAGRA is supported by the Ministry of Education, Culture, Sports, Science and Technology (MEXT) and the Japan Society for the Promotion of Science (JSPS) in Japan; the National Research Foundation (NRF) and the Ministry of Science and ICT (MSIT) in Korea; and Academia Sinica (AS) and the National Science and Technology Council (NSTC) in Taiwan.

ODA thanks the Brazilian Ministry of Science, Technology and Innovation and the Brazilian Space Agency (AEB), which supported the present work under PO 20VB.0009. He also thanks CNPq for grant number 310087/2021-0.

We also acknowledge the financial support provided by the São Paulo Research Foundation (FAPESP), grant number 2024/06617-4. This study was financed in part by the Coordenação de Aperfeiçoamento de Pessoal de Nível Superior - Brasil (CAPES) - Finance Code 001.

\appendix

\section{Data Distribution}
\label{dataapendice}

\begin{figure}[H]
    \centering
    \begin{subfigure}{0.49\linewidth}
        \includegraphics[width=\linewidth]{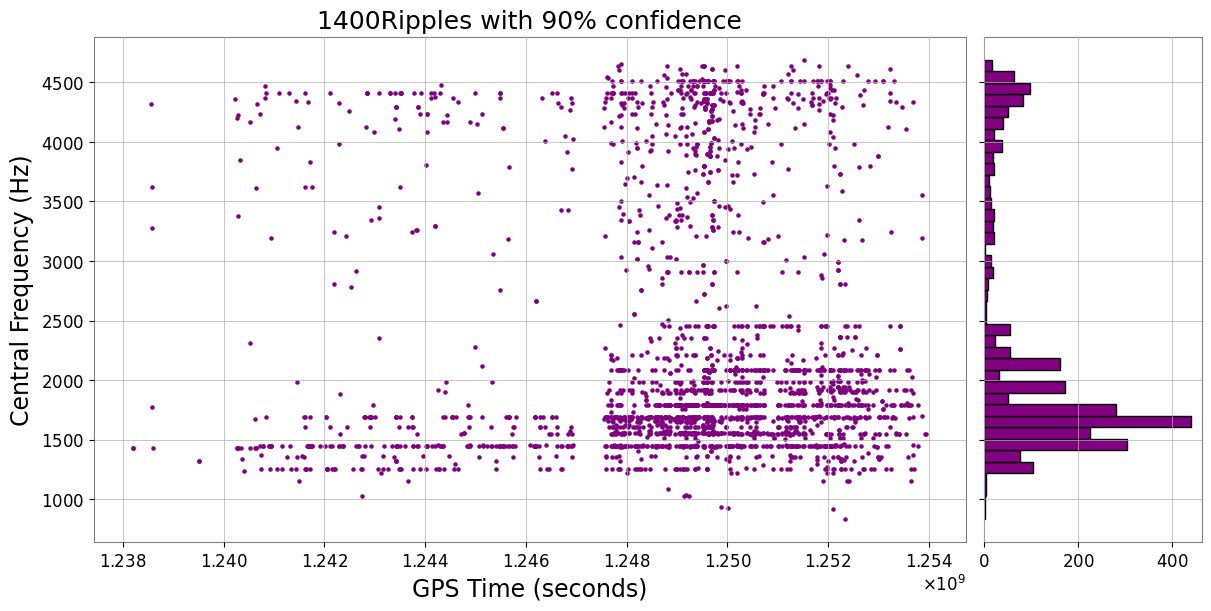}
    \end{subfigure}
    \begin{subfigure}{0.49\linewidth}
        \includegraphics[width=\linewidth]{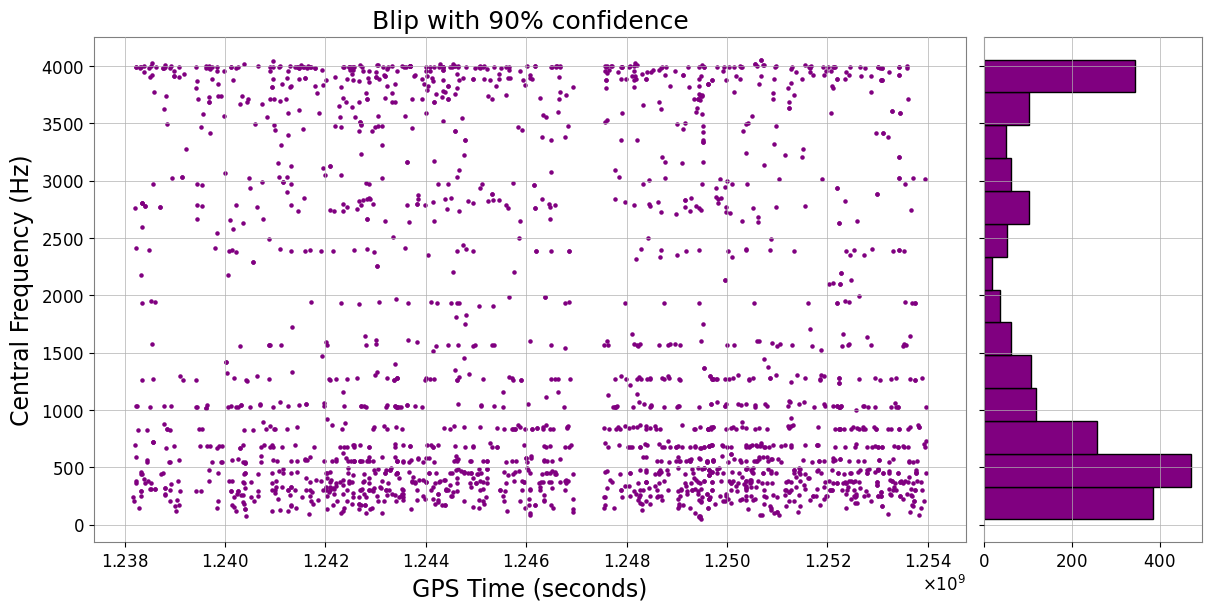}
    \end{subfigure}

    \begin{subfigure}{0.49\linewidth}
        \includegraphics[width=\linewidth]{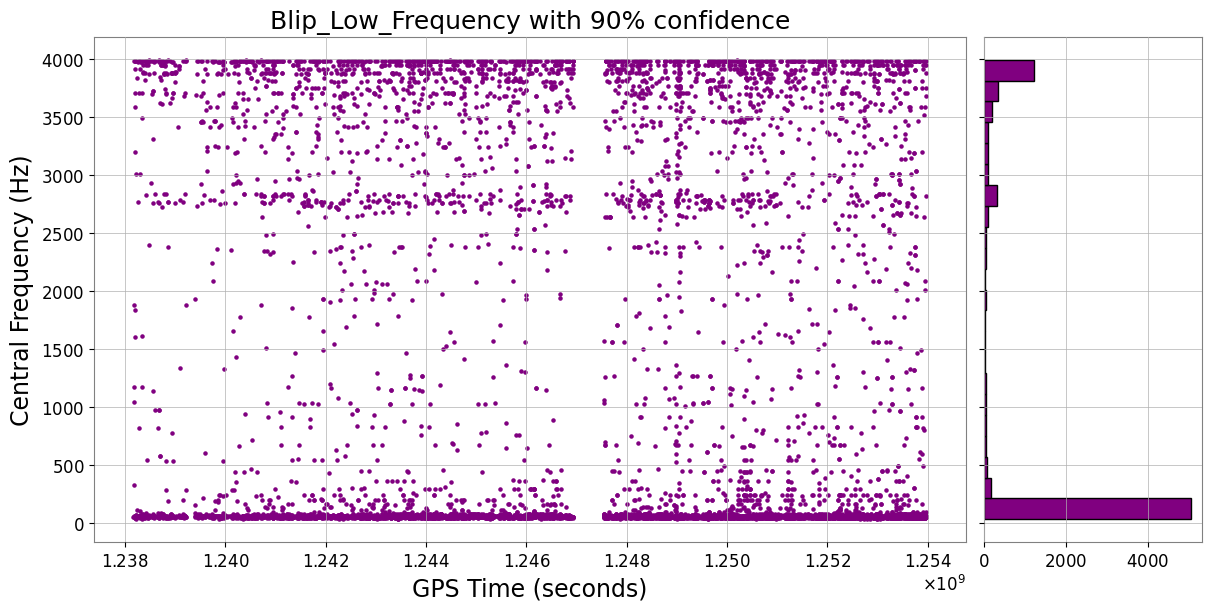}
    \end{subfigure}
    \begin{subfigure}{0.49\linewidth}
        \includegraphics[width=\linewidth]{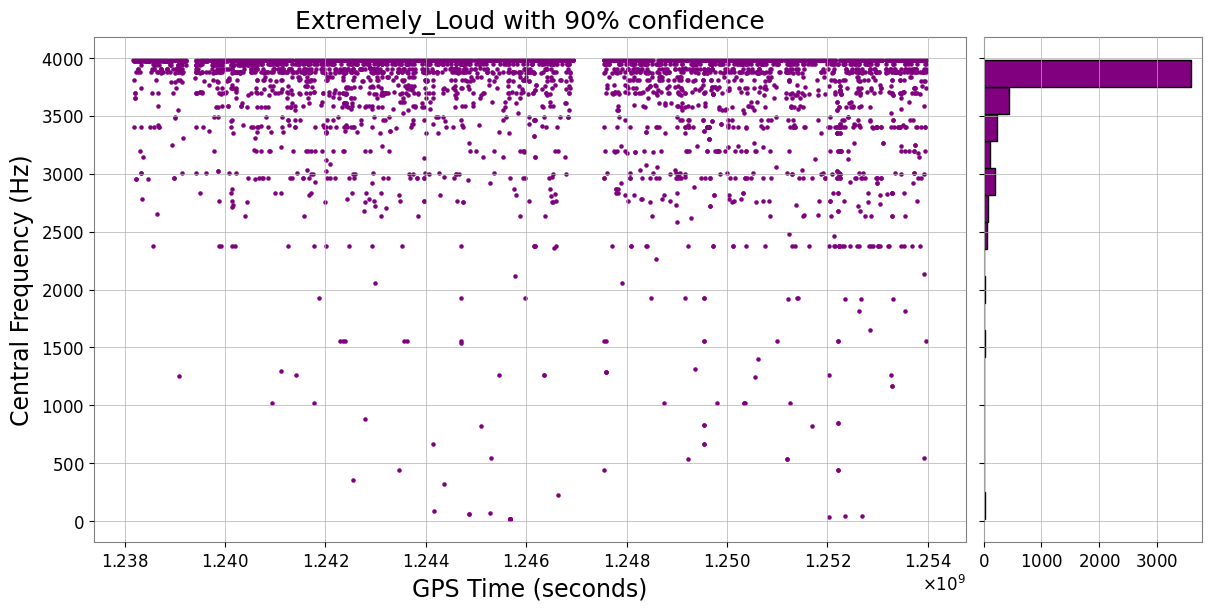}
    \end{subfigure}

    \begin{subfigure}{0.49\linewidth}
        \includegraphics[width=\linewidth]{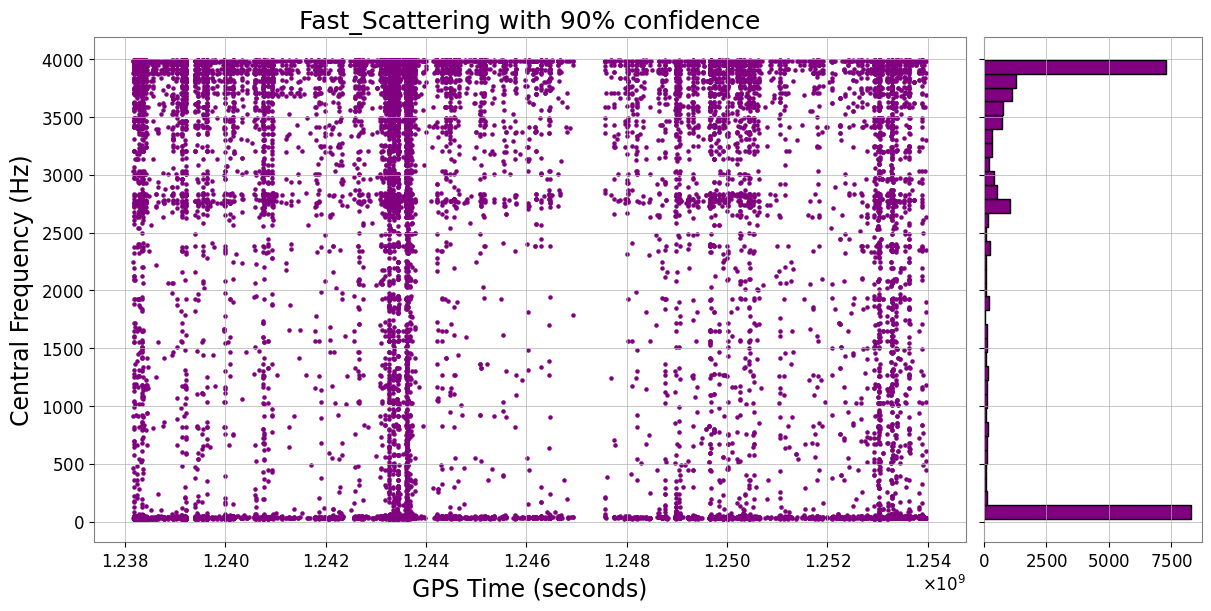}
    \end{subfigure}
    \begin{subfigure}{0.49\linewidth}
        \includegraphics[width=\linewidth]{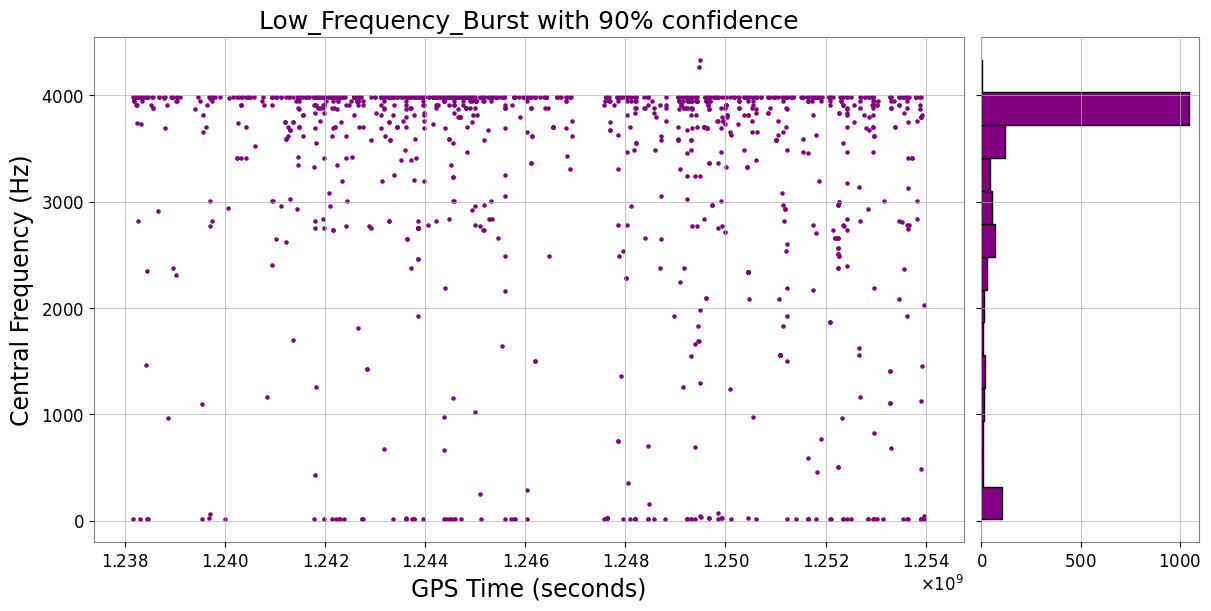}
    \end{subfigure}

    \begin{subfigure}{0.49\linewidth}
        \includegraphics[width=\linewidth]{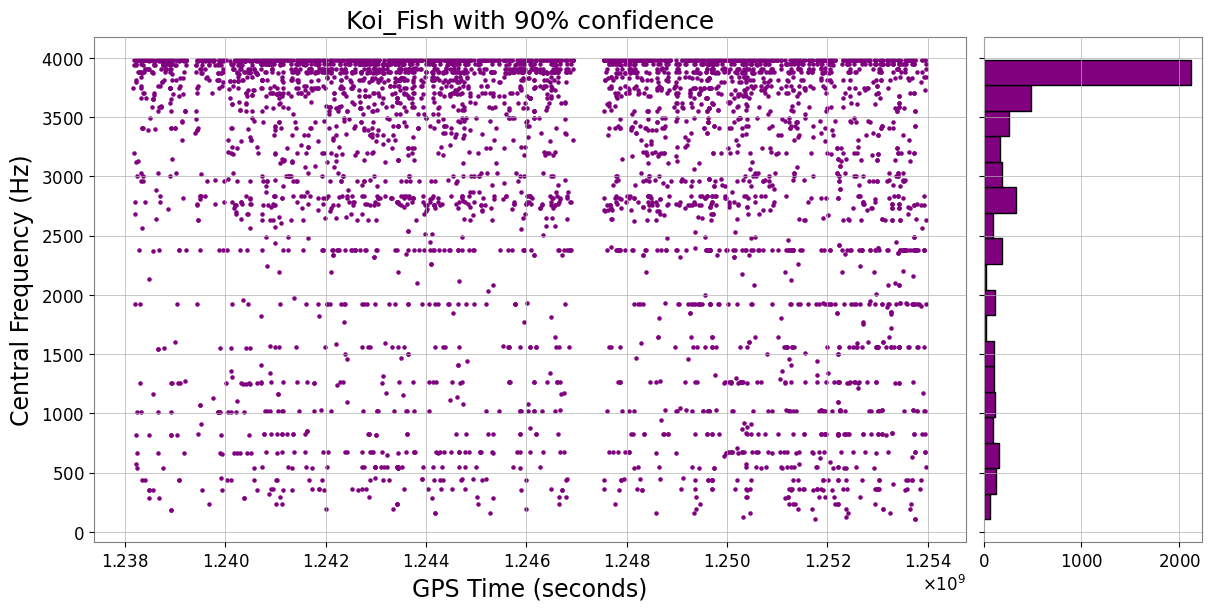}
    \end{subfigure}
    \begin{subfigure}{0.49\linewidth}
        \includegraphics[width=\linewidth]{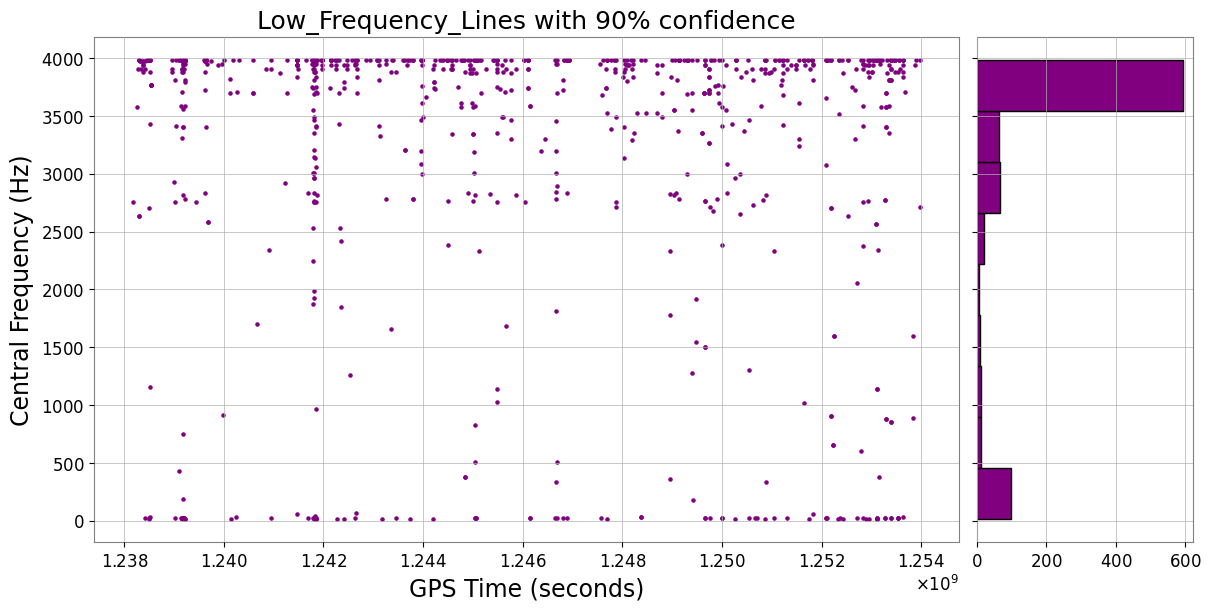}
    \end{subfigure}

    \begin{subfigure}{0.49\linewidth}
        \includegraphics[width=\linewidth]{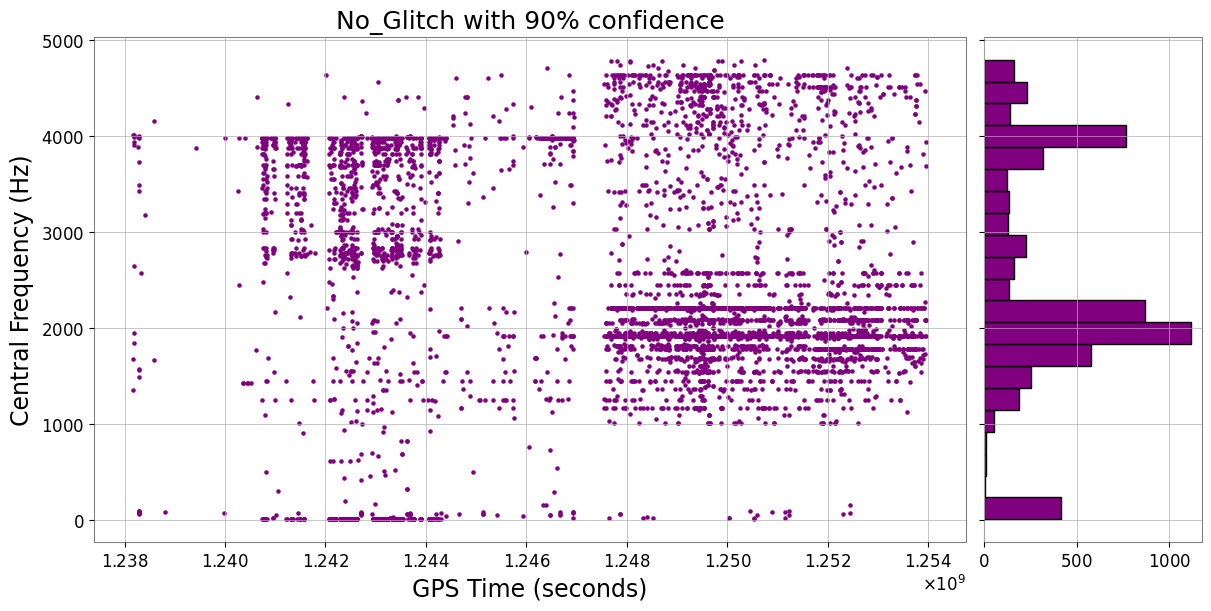}
    \end{subfigure}
    \begin{subfigure}{0.49\linewidth}
        \includegraphics[width=\linewidth]{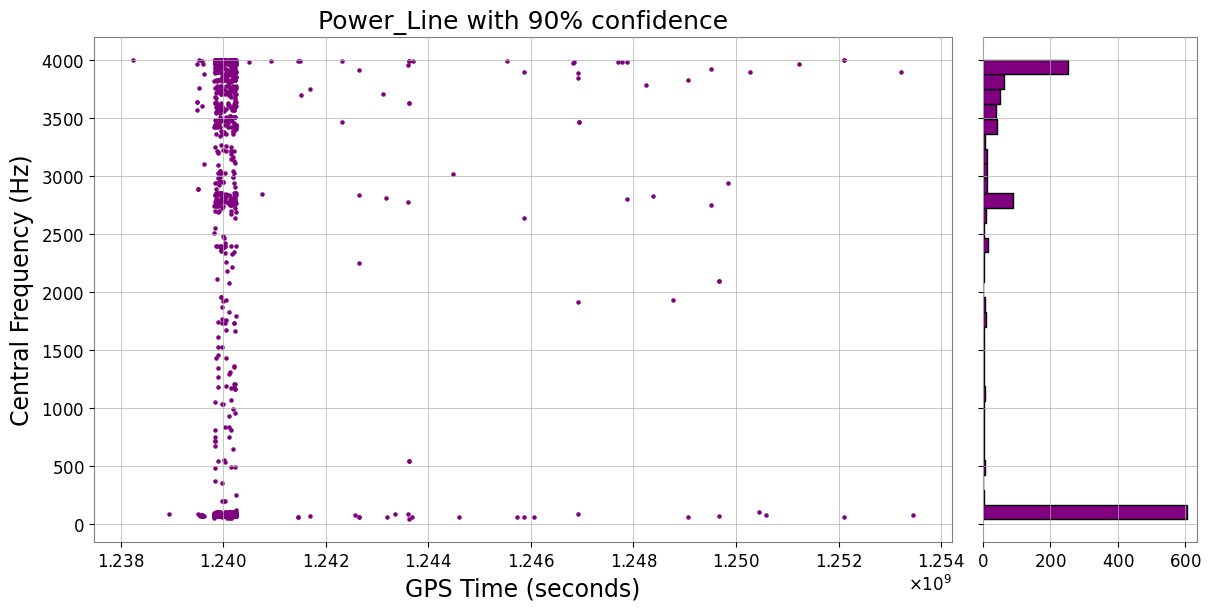}
    \end{subfigure}

    \begin{subfigure}{0.49\linewidth}
        \includegraphics[width=\linewidth]{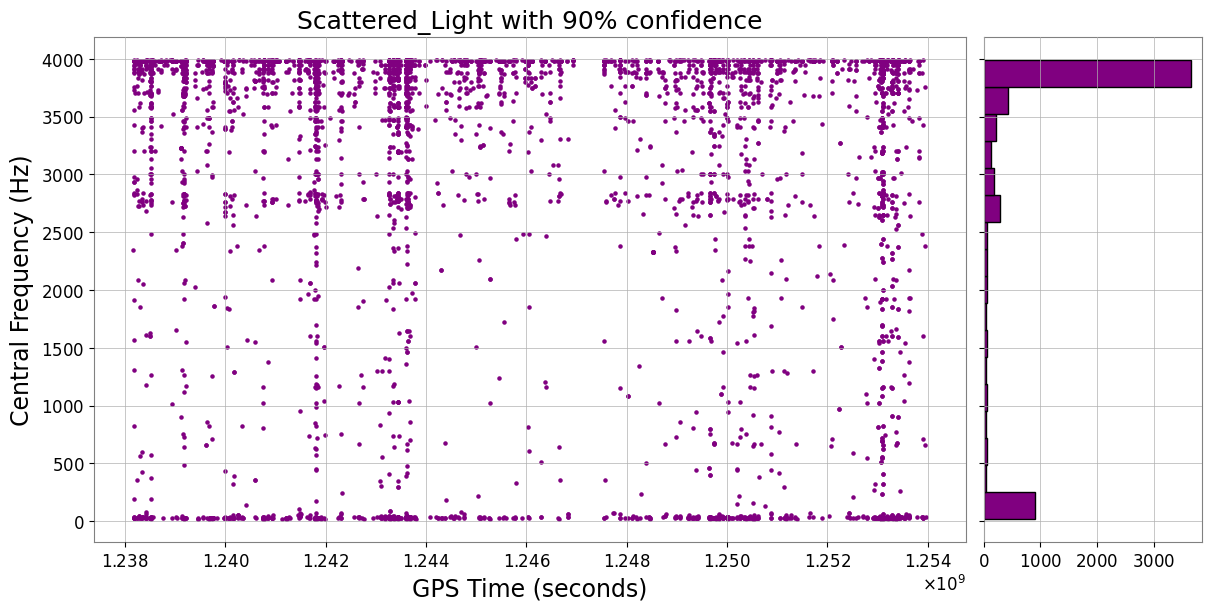}
    \end{subfigure}
    \begin{subfigure}{0.49\linewidth}
        \includegraphics[width=\linewidth]{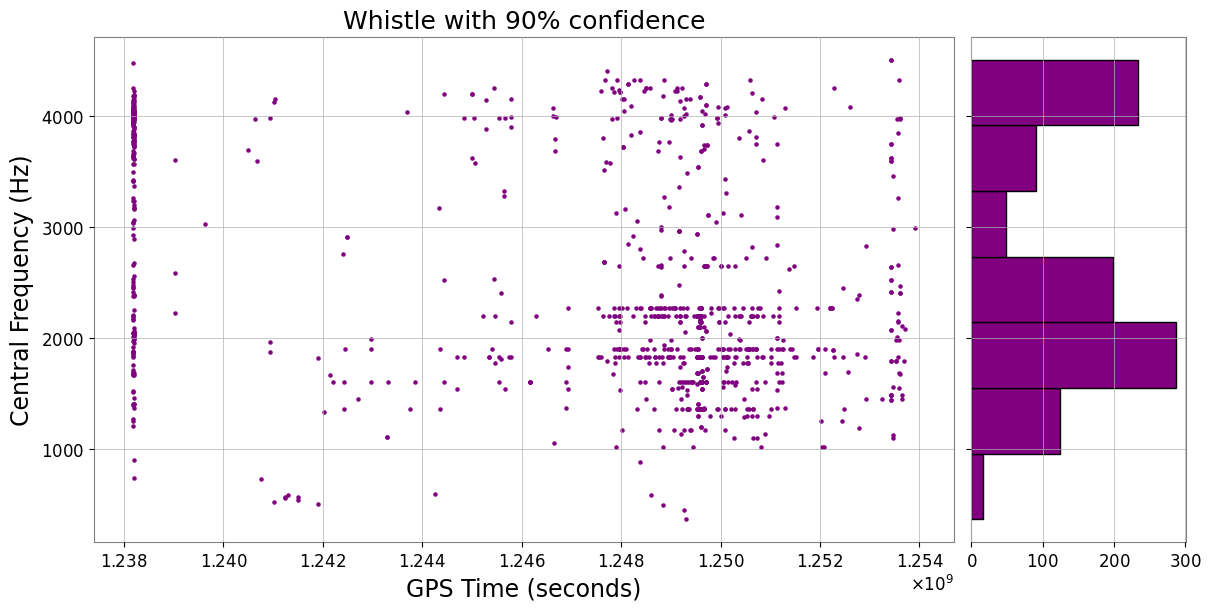}
    \end{subfigure}

    \caption{Frequency distributions of different glitch classes observed in the Livingston interferometer during O3a.}
    \label{fig:freqO3a}
\end{figure}


\begin{figure}[H]
    \centering
    \begin{subfigure}{0.49\linewidth}
        \includegraphics[width=\linewidth]{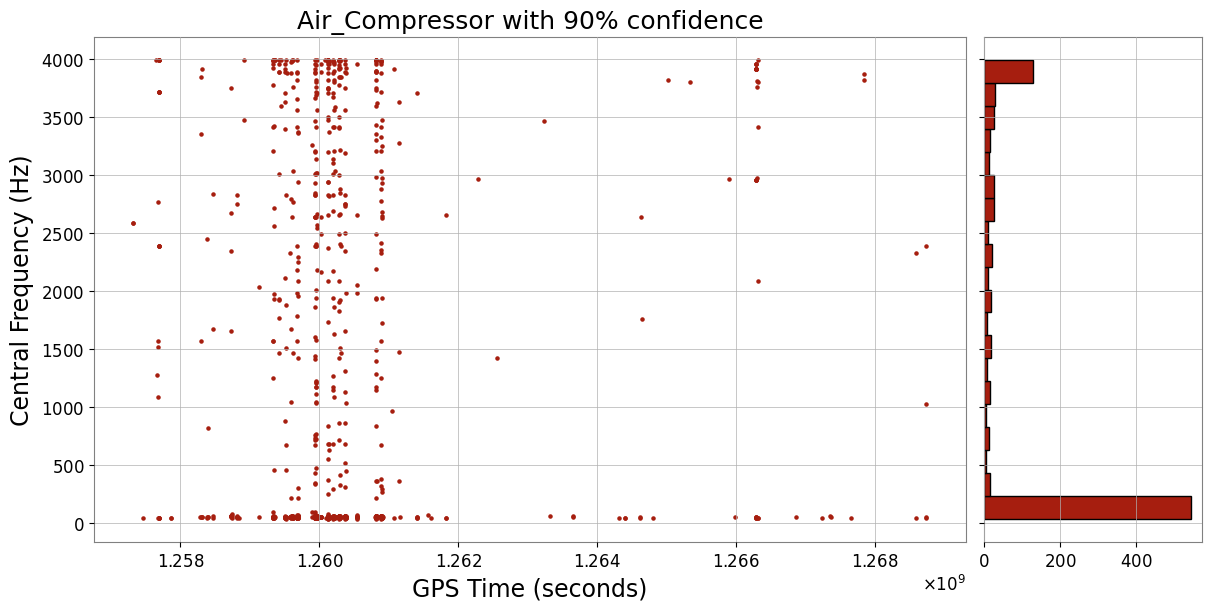}
    \end{subfigure}
    \begin{subfigure}{0.49\linewidth}
        \includegraphics[width=\linewidth]{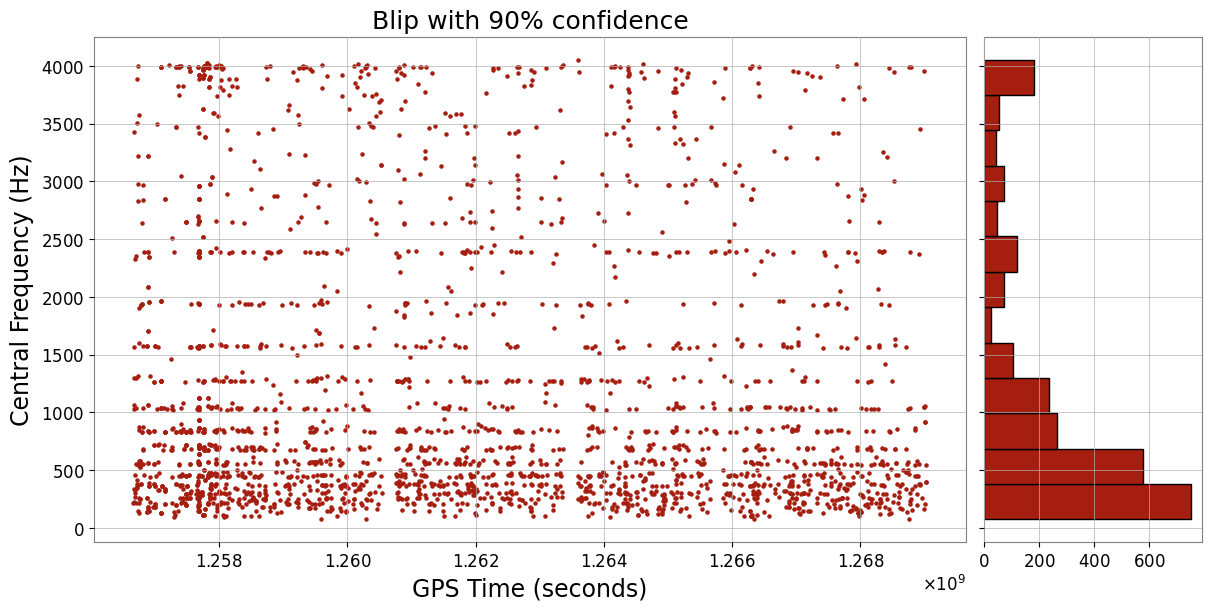}
    \end{subfigure}

    \begin{subfigure}{0.49\linewidth}
        \includegraphics[width=\linewidth]{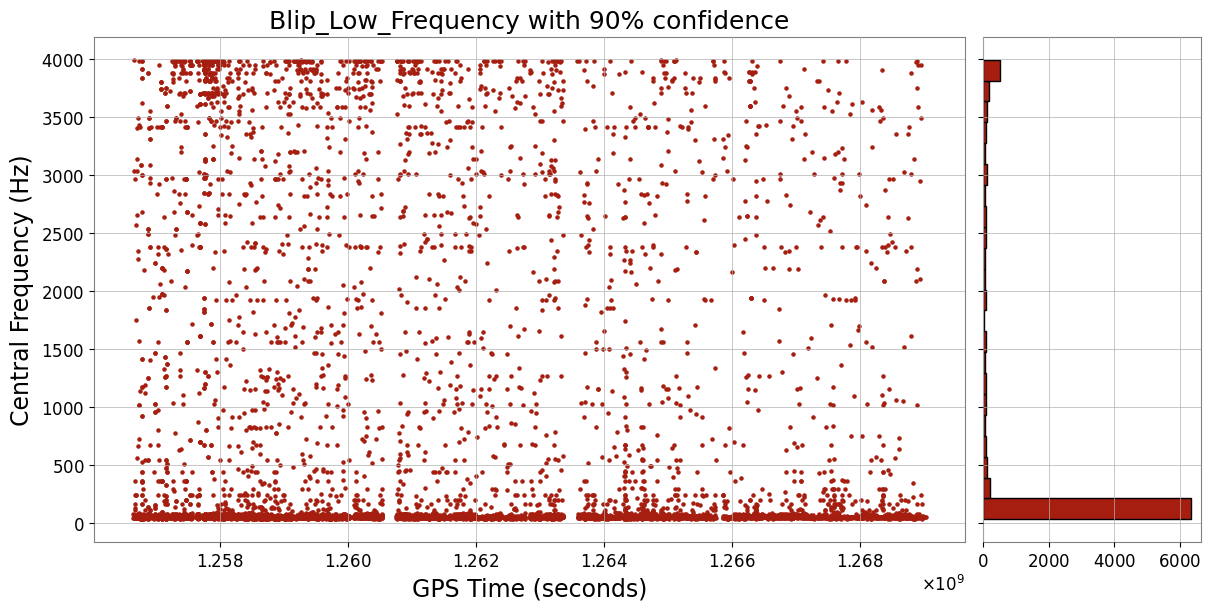}
    \end{subfigure}
    \begin{subfigure}{0.49\linewidth}
        \includegraphics[width=\linewidth]{freqloud03bliv.png}
    \end{subfigure}

    \begin{subfigure}{0.49\linewidth}
        \includegraphics[width=\linewidth]{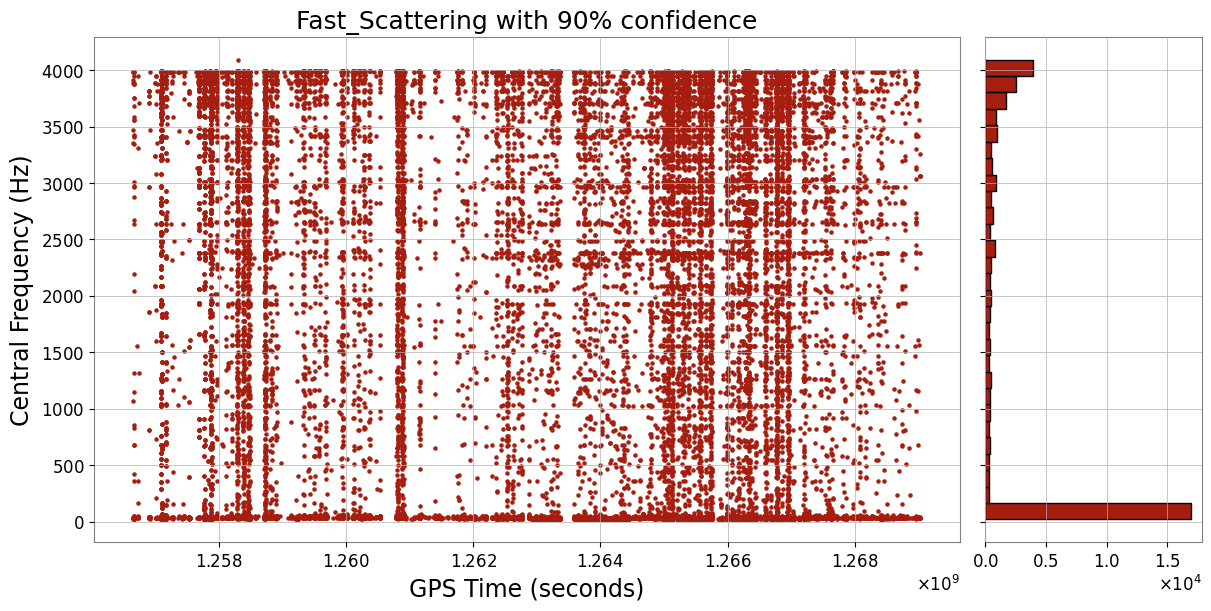}
    \end{subfigure}
    \begin{subfigure}{0.49\linewidth}
        \includegraphics[width=\linewidth]{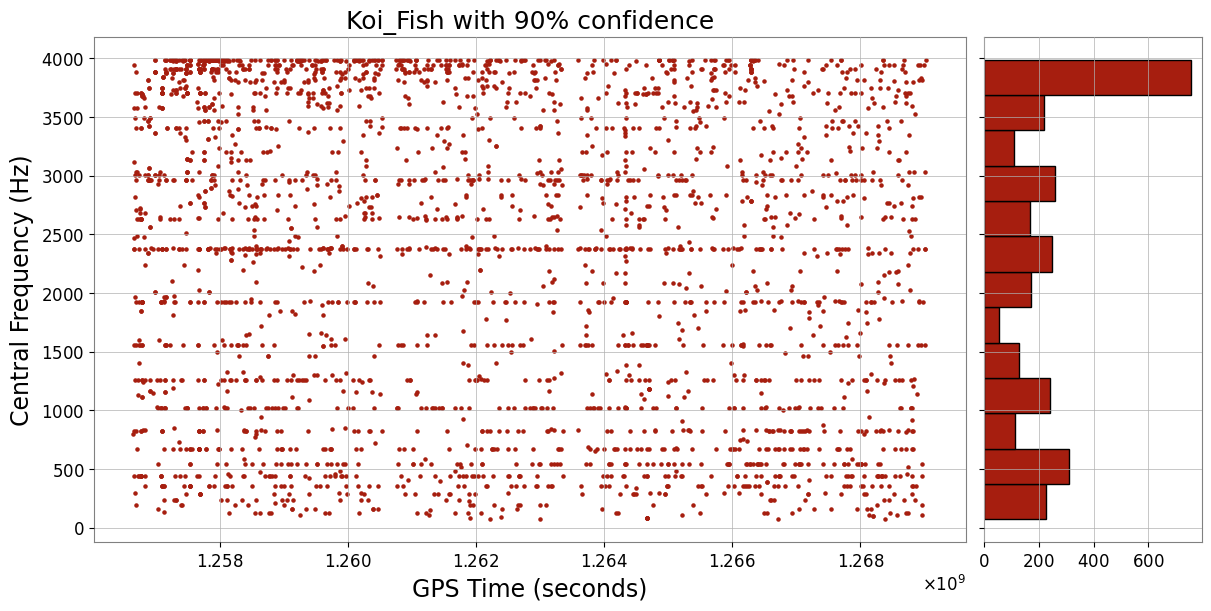}
    \end{subfigure}

    \begin{subfigure}{0.49\linewidth}
        \includegraphics[width=\linewidth]{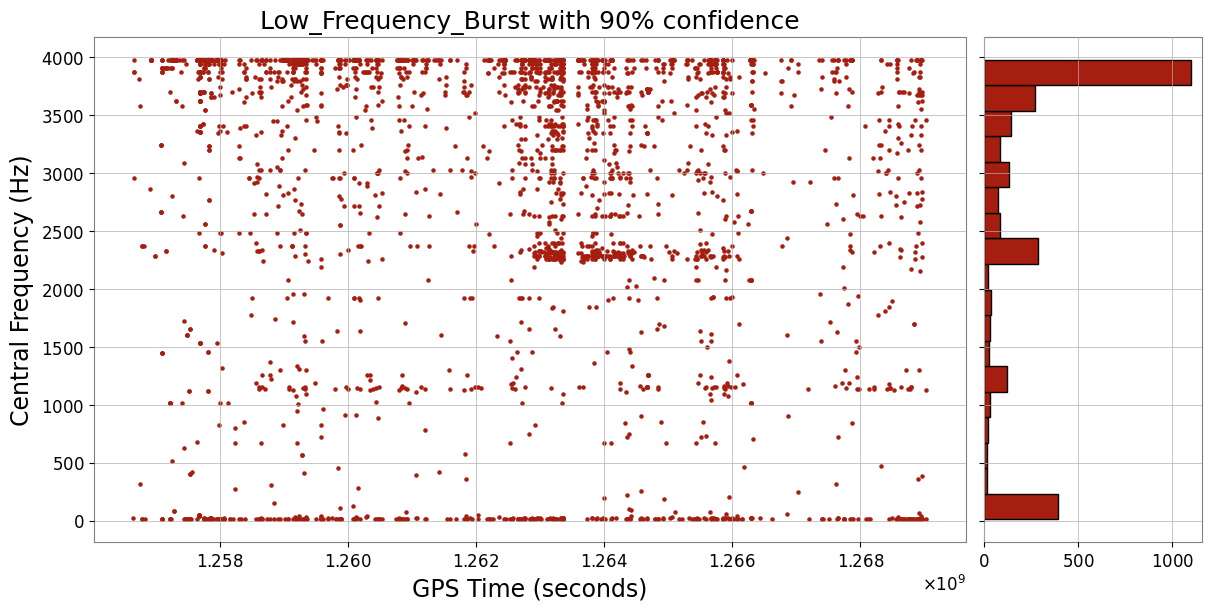}
    \end{subfigure}
    \begin{subfigure}{0.49\linewidth}
        \includegraphics[width=\linewidth]{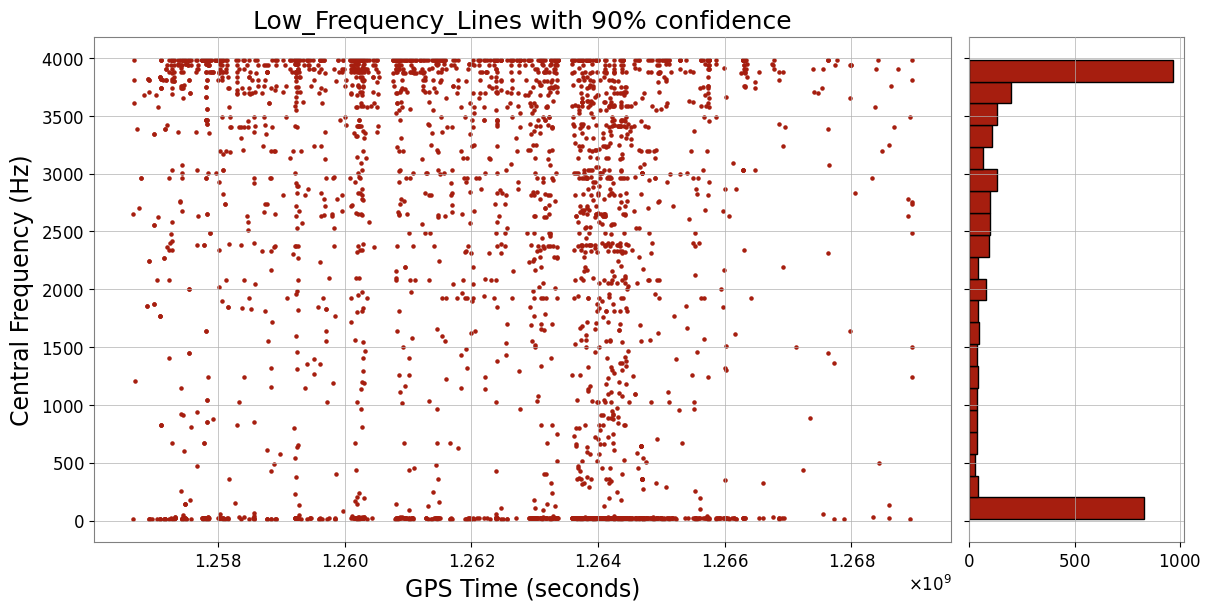}
    \end{subfigure}

    \begin{subfigure}{0.49\linewidth}
        \includegraphics[width=\linewidth]{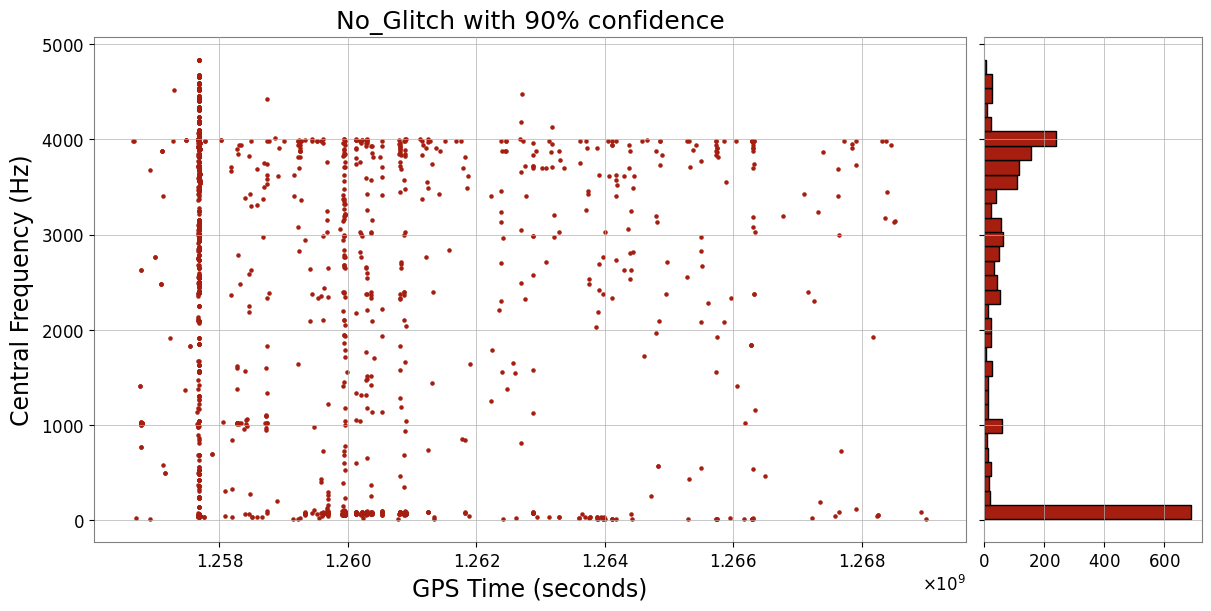}
    \end{subfigure}
    \begin{subfigure}{0.49\linewidth}
        \includegraphics[width=\linewidth]{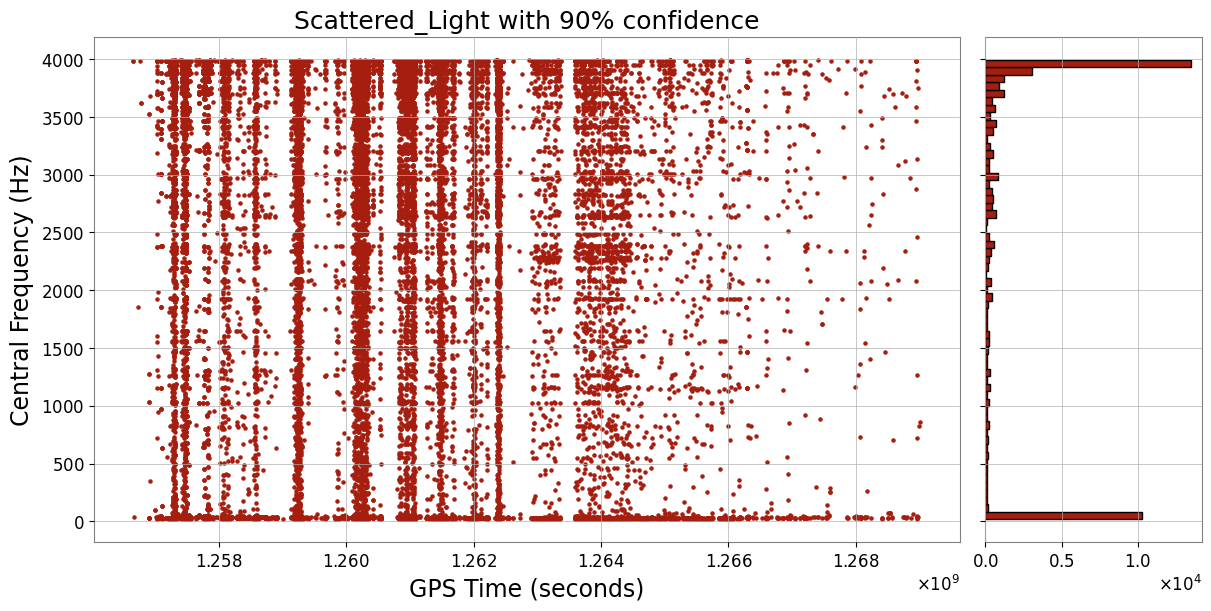}
    \end{subfigure}

    \begin{subfigure}{0.49\linewidth}
        \includegraphics[width=\linewidth]{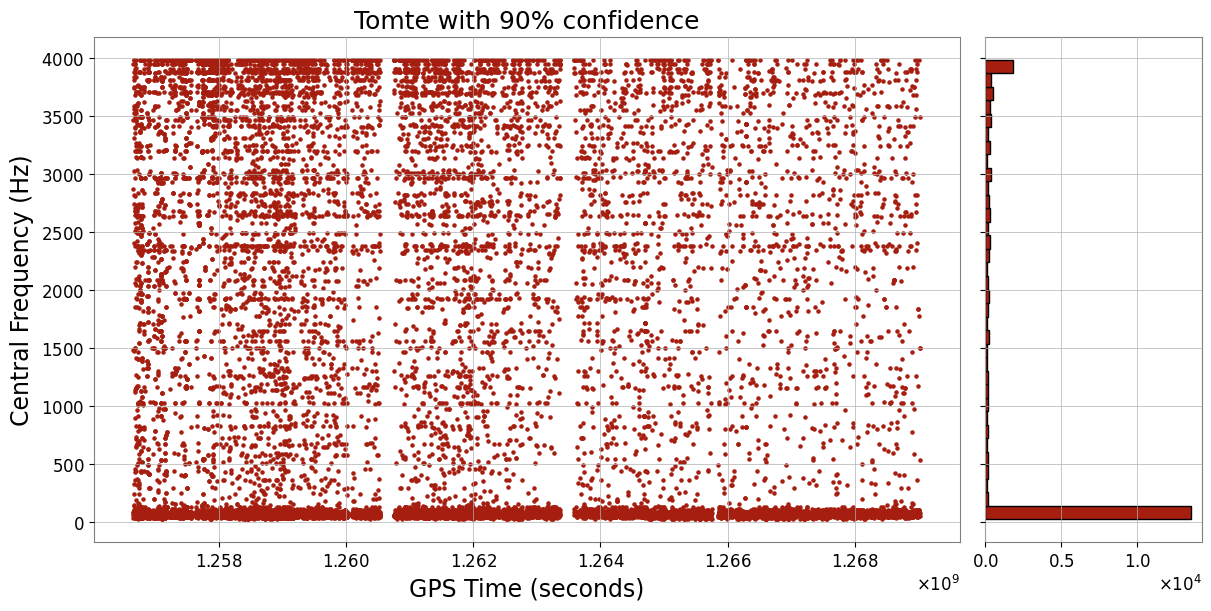}
    \end{subfigure}

    \caption{Frequency distributions of different glitch classes observed in the Livingston interferometer during O3b.}
    \label{fig:freqO3b}
\end{figure}

The same procedure was applied to the data obtained from the Hanford interferometer during O3a and O3b. 

\begin{figure}[H]
    \centering
    \begin{subfigure}{0.49\linewidth}
        \includegraphics[width=\linewidth]{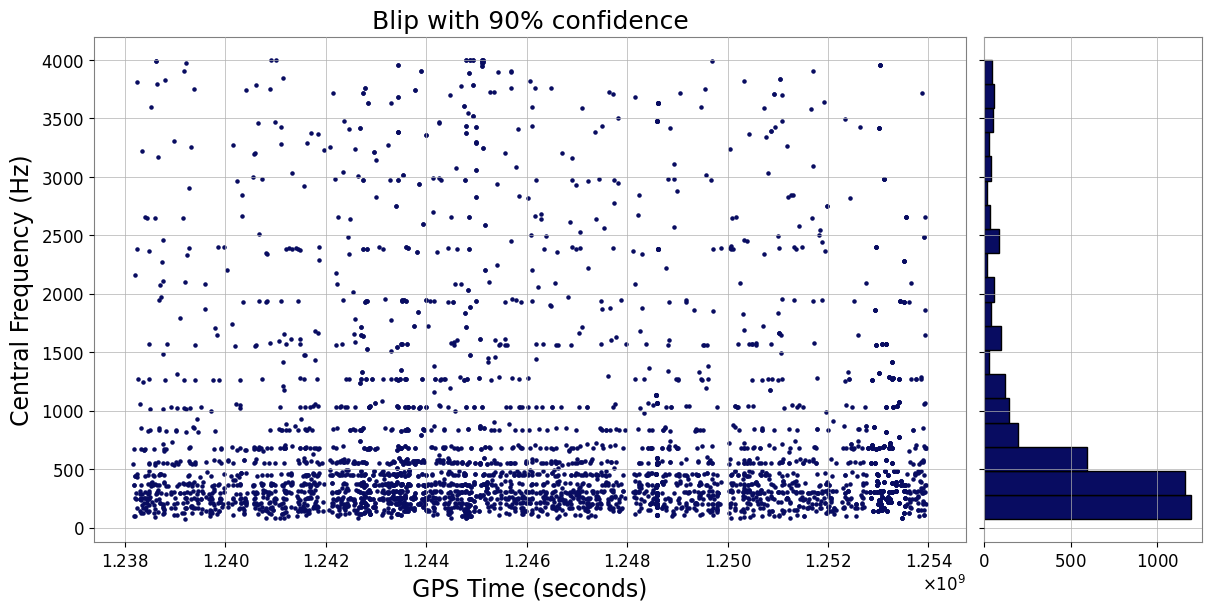}
    \end{subfigure}
    \begin{subfigure}{0.49\linewidth}
        \includegraphics[width=\linewidth]{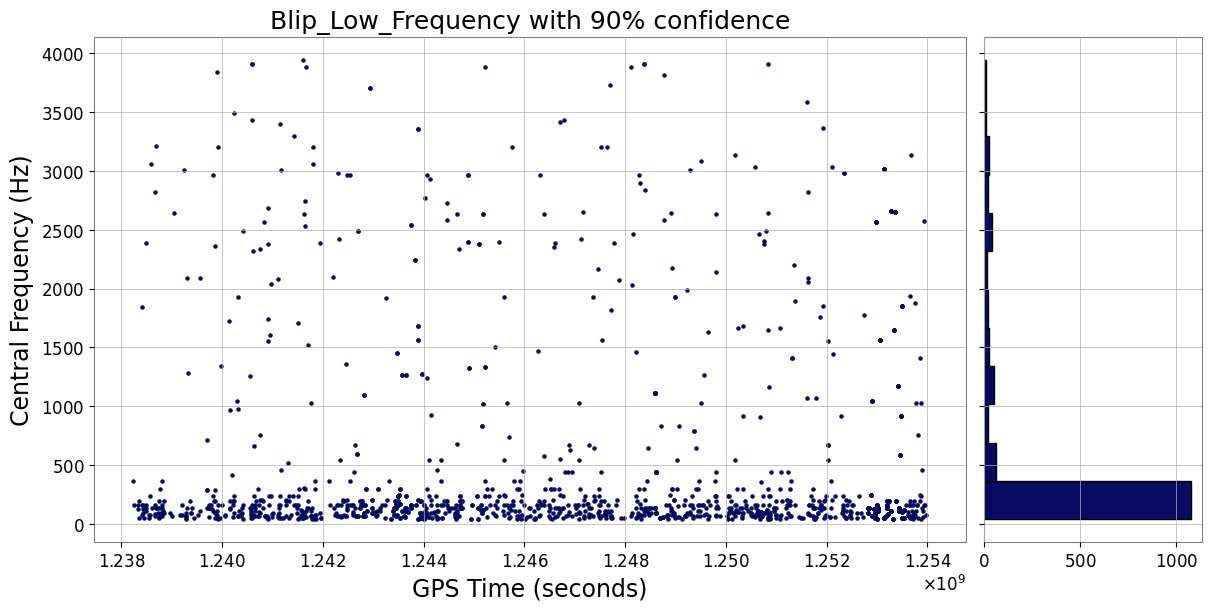}
    \end{subfigure}

    \begin{subfigure}{0.49\linewidth}
        \includegraphics[width=\linewidth]{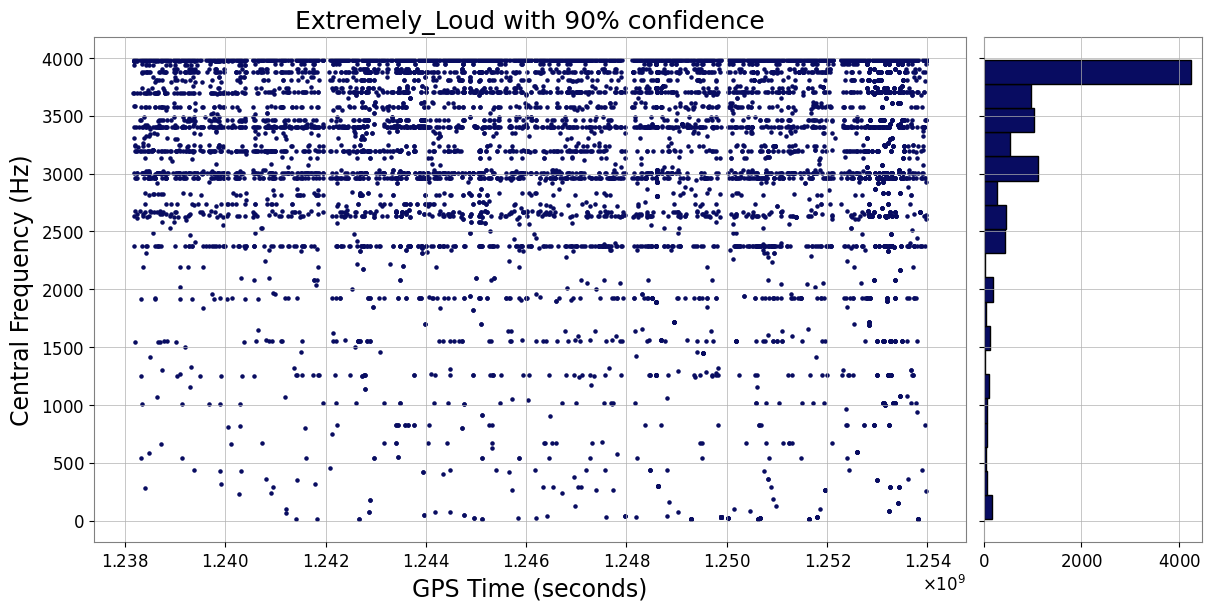}
    \end{subfigure}
    \begin{subfigure}{0.49\linewidth}
        \includegraphics[width=\linewidth]{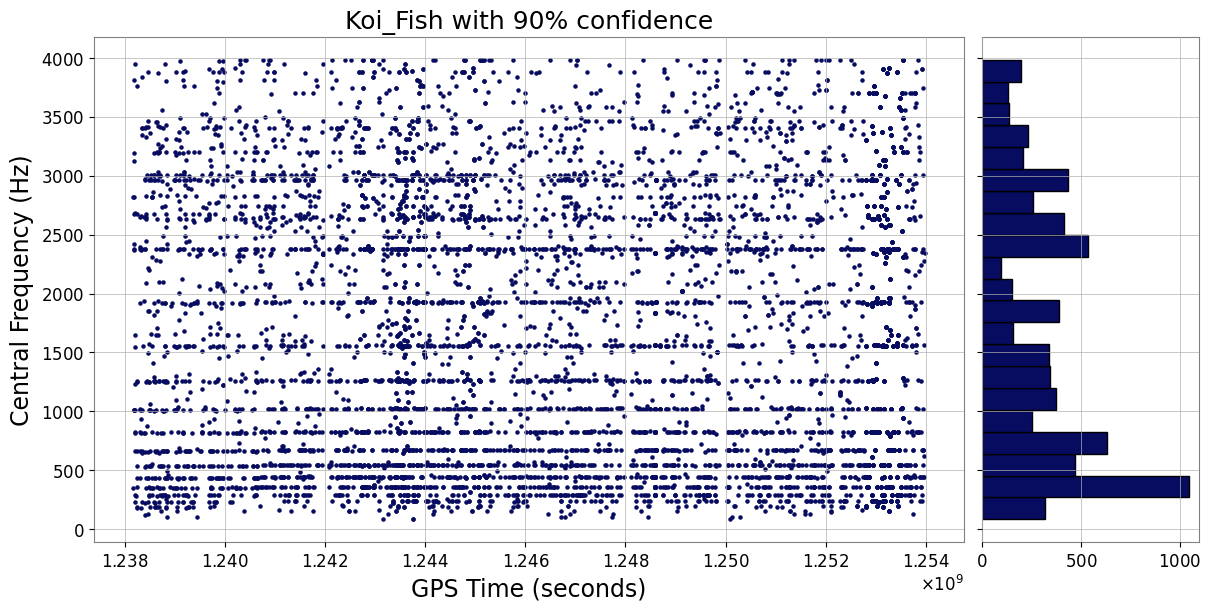}
    \end{subfigure}

    \begin{subfigure}{0.49\linewidth}
        \includegraphics[width=\linewidth]{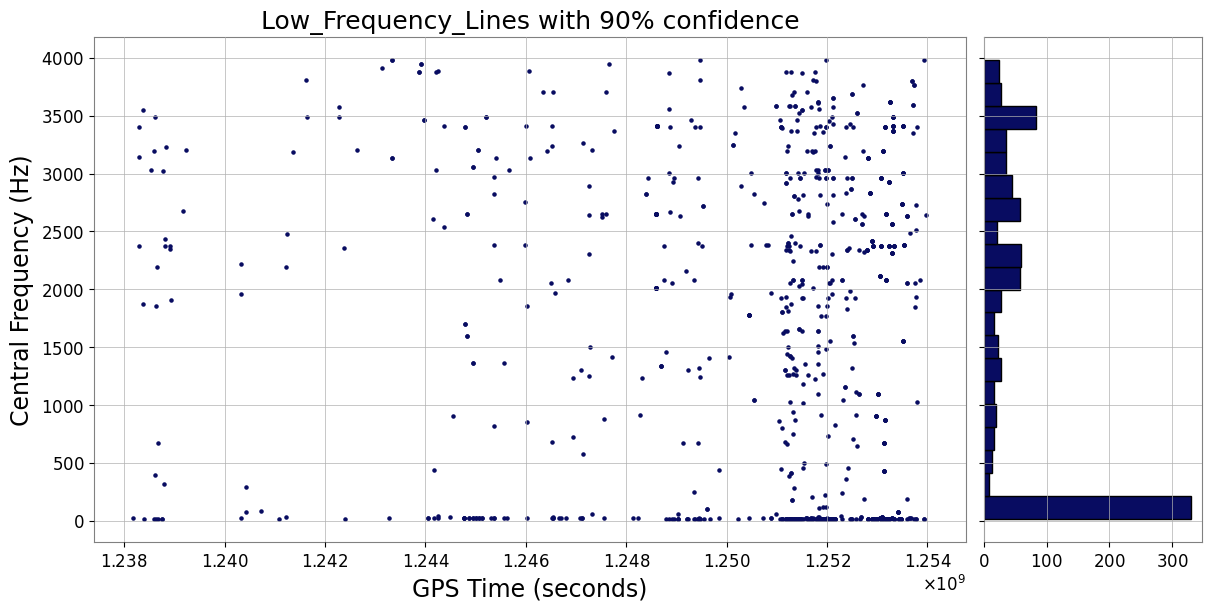}
    \end{subfigure}
    \begin{subfigure}{0.49\linewidth}
        \includegraphics[width=\linewidth]{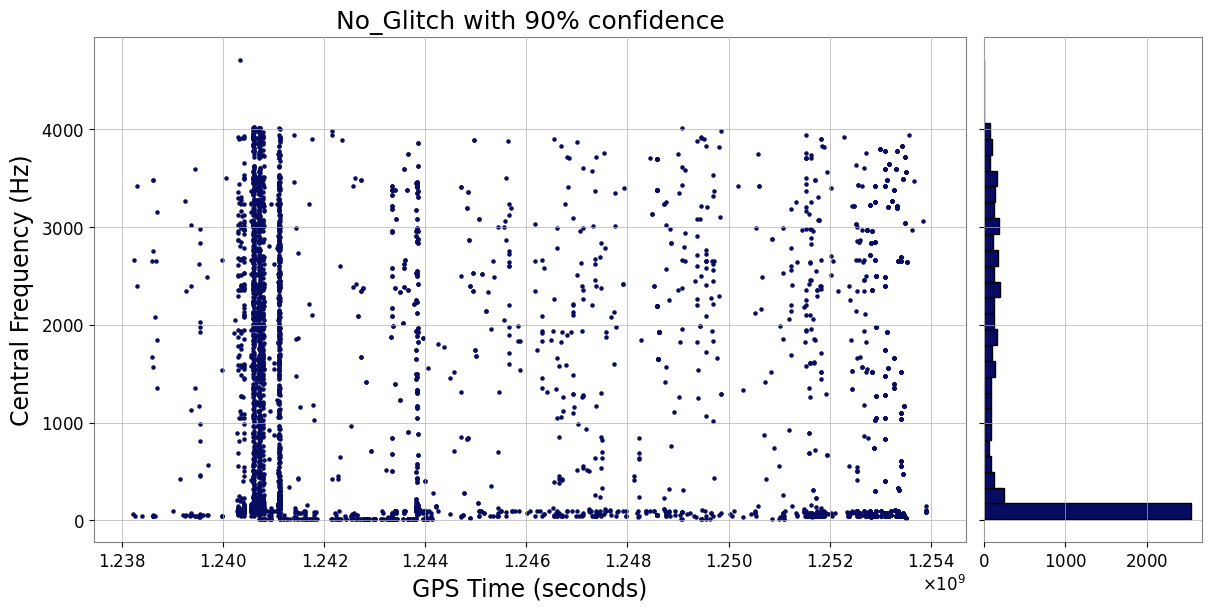}
    \end{subfigure}

    \begin{subfigure}{0.49\linewidth}
        \includegraphics[width=\linewidth]{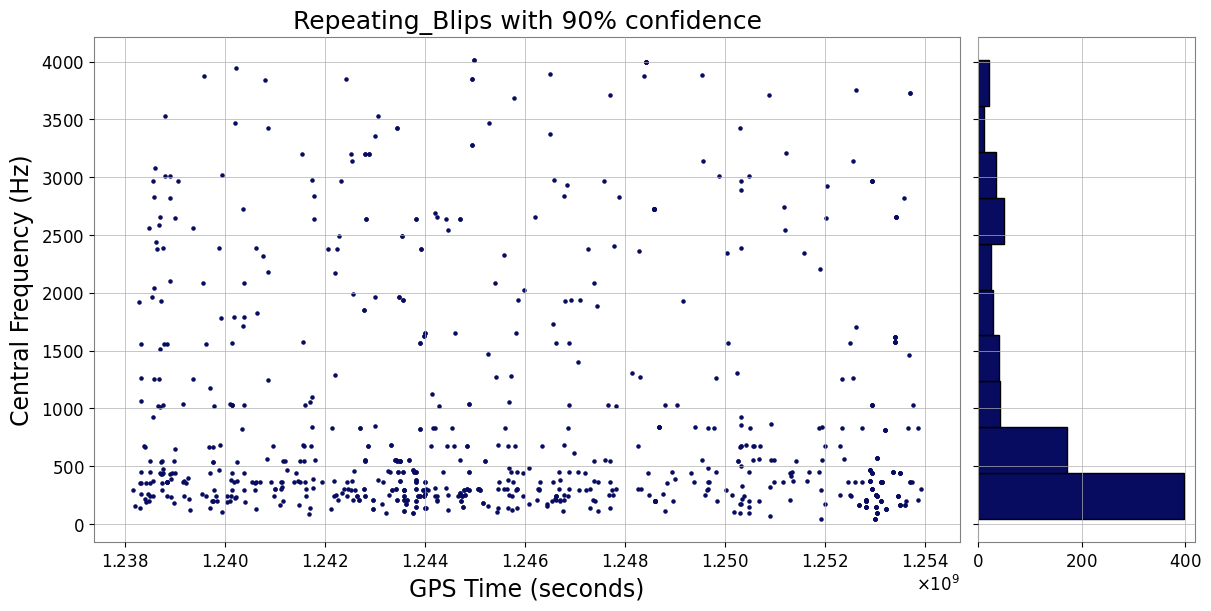}
    \end{subfigure}
    \begin{subfigure}{0.49\linewidth}
        \includegraphics[width=\linewidth]{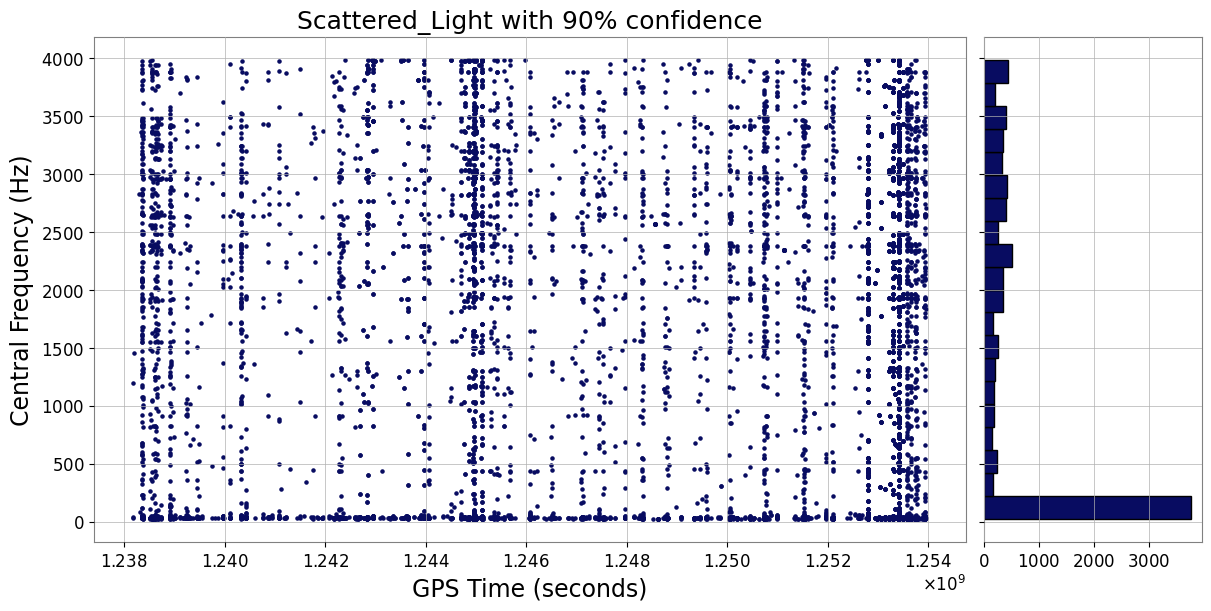}
    \end{subfigure}

    \begin{subfigure}{0.49\linewidth}
        \includegraphics[width=\linewidth]{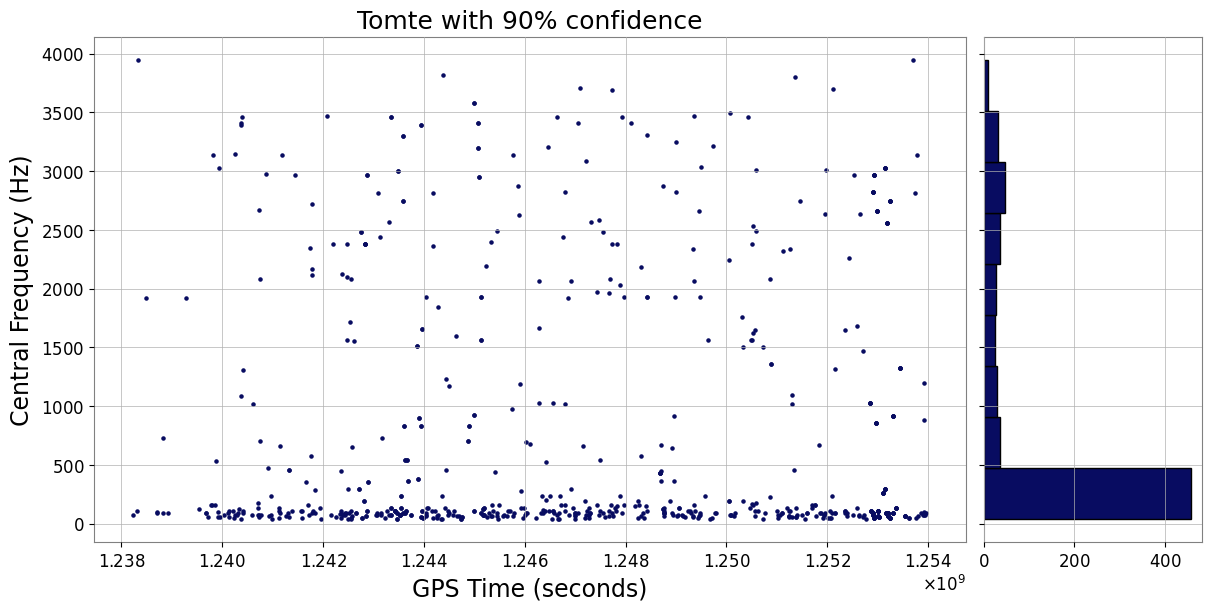}
    \end{subfigure}
    \begin{subfigure}{0.49\linewidth}
        \includegraphics[width=\linewidth]{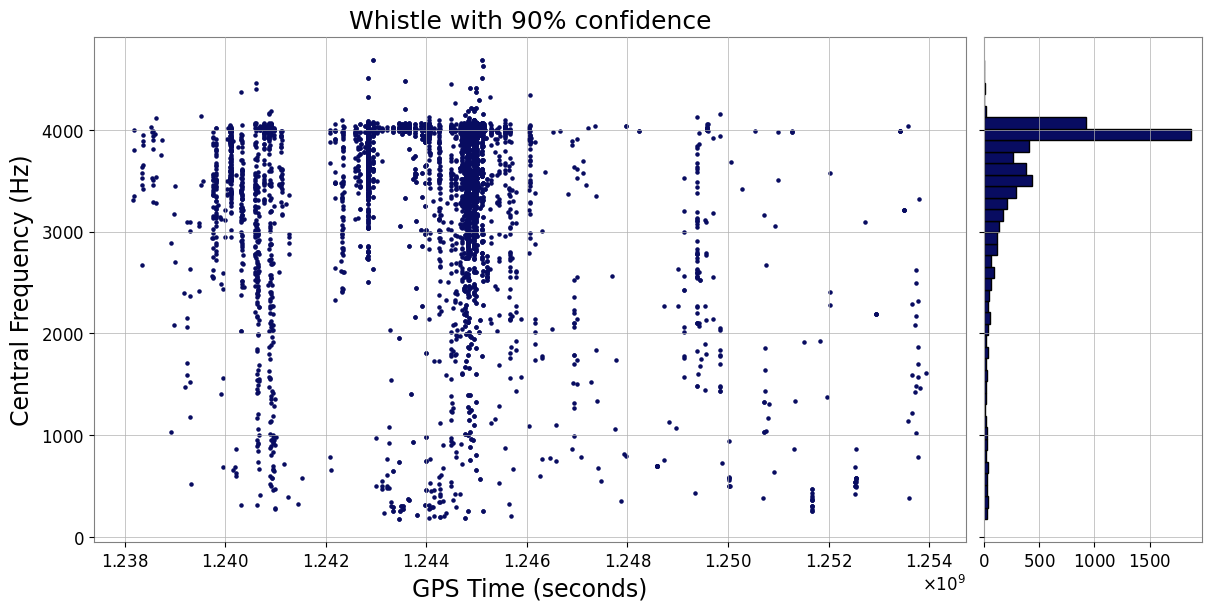}
    \end{subfigure}

    \caption{Frequency distributions of different glitch classes observed in the Hanford interferometer during O3a.}
    \label{fig:freqO3aHanford}
\end{figure}

\vspace{3mm}

\begin{figure}[H]
    \centering
    \begin{subfigure}{0.49\linewidth}
        \includegraphics[width=\linewidth]{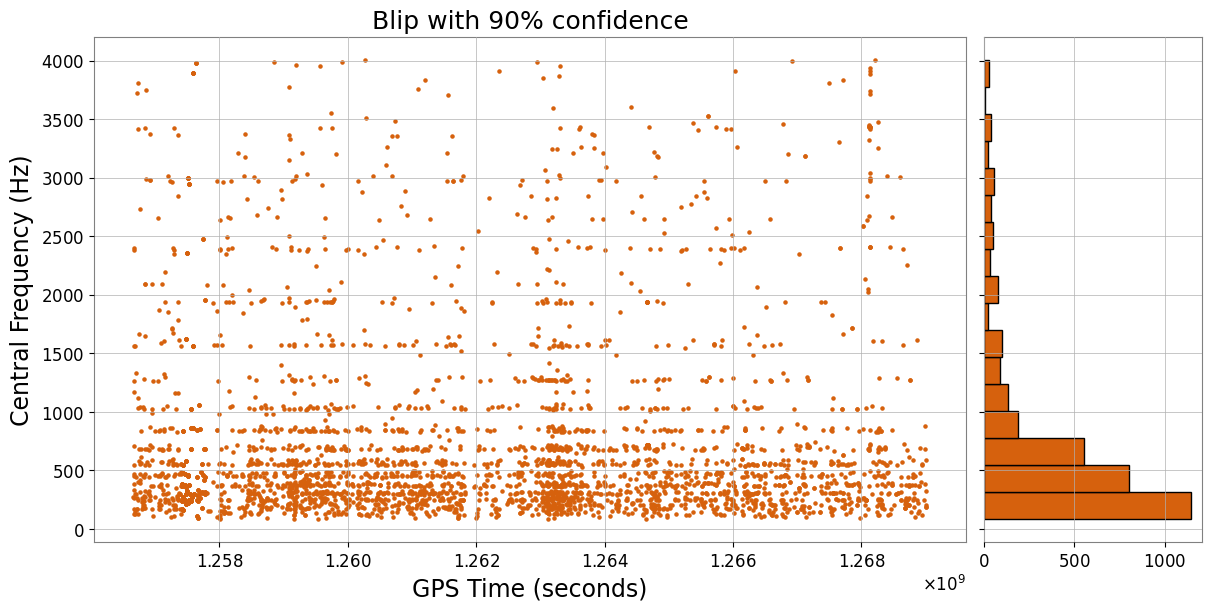}
    \end{subfigure}
    \begin{subfigure}{0.49\linewidth}
        \includegraphics[width=\linewidth]{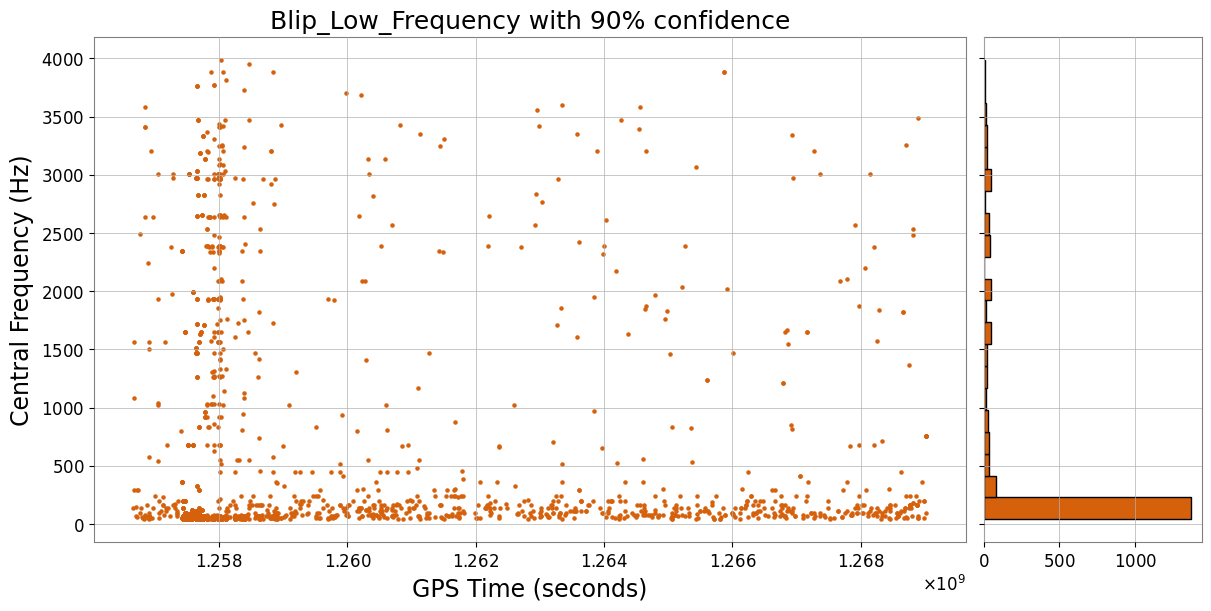}
    \end{subfigure}

    \begin{subfigure}{0.49\linewidth}
        \includegraphics[width=\linewidth]{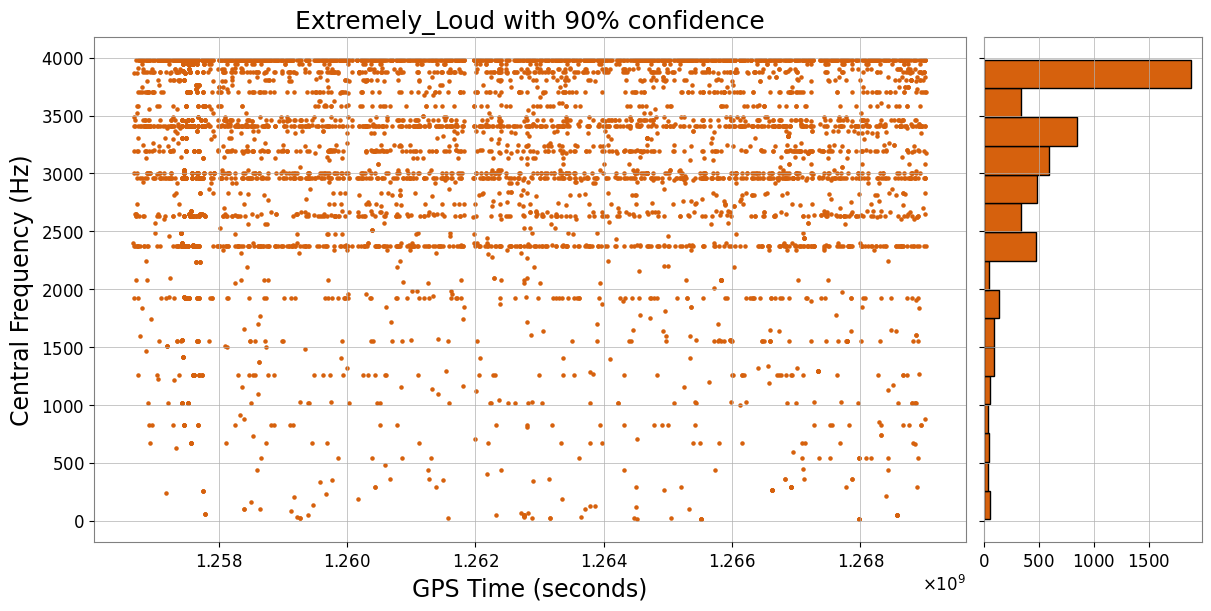}
    \end{subfigure}
    \begin{subfigure}{0.49\linewidth}
        \includegraphics[width=\linewidth]{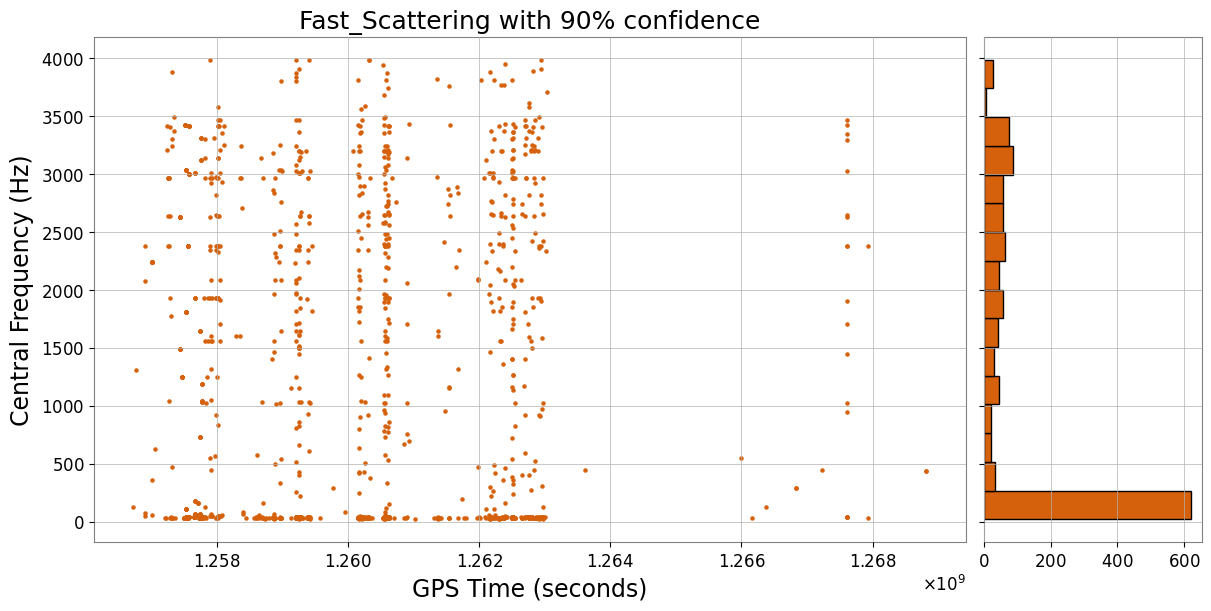}
    \end{subfigure}

    \begin{subfigure}{0.49\linewidth}
        \includegraphics[width=\linewidth]{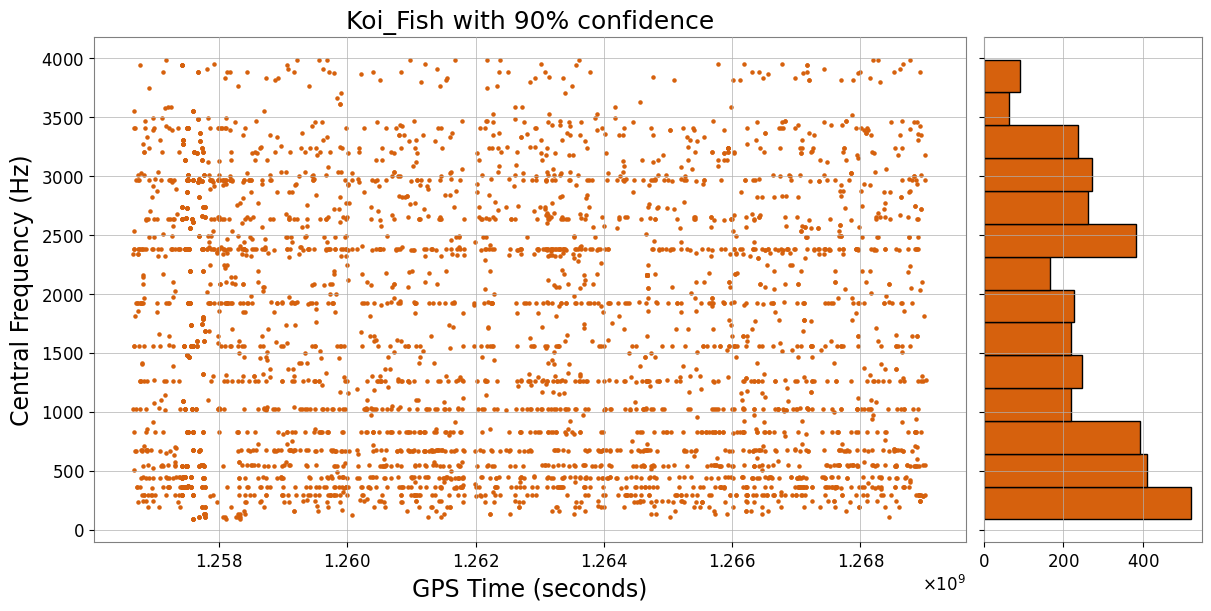}
    \end{subfigure}
    \begin{subfigure}{0.49\linewidth}
        \includegraphics[width=\linewidth]{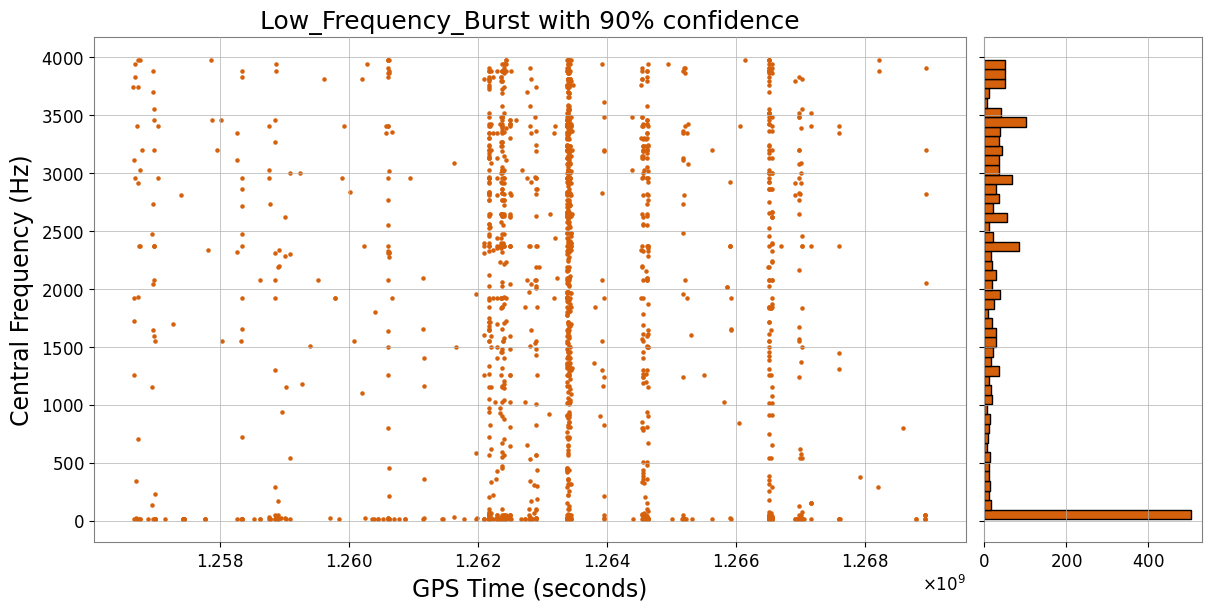}
    \end{subfigure}

    \begin{subfigure}{0.49\linewidth}
        \includegraphics[width=\linewidth]{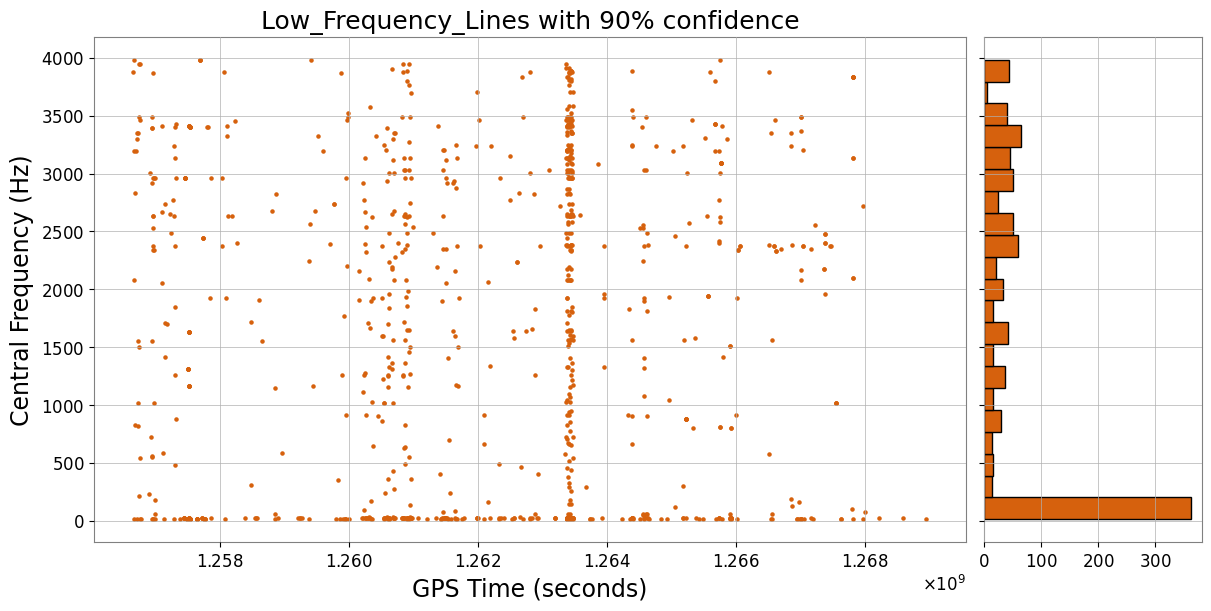}
    \end{subfigure}
    \begin{subfigure}{0.49\linewidth}
        \includegraphics[width=\linewidth]{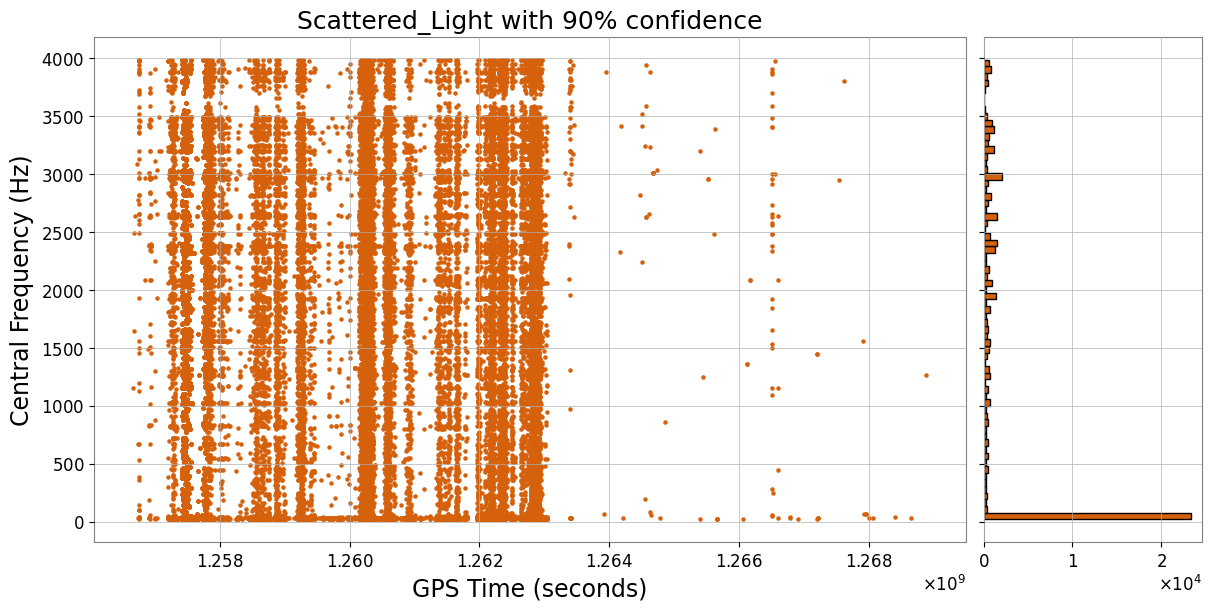}
    \end{subfigure}

    \begin{subfigure}{0.49\linewidth}
        \includegraphics[width=\linewidth]{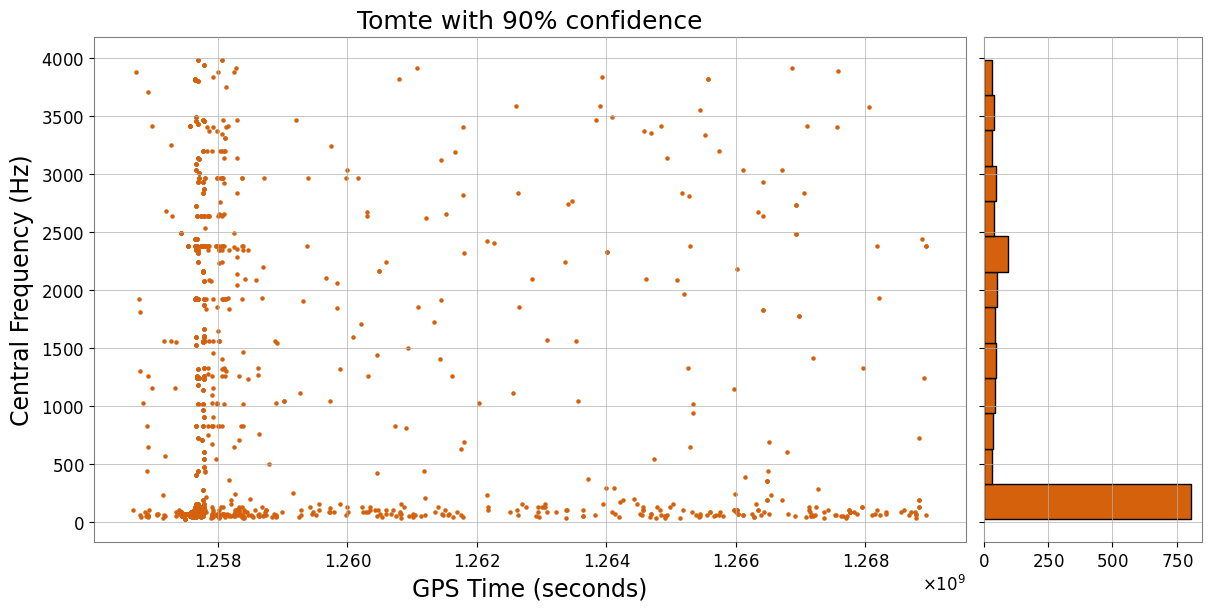}
    \end{subfigure}
    \begin{subfigure}{0.49\linewidth}
        \includegraphics[width=\linewidth]{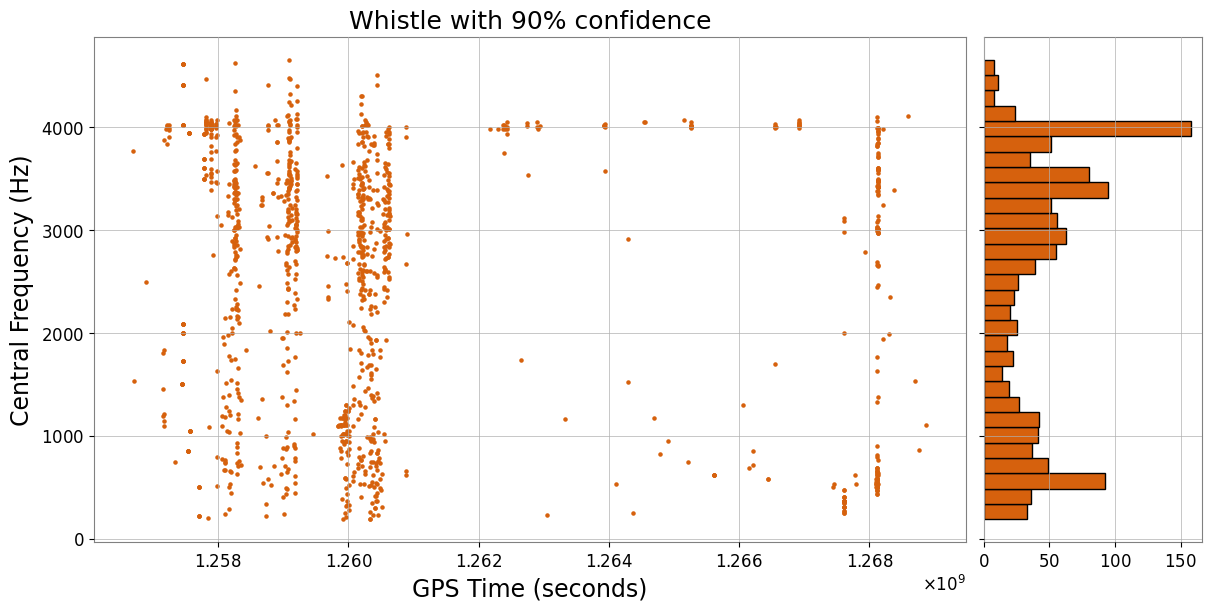}
    \end{subfigure}

    \caption{Frequency distributions of different glitch classes observed in the Hanford interferometer during O3b.}
    \label{fig:freqO3bHanford}
\end{figure}

\section{Histograms}
\label{histo}

\begin{figure}[H]
    \centering
    \begin{minipage}{0.49\linewidth}
        \includegraphics[width=\linewidth]{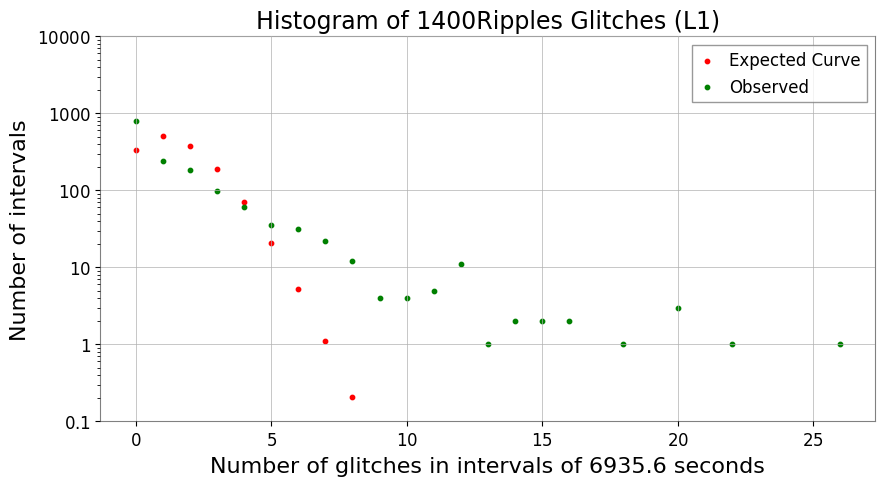}
    \end{minipage}
    \begin{minipage}{0.49\linewidth}
        \includegraphics[width=\linewidth]{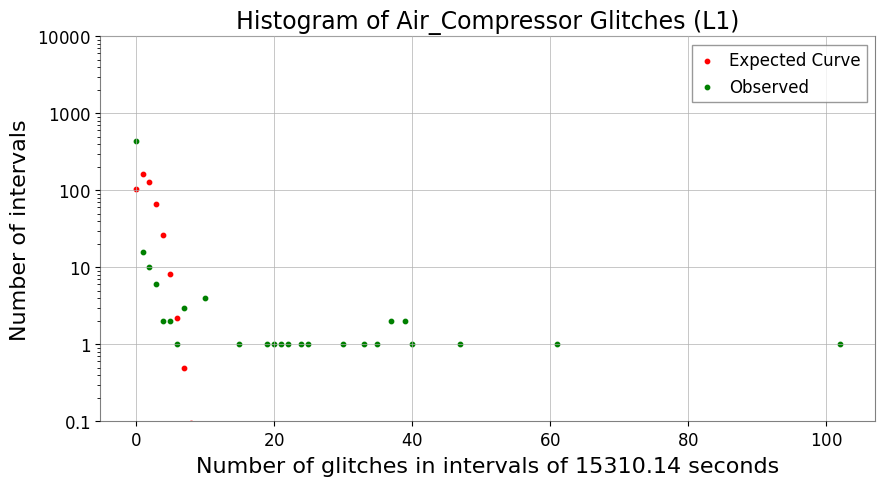}
    \end{minipage}
    \caption{Histograms of Ripples (left) and Air Compressor (right) glitches during O3a–O3b (L1).}
    \label{fig:ripples_aircompressor}
\end{figure}

\begin{figure}[H]
    \centering
    \begin{minipage}{0.49\linewidth}
        \includegraphics[width=\linewidth]{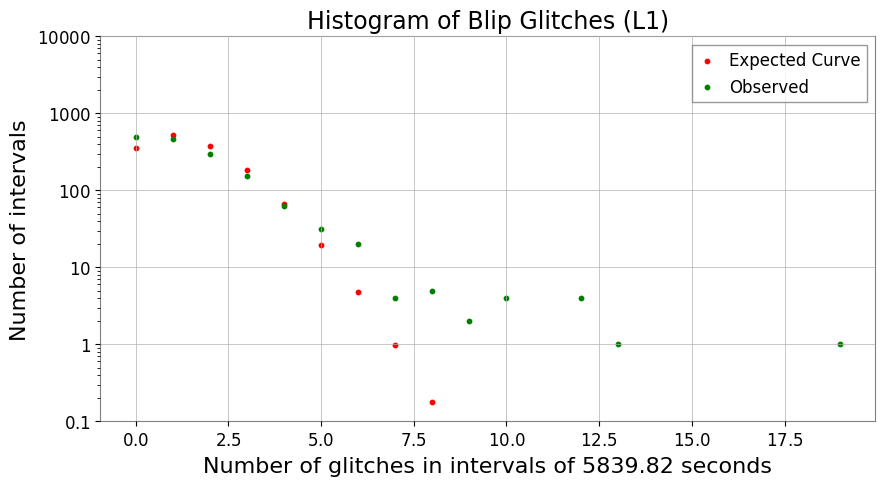}
    \end{minipage}
    \begin{minipage}{0.49\linewidth}
        \includegraphics[width=\linewidth]{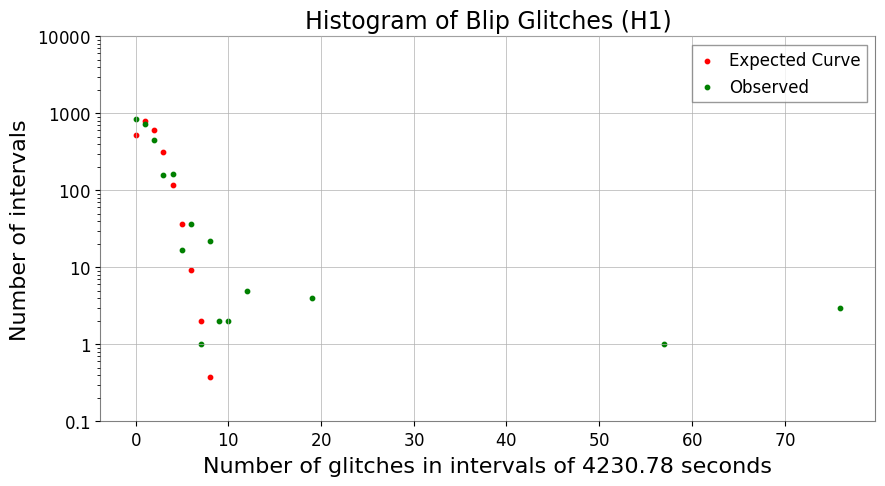}
    \end{minipage}
    \caption{Histograms of Blip glitches during O3b (L1, left) and O3a (H1, right).}
    \label{fig:blip_l1h1}
\end{figure}

\begin{figure}[H]
    \centering
    \begin{minipage}{0.49\linewidth}
        \includegraphics[width=\linewidth]{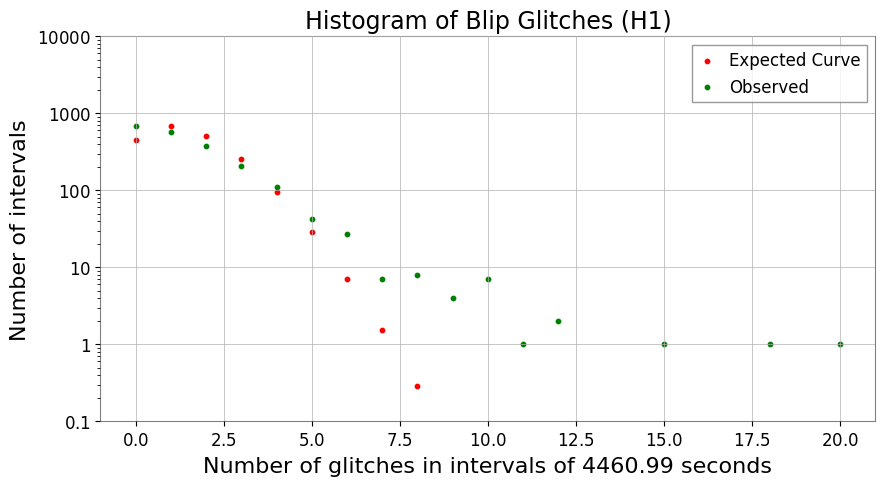}
    \end{minipage}
    \begin{minipage}{0.49\linewidth}
        \includegraphics[width=\linewidth]{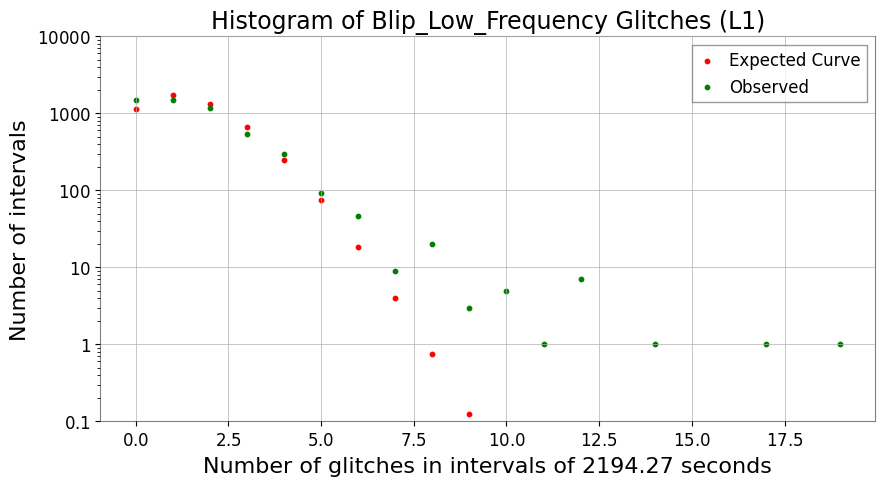}
    \end{minipage}
    \caption{Histograms of Blip (left) and Blip Low Frequency (right) glitches during O3b–O3a.}
    \label{fig:blip_bliplf_1}
\end{figure}

\begin{figure}[H]
    \centering
    \begin{minipage}{0.49\linewidth}
        \includegraphics[width=\linewidth]{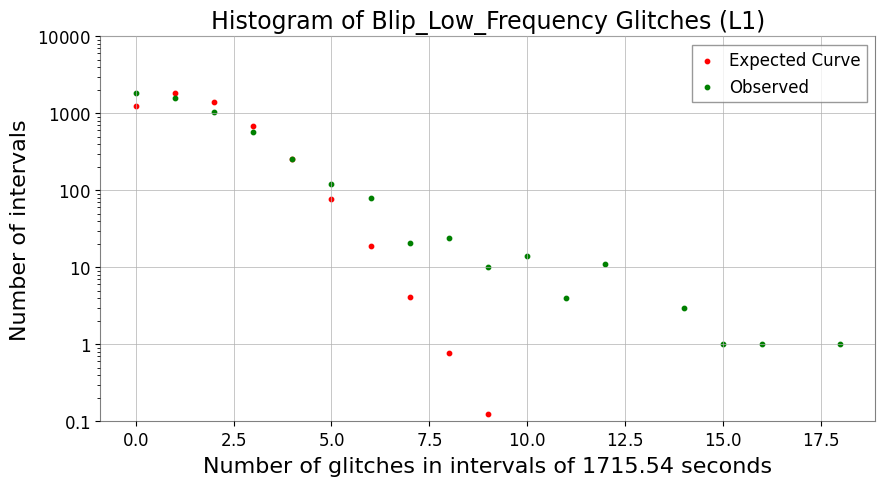}
    \end{minipage}
    \begin{minipage}{0.49\linewidth}
        \includegraphics[width=\linewidth]{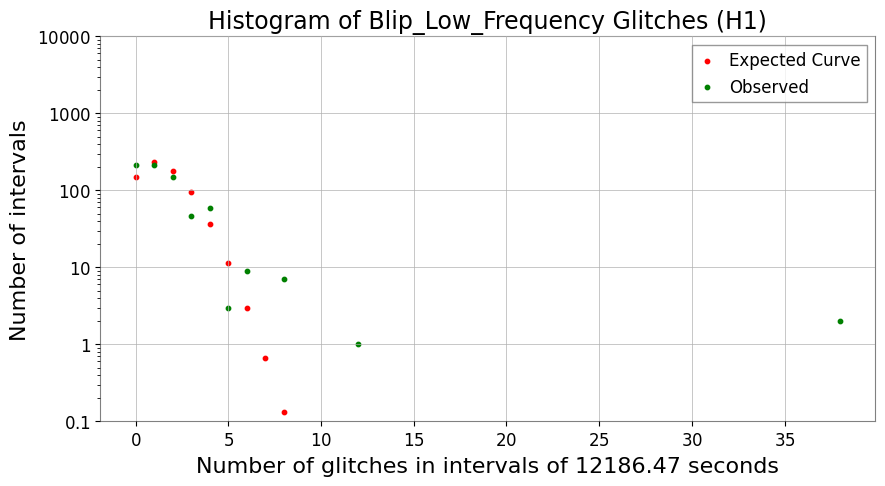}
    \end{minipage}
    \caption{Histograms of Blip Low Frequency glitches during O3b (L1, left) and O3a (H1, right).}
    \label{fig:bliplf_l1h1}
\end{figure}

\begin{figure}[H]
    \centering
    \begin{minipage}{0.49\linewidth}
        \includegraphics[width=\linewidth]{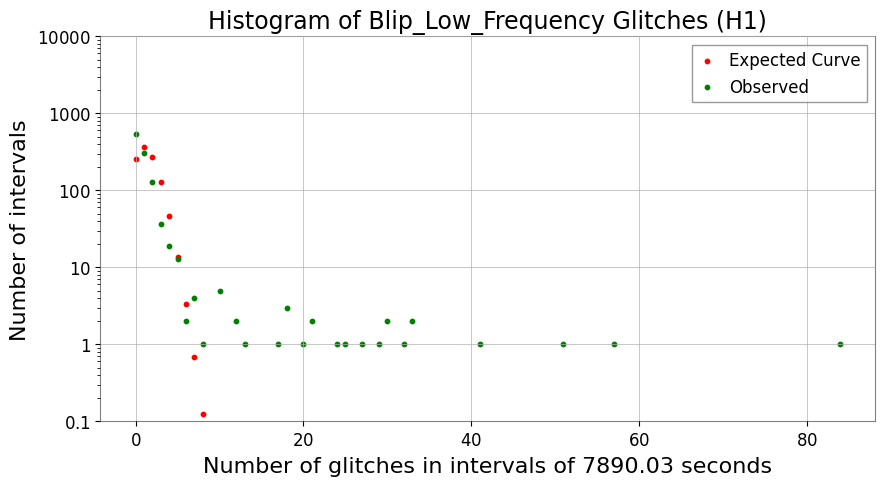}
    \end{minipage}
    \begin{minipage}{0.49\linewidth}
        \includegraphics[width=\linewidth]{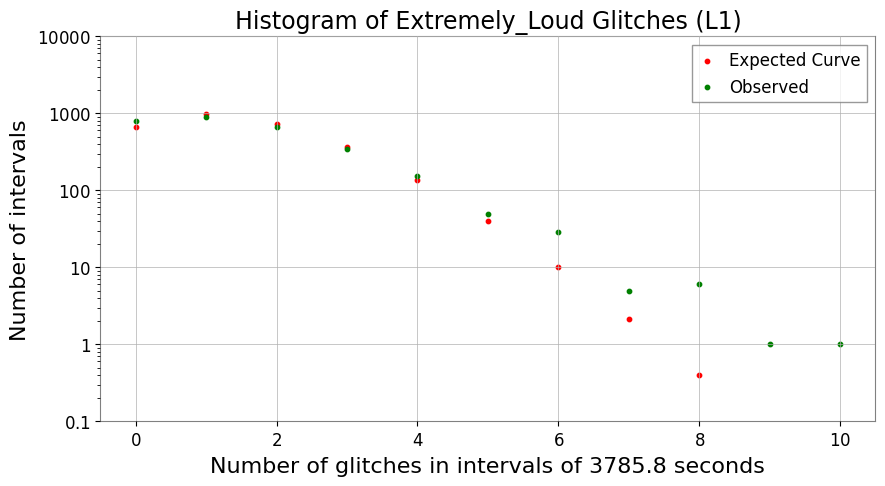}
    \end{minipage}
    \caption{Histograms of Blip Low Frequency (left) and Extremely Loud (right) glitches during O3b–O3a.}
    \label{fig:bliplf_extremely}
\end{figure}

\begin{figure}[H]
    \centering
    \begin{minipage}{0.49\linewidth}
        \includegraphics[width=\linewidth]{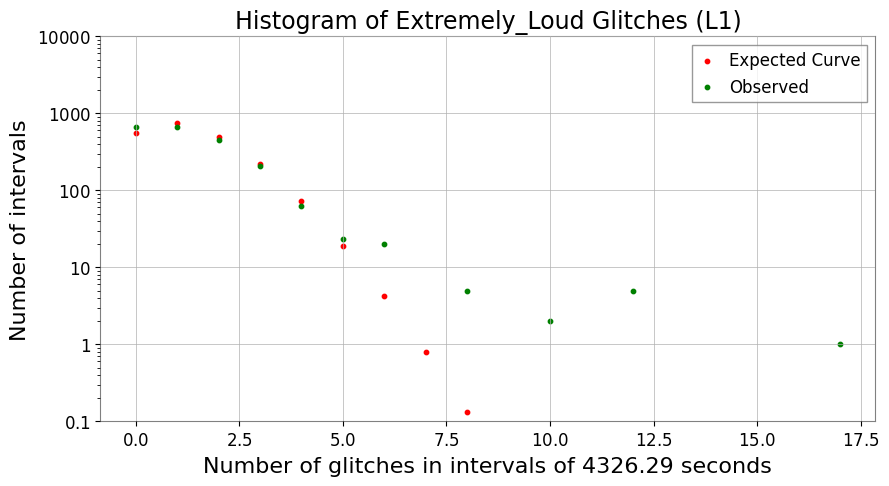}
    \end{minipage}
    \begin{minipage}{0.49\linewidth}
        \includegraphics[width=\linewidth]{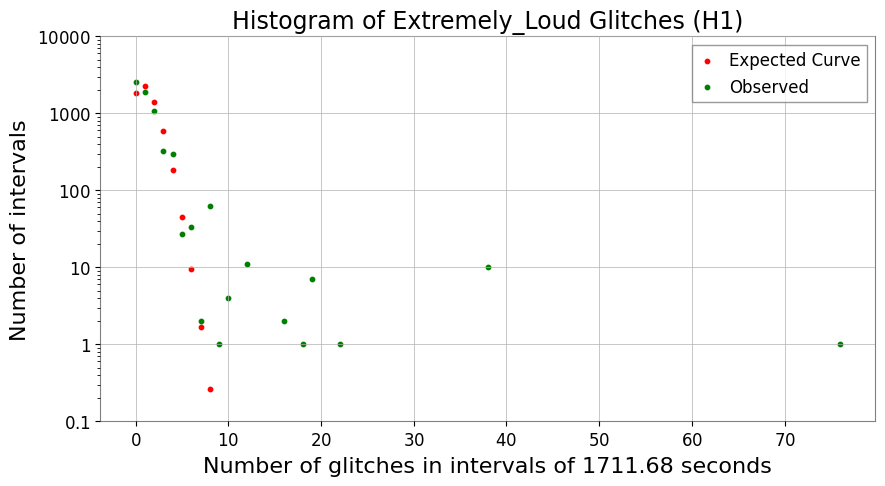}
    \end{minipage}
    \caption{Histograms of Extremely Loud glitches during O3b (L1, left) and O3a (H1, right).}
    \label{fig:extremely_l1h1}
\end{figure}

\begin{figure}[H]
    \centering
    \begin{minipage}{0.49\linewidth}
        \includegraphics[width=\linewidth]{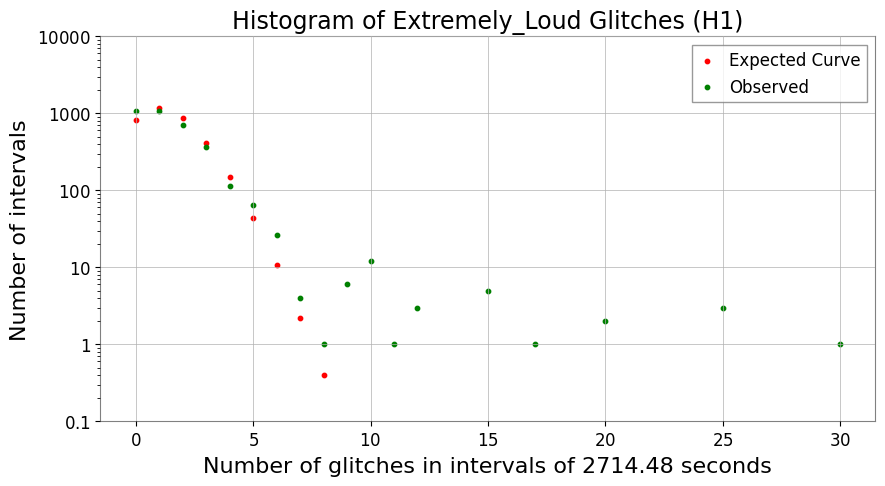}
    \end{minipage}
    \begin{minipage}{0.49\linewidth}
        \includegraphics[width=\linewidth]{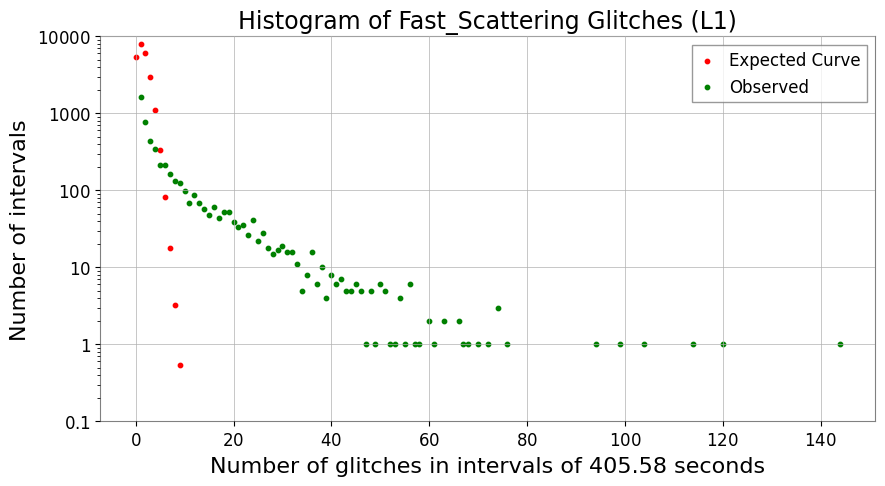}
    \end{minipage}
    \caption{Histograms of Extremely Loud (left) and Fast Scattering (right) glitches during O3b.}
    \label{fig:extremely_fast}
\end{figure}

\begin{figure}[H]
    \centering
    \begin{minipage}{0.49\linewidth}
        \includegraphics[width=\linewidth]{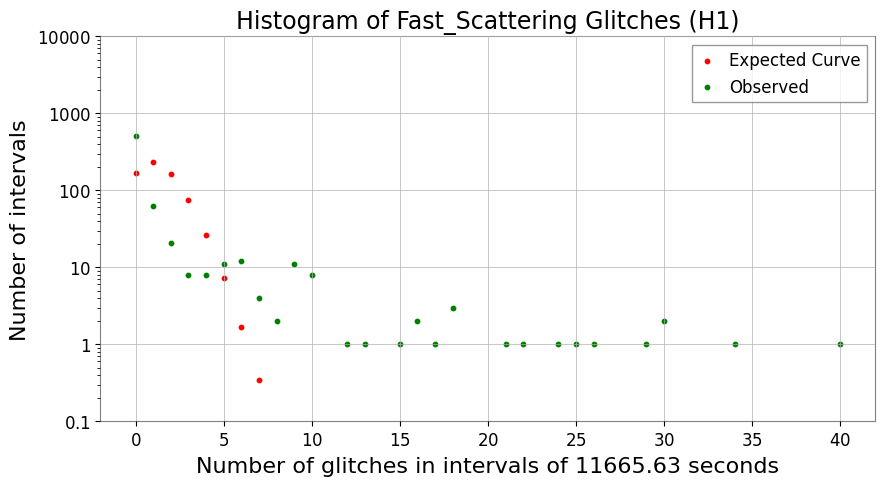}
    \end{minipage}
    \begin{minipage}{0.49\linewidth}
        \includegraphics[width=\linewidth]{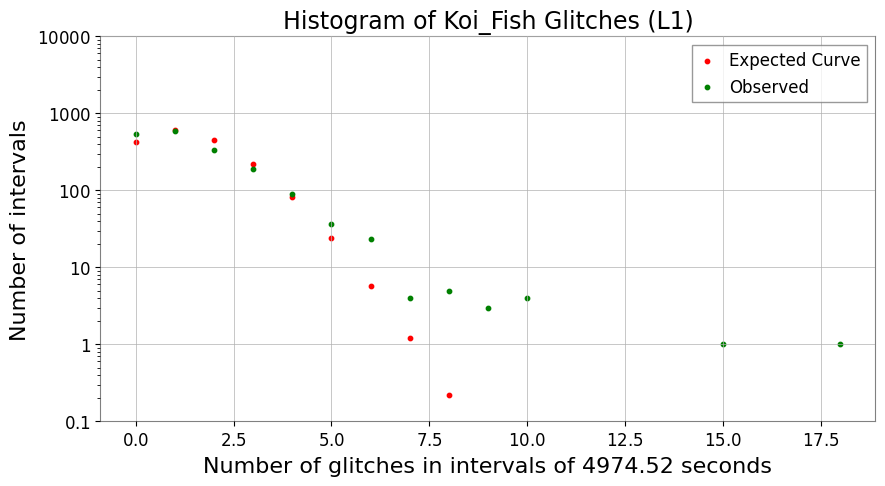}
    \end{minipage}
    \caption{Histograms of Fast Scattering (left) and Koi Fish (right) glitches during O3b.}
    \label{fig:fast_koi}
\end{figure}

\begin{figure}[H]
    \centering
    \begin{minipage}{0.49\linewidth}
        \includegraphics[width=\linewidth]{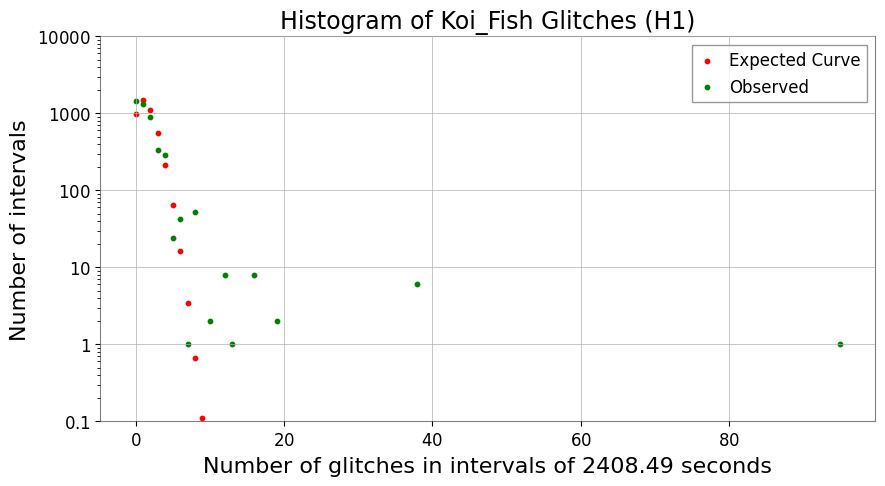}
    \end{minipage}
    \begin{minipage}{0.49\linewidth}
        \includegraphics[width=\linewidth]{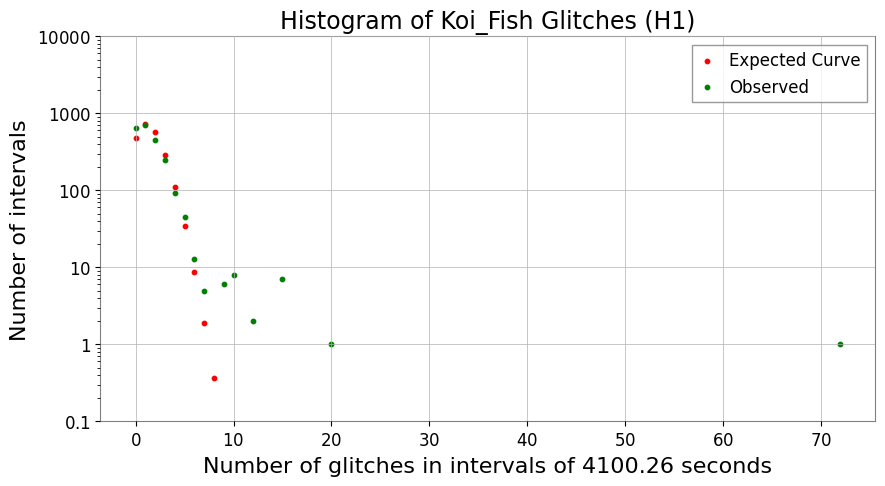}
    \end{minipage}
    \caption{Histograms of Koi Fish glitches during O3a (H1, left) and O3b (H1, right).}
    \label{fig:koi_h1}
\end{figure}

\begin{figure}[H]
    \centering
    \begin{minipage}{0.49\linewidth}
        \includegraphics[width=\linewidth]{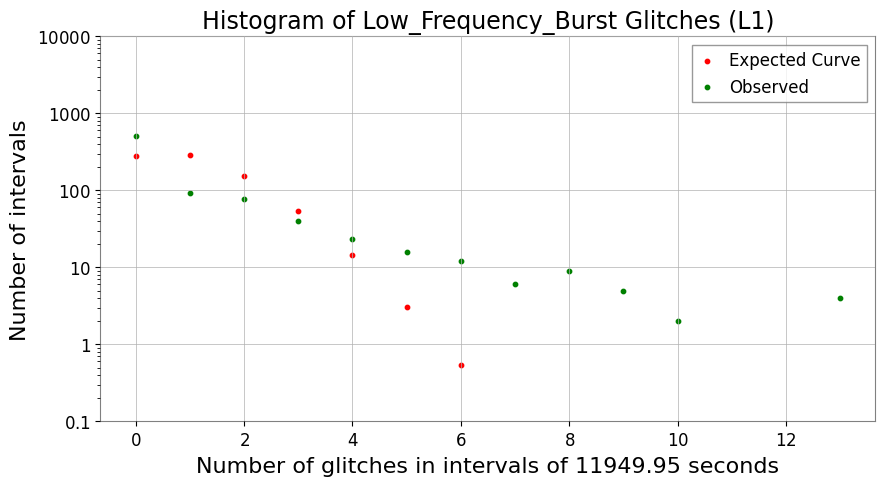}
    \end{minipage}
    \begin{minipage}{0.49\linewidth}
        \includegraphics[width=\linewidth]{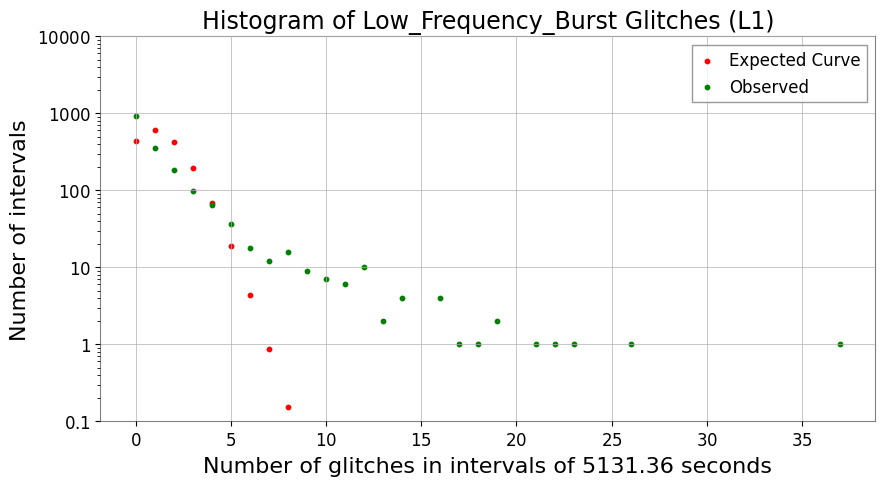}
    \end{minipage}
    \caption{Histograms of Low Frequency Burst glitches during O3a–O3b (L1).}
    \label{fig:burst_l1}
\end{figure}

\begin{figure}[H]
    \centering
    \begin{minipage}{0.49\linewidth}
        \includegraphics[width=\linewidth]{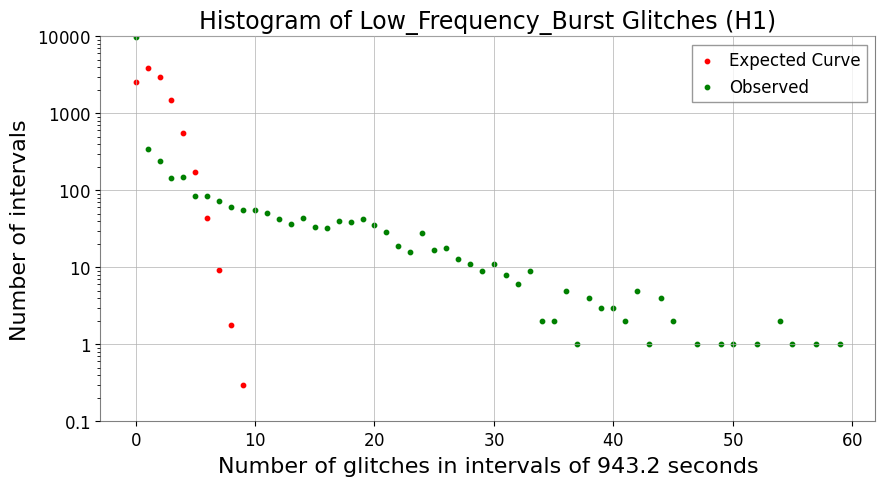}
    \end{minipage}
    \begin{minipage}{0.49\linewidth}
        \includegraphics[width=\linewidth]{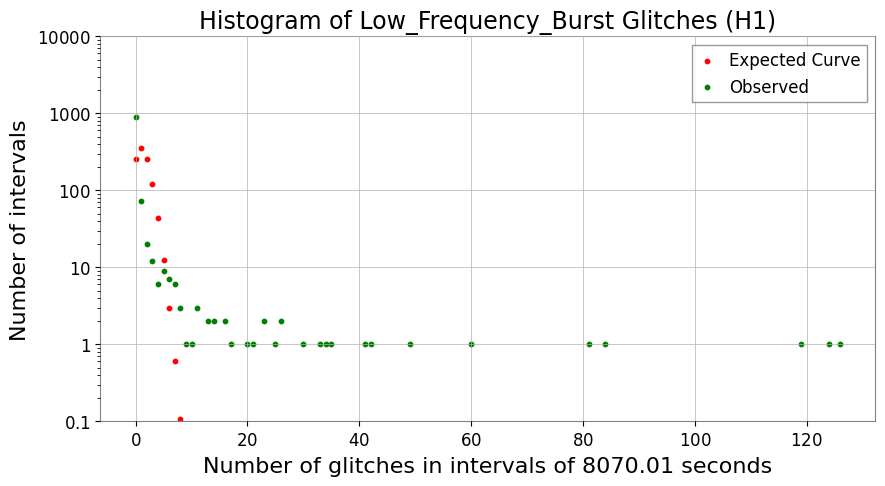}
    \end{minipage}
    \caption{Histograms of Low Frequency Burst glitches during O3a–O3b (H1).}
    \label{fig:burst_h1}
\end{figure}

\begin{figure}[H]
    \centering
    \begin{minipage}{0.49\linewidth}
        \includegraphics[width=\linewidth]{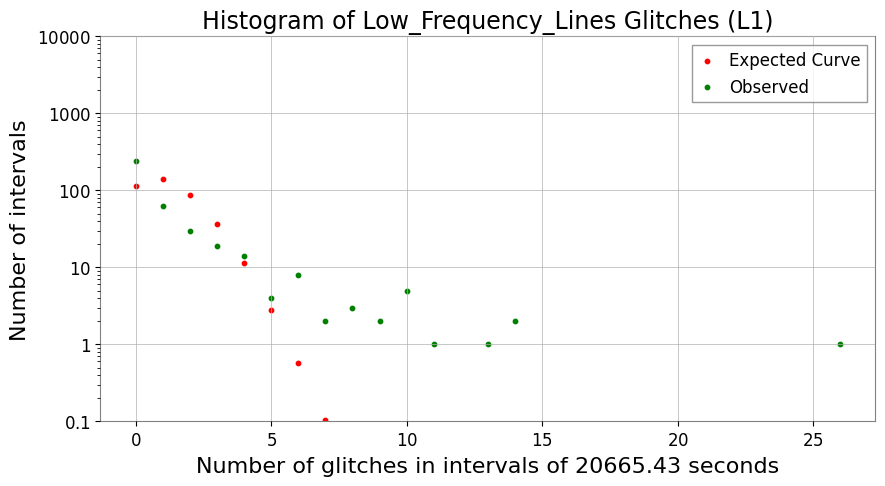}
    \end{minipage}
    \begin{minipage}{0.49\linewidth}
        \includegraphics[width=\linewidth]{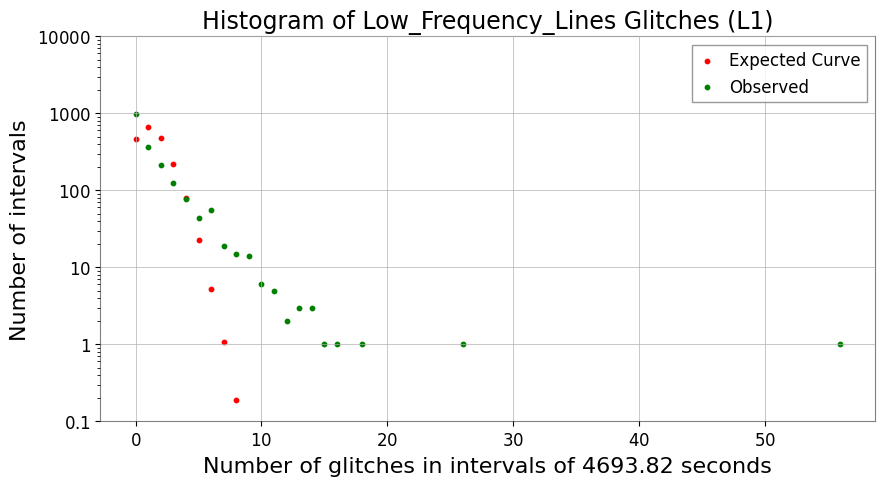}
    \end{minipage}
    \caption{Histograms of Low Frequency Lines glitches during O3a–O3b (L1).}
    \label{fig:lines_l1}
\end{figure}

\begin{figure}[H]
    \centering
    \begin{minipage}{0.49\linewidth}
        \includegraphics[width=\linewidth]{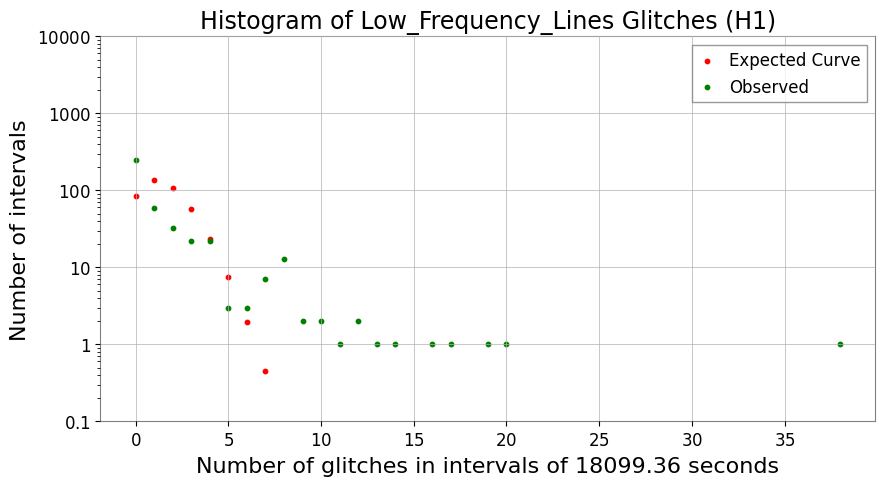}
    \end{minipage}
    \begin{minipage}{0.49\linewidth}
        \includegraphics[width=\linewidth]{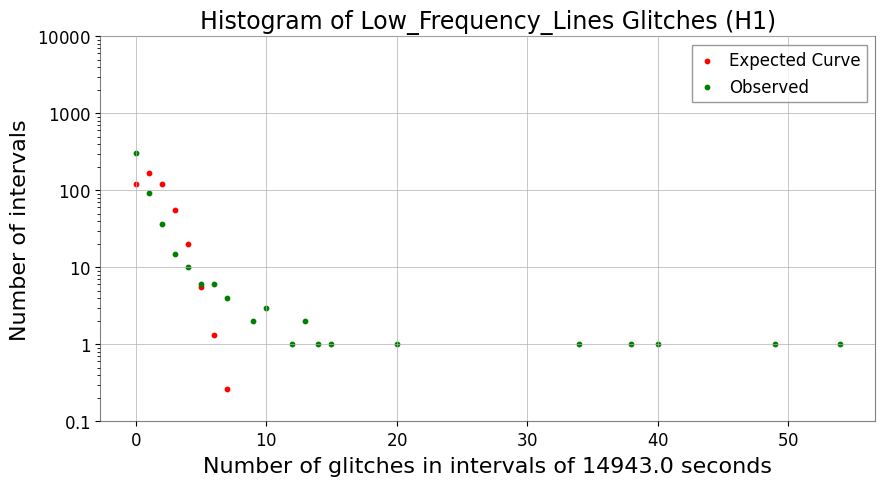}
    \end{minipage}
    \caption{Histograms of Low Frequency Lines glitches during O3a–O3b (H1).}
    \label{fig:lines_h1}
\end{figure}

\begin{figure}[H]
    \centering
    \begin{minipage}{0.49\linewidth}
        \includegraphics[width=\linewidth]{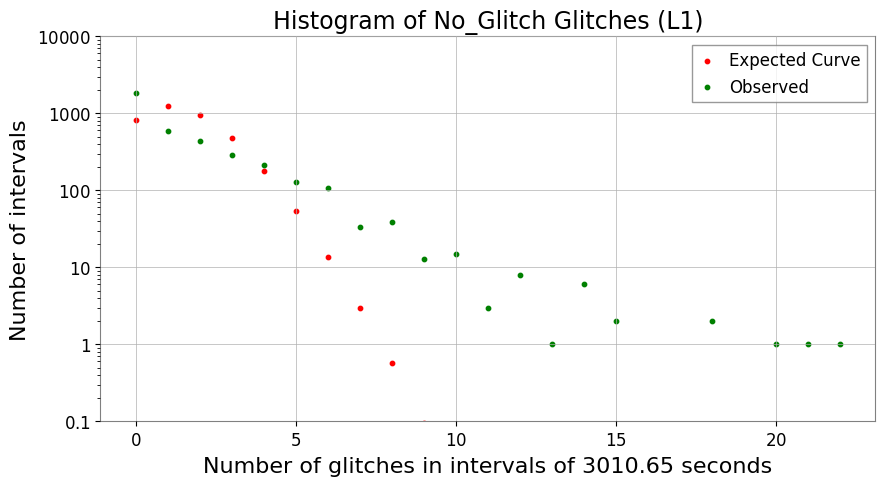}
    \end{minipage}
    \begin{minipage}{0.49\linewidth}
        \includegraphics[width=\linewidth]{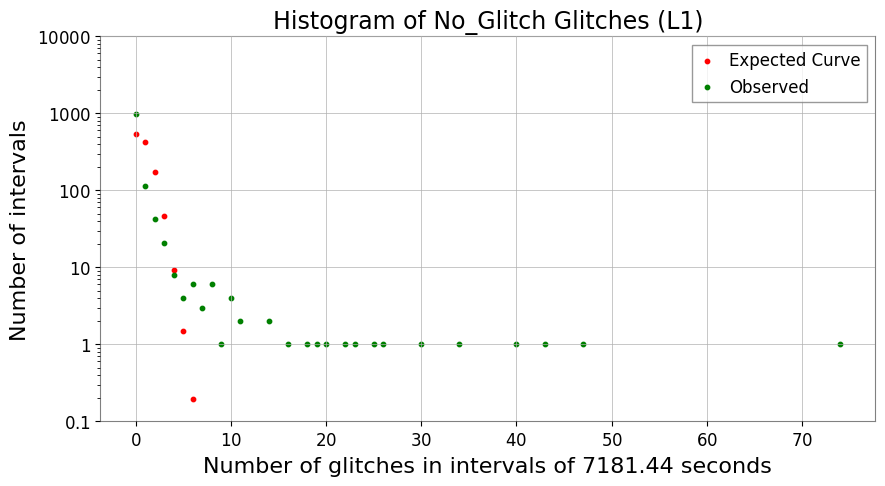}
    \end{minipage}
    \caption{Histograms of No Glitch counts during O3a–O3b (L1).}
    \label{fig:noglitch_l1}
\end{figure}

\begin{figure}[H]
    \centering
    \begin{minipage}{0.49\linewidth}
        \includegraphics[width=\linewidth]{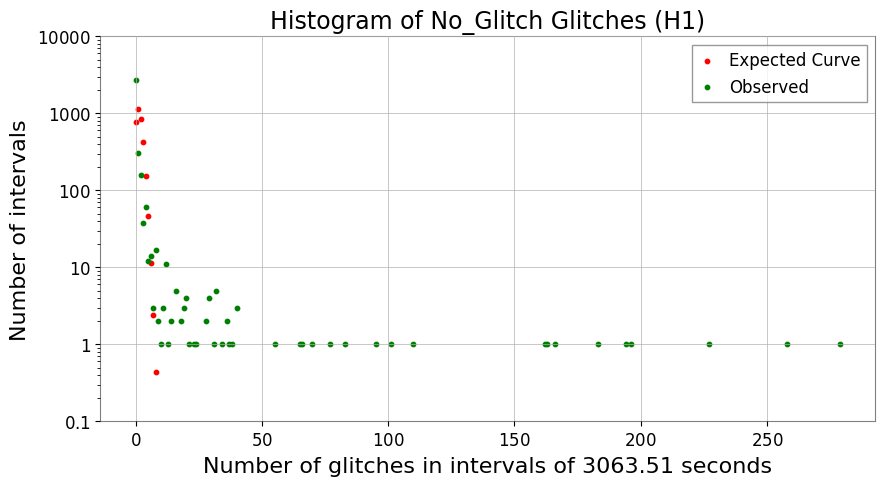}
    \end{minipage}
    \begin{minipage}{0.49\linewidth}
        \includegraphics[width=\linewidth]{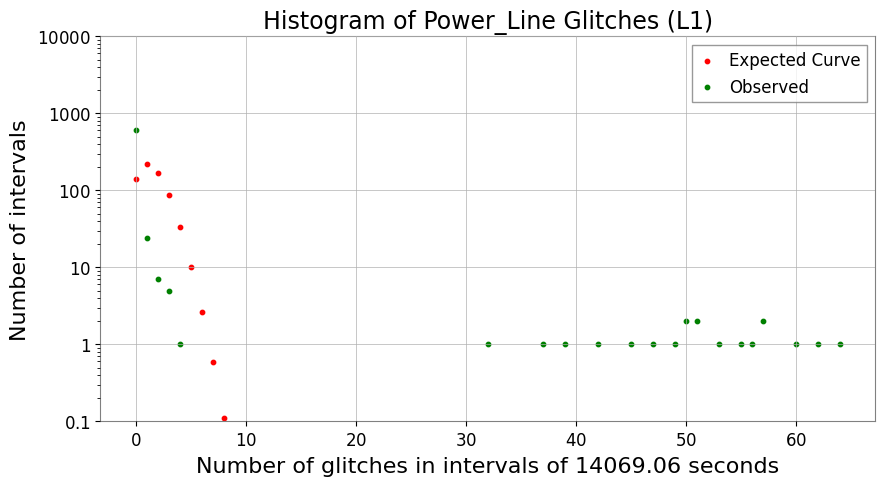}
    \end{minipage}
    \caption{Histograms of No Glitch (left) and Power Line (right) during O3a.}
    \label{fig:noglitch_powerline}
\end{figure}

\begin{figure}[H]
    \centering
    \begin{minipage}{0.49\linewidth}
        \includegraphics[width=\linewidth]{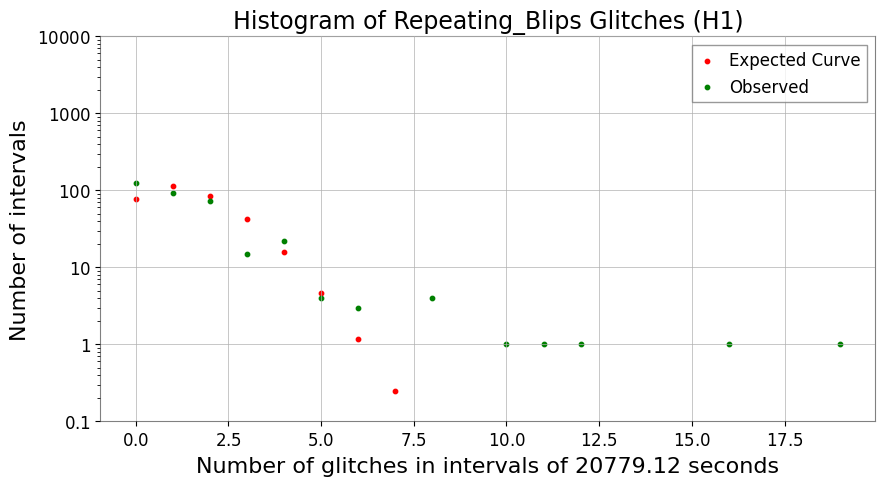}
    \end{minipage}
    \begin{minipage}{0.49\linewidth}
        \includegraphics[width=\linewidth]{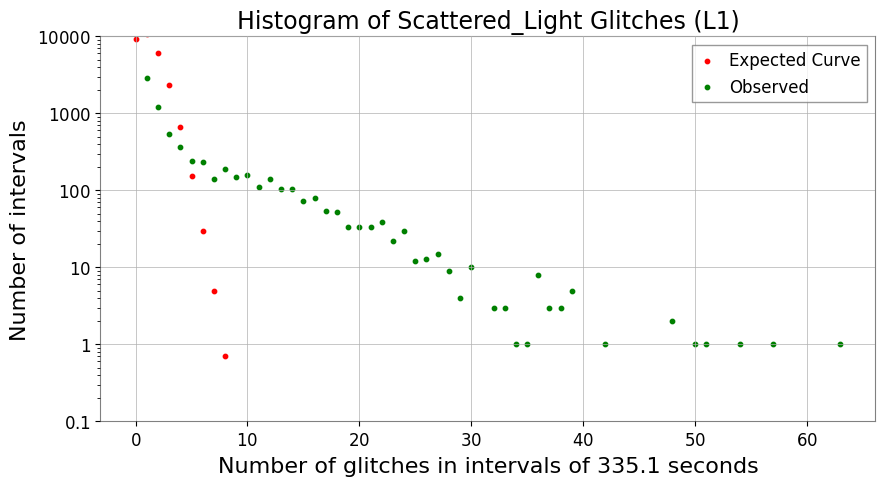}
    \end{minipage}
    \caption{Histograms of Repeating Blips (left) and Scattered Light (right) glitches.}
    \label{fig:repeating_scattered}
\end{figure}

\begin{figure}[H]
    \centering
    \begin{minipage}{0.49\linewidth}
        \includegraphics[width=\linewidth]{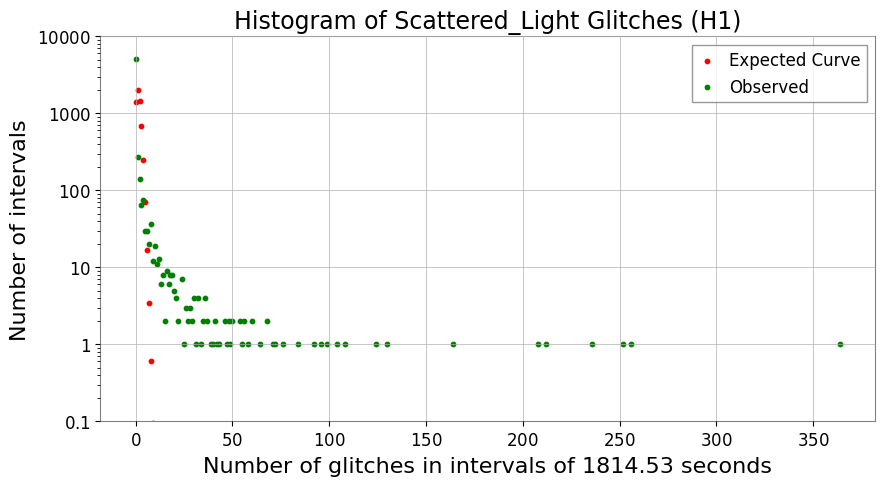}
    \end{minipage}
    \begin{minipage}{0.49\linewidth}
        \includegraphics[width=\linewidth]{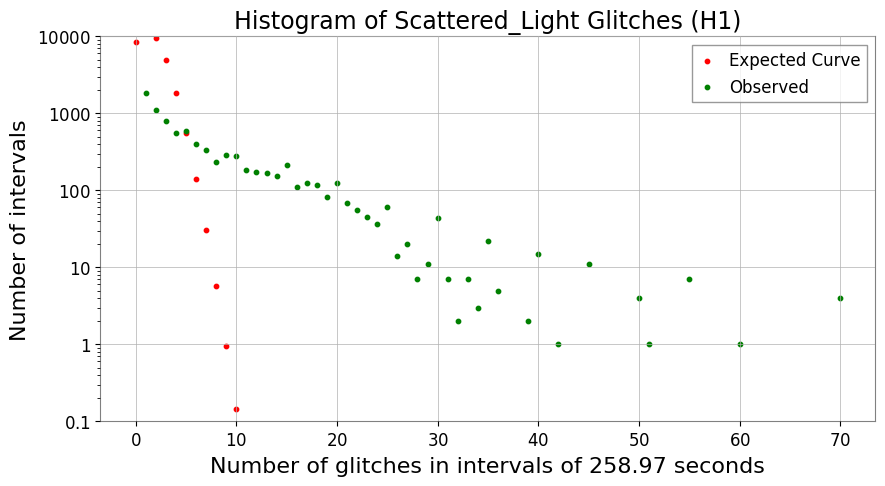}
    \end{minipage}
    \caption{Histograms of Scattered Light glitches during O3a–O3b (H1).}
    \label{fig:scattered_h1}
\end{figure}

\begin{figure}[H]
    \centering
    \begin{subfigure}{0.49\linewidth}
        \includegraphics[width=\linewidth]{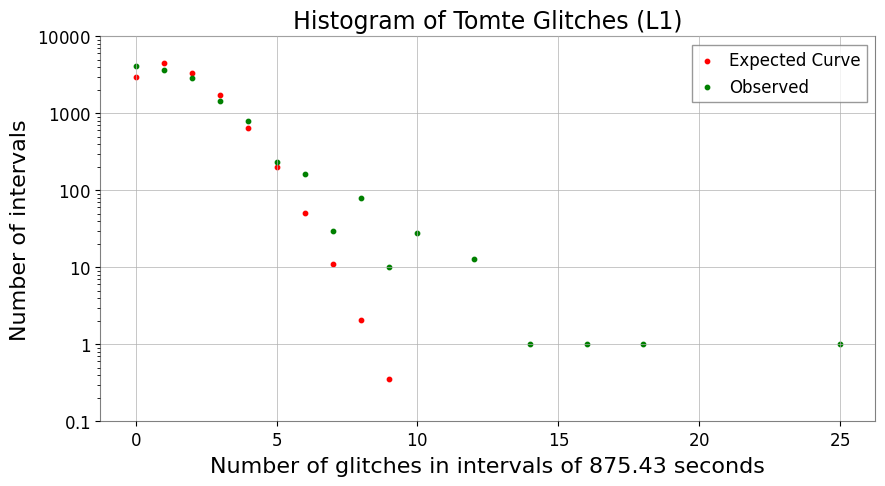}
    \end{subfigure}
    \begin{subfigure}{0.49\linewidth}
        \includegraphics[width=\linewidth]{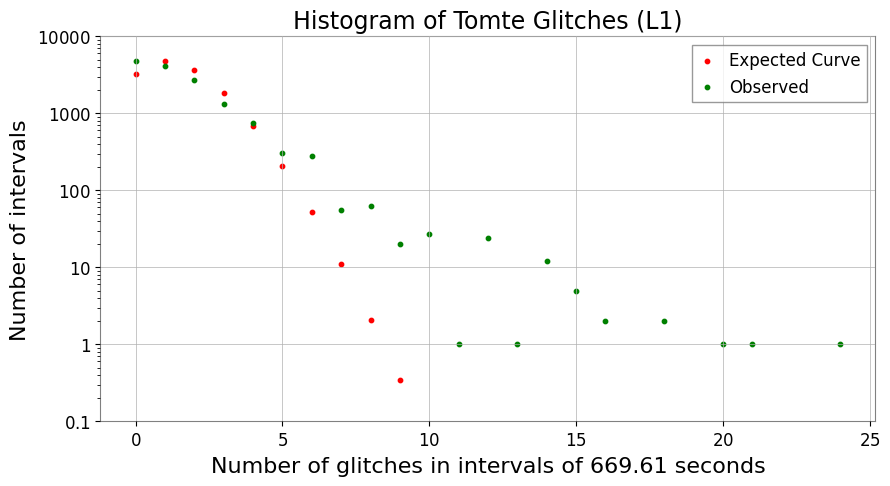}
    \end{subfigure}
    \caption{Histograms of Tomte glitch during O3a-O3b (L1).}
    \label{fig:tomte_l1}
\end{figure}

\begin{figure}[H]
    \centering
    \begin{subfigure}{0.49\linewidth}
        \includegraphics[width=\linewidth]{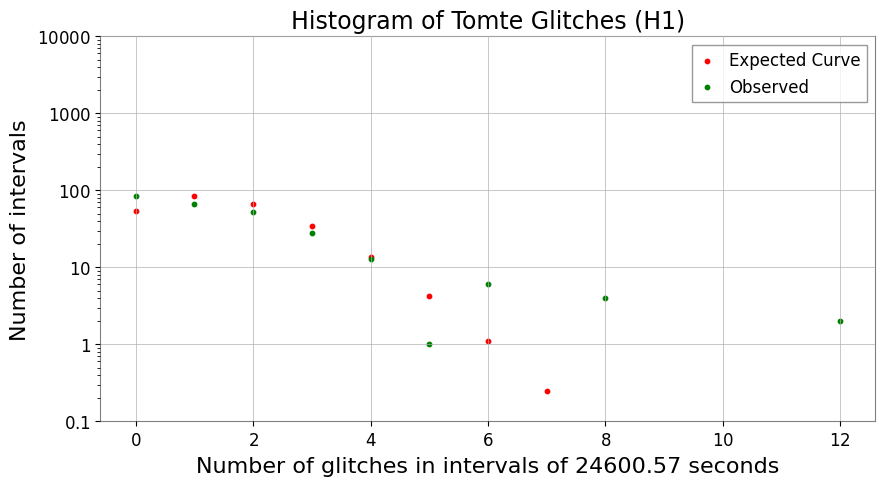}
    \end{subfigure}
    \begin{subfigure}{0.49\linewidth}
        \includegraphics[width=\linewidth]{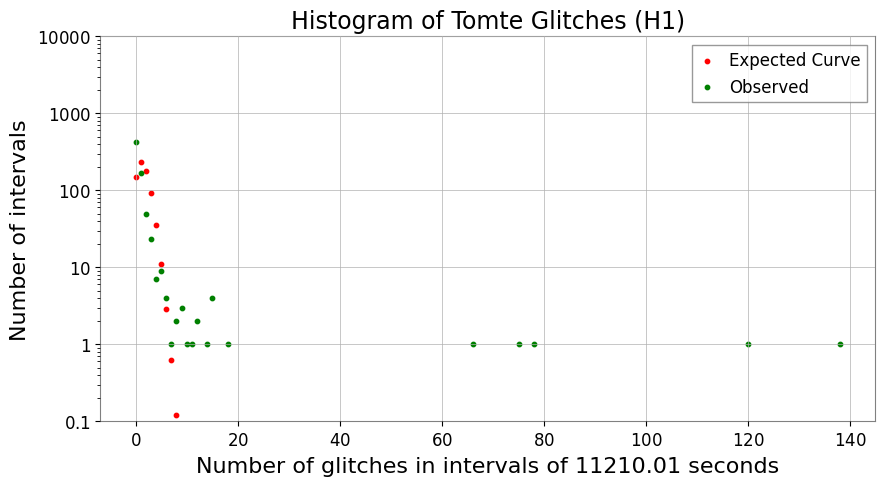}
    \end{subfigure}
    \caption{Histograms of Tomte glitch during O3a-O3b (H1).}
    \label{fig:tomte_h1}
\end{figure}

\begin{figure}[H]
    \centering
    \begin{subfigure}{0.49\linewidth}
        \includegraphics[width=\linewidth]{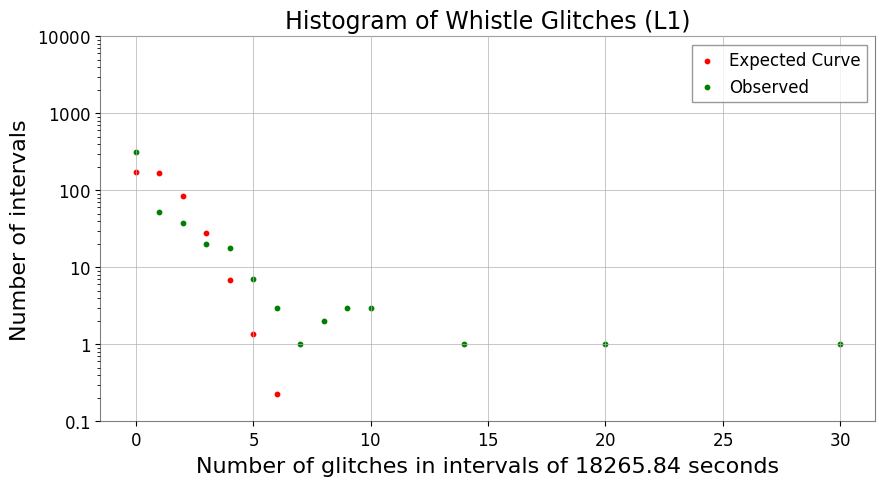}
    \end{subfigure}
    \begin{subfigure}{0.49\linewidth}
        \includegraphics[width=\linewidth]{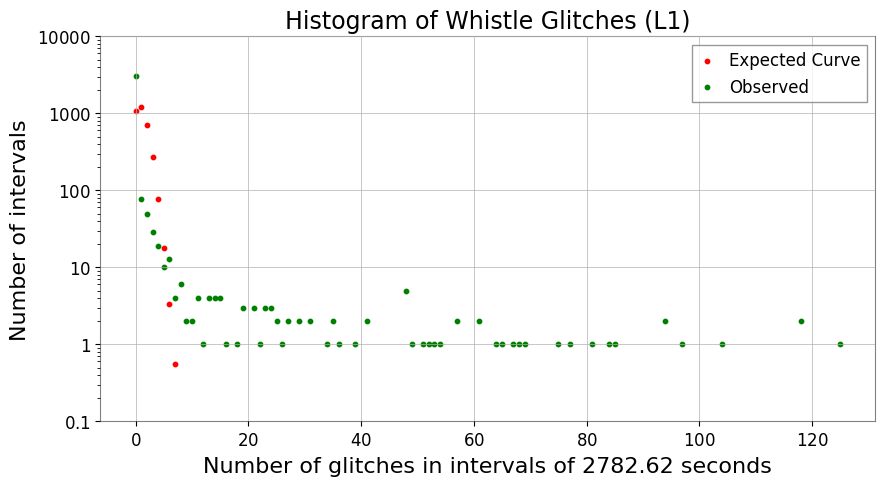}
    \end{subfigure}
    \caption{Histograms of Whistle glitch during O3a-O3b (L1).}
    \label{fig:whistle_l1}
\end{figure}

\begin{figure}[H]
    \centering
    \begin{subfigure}{0.49\linewidth}
        \includegraphics[width=\linewidth]{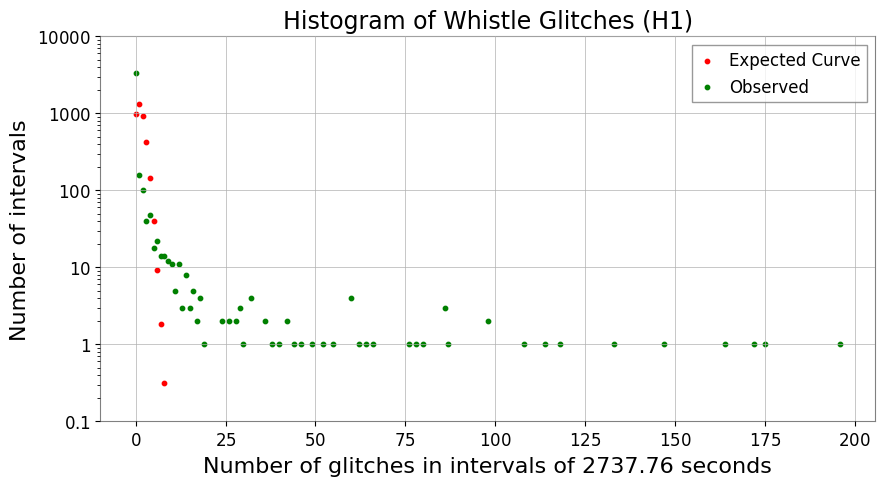}
    \end{subfigure}
    \begin{subfigure}{0.49\linewidth}
        \includegraphics[width=\linewidth]{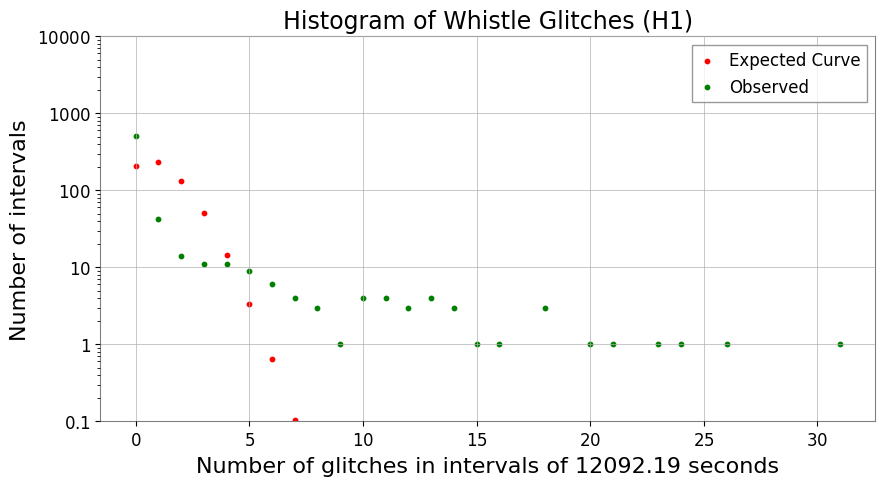}
    \end{subfigure}
    \caption{Histograms of Whistle glitch during O3a-O3b (H1).}
    \label{fig:whistle_h1}
\end{figure}

\nocite{*}
\bibliographystyle{iopart-num}
\bibliography{areferences}

\end{document}